\documentclass[structabstract]{aa}  
\usepackage{graphicx}
\usepackage{txfonts}
\usepackage[T1]{fontenc}
\usepackage{natbib}
\usepackage{hyperref}
\usepackage{siunitx}  
\usepackage{xcolor}
\usepackage{booktabs}   
\usepackage{placeins}   
\usepackage[safe]{tipa}

\newcommand{\mci}[1]{\multicolumn{1}{c}{#1}}
\newcommand{\mcii}[1]{\multicolumn{2}{c}{#1}}

\newcommand{\mciv}[1]{\multicolumn{4}{c}{#1}}

\newcommand{\Teff}{\mbox{$T_{\rm eff}$}}
\newcommand{\logg}{\mbox{$\log g$}}
\newcommand{\logQ}{\mbox{$\log Q$}}

\newcommand{\vsini}{\mbox{$v\,\sin i$}}

\newcommand{\Msun}{\mbox{M$_\odot$}}
\newcommand{\Rsun}{\mbox{R$_\odot$}}

\newcommand{\logL}{\mbox{$\log (L/$L$_\odot)$}}
\newcommand{\logLL}{\mbox{$\log (\mathcal{L}/\mathcal{L}_\odot)$}}
\newcommand{\Mspec}{\mbox{$M_{\rm spec}$}}
\newcommand{\Mevol}{\mbox{$M_{\rm evol}$}}

\newcommand{\MTW}{M$^3$W}

\newcommand{\cre}{\color{red}}
\newcommand{\cgr}{\color{green}}
\newcommand{\cbl}{\color{blue}}
\newcommand{\cog}{\color{orange}}
\newcommand{\cma}{\color{magenta}}
\newcommand{\lili}{LiLiMaRlin}
\newcommand{\Gaia}{\textit{Gaia}}

\begin{document}

   \title{Multiplicity of Massive stars in the Milky Way (\MTW)}
   \subtitle{I. Project description, UNWIND, application \linebreak
             to GLS~\num[detect-all]{11448} (= LS~III~+46~11), and DIB catalog}
   \titlerunning{\MTW-I. Description, UNWIND, application to GLS~\num[detect-all]{11448}, and DIB catalog}
   \authorrunning{Ma\'{\i}z Apell\'aniz et al.}

   \author{J. Ma\'{\i}z Apell\'aniz\inst{1}
          \and
          R. C. Gamen\inst{2,3}
          \and
          G. Holgado\inst{4,5}
          \and
          S. Rosu\inst{6}
          \and
          J. I. Arias\inst{7}
          \and
          S. Sim\'on-D\'{\i}az\inst{4,5}
          \and
          A. Pellerin\inst{8}
          \and
          \\
          M. Abdul-Masih\inst{4,5}
          \and
          E. Madero Fuentes\inst{1,9}
          \and
          J. A. Molina-Calzada\inst{1,9}
          \and 
          R. H. Barb\'a\thanks{Deceased.}
          }

   \institute{Centro de Astrobiolog\'{\i}a, CSIC-INTA. Campus ESAC. 
              C. bajo del castillo s/n. 
              E-\num[detect-all]{28692} Villanueva de la Ca\~nada, Madrid, Spain.\\
              \email{jmaiz@cab.inta-csic.es}
         \and
              Instituto de Astrof\'isica de La Plata, CONICET--UNLP.
              Paseo del Bosque s/n.
              La Plata, Argentina.
         \and
              Facultad de Ciencias Astron\'omicas y Geof\'isicas, UNLP. 
              Paseo del Bosque s/n. 
              La Plata, Argentina.
         \and
              Instituto de Astrof\'isica de Canarias. 
              E-\num[detect-all]{38200} La Laguna, Tenerife, Spain.
         \and
              Departamento de Astrof\'isica, Universidad de La Laguna.
              E-\num[detect-all]{38205} La Laguna, Tenerife, Spain.
         \and
              D\'epartement d'Astronomie, Universit\'e de Gen\`eve.
              Chemin Pegasi 51.
              CH-1290, Versoix, Switzerland.
         \and
              Departamento de Astronom\'ia, Universidad de La Serena,
              Av. Juan Cisternas 1200 Norte.
              La Serena, Chile.
         \and
              Departament of Physics and Astronomy, SUNY Geneseo.
              Geneseo, NY \num{14454}, United States of America.
         \and
              Departamento de Astrof{\'\i}sica y F{\'\i}sica de la Atm\'osfera. 
              Universidad Complutense de Madrid. 
              E-\num[detect-all]{28040} Madrid, Spain. 
             }

   \date{Received 2 April 2026; accepted 17 July 2026}

  \abstract
   {Multiplicity is ubiquitous among massive stars and its understanding is constrained by the limited sample of well-determined 
    orbits. This hampers the progress of astronomical fields ranging from the initial mass function to 
    gravitational-wave sources.}
   {The immediate goal of Multiplicity of Massive stars in the Milky Way (hereafter, \MTW) is to significantly increase the number 
    of massive multiple systems with well-determined orbits and masses. With that information in hand, we will address issues such 
    as multiplicity statistics, the mass function in clusters and the field, the properties of binaries with 
    compact companions and gravitational-wave progenitors, the origin and characteristics of runaways and their 3-D motions, 
    the use of apsidal motion as a probe of stellar interiors, and the mass discrepancy between different methods (evolutionary,
    spectroscopic, and Keplerian).}
   {In this first paper, we present the project; describe the data and tools that will be used,
    including the disentangling UNWIND tool; analyse the very massive twin binary system GLS~\num{11448}; and 
    describe different issues related to massive-star multiplicity that will be analysed in the series. 
    GLS~\num[detect-all]{11448} was chosen as the first object to be analysed for three reasons: the extreme high mass of their 
    components, the difficulties associated with its high extinction that can be addressed with our technique, and its status as an 
    ISM standard star.}
   {We present a new orbital solution for GLS~\num{11448}, using UNWIND to obtain for the first time disentangled spectra for the 
    full 3820-\num{11000}~\AA\ range for an OB spectroscopic binary. We derive the stellar parameters and present how UNWIND makes 
    new stellar lines available for the study of O stars. The Aa and Ab components of GLS~\num{11448}, both classified as
    O3.5~II(f*), are the two most massive O stars ever detected according to the evolutionary masses of $70\pm10$~\Msun\ and 
    $76\pm11$~\Msun\ determined in this paper. We also report the first-ever detection of the interstellar 
    \ion{He}{i}~$\lambda$\num{10830} triplet in absorption in an OB-star sightline. As a by-product of the 
    interstellar-medium model derived for UNWIND using GLS~\num{11448} and five other standard stars, we present the most detailed 
    diffuse-interstellar-band (DIB) library ever built, with a total of 631 DIBs in the 4000-\num{17100}~\AA\ range, of which 37 
    are fitted with multiple-Gaussian profiles and 
    116 had never been identified before.}
   {}


   \maketitle

\section{Introduction}

$\,\!$\indent It has been known for several decades that massive stars have a high degree of multiplicity. The first large-scale 
systematic study of massive-star multiplicity was that of \citet{Masoetal98}, see also \citet{Garmetal80,Abt83}.
\citet{Sotaetal14} showed that few (if any) massive binaries are born in single 
systems and \citet{Maizetal19b} that triple- and higher-order systems are even more common than regular binaries. Massive-star
multiplicity is a blessing because it allows to determine component masses and other characteristics, but is also a curse because it 
complicates the understanding of star formation and the determination of mass functions. Furthermore, a large fraction of stars 
in multiple systems are close enough to mutually influence their evolution \citep{Sanaetal12a} and, eventually, their end products, 
making massive-star multiplicity one of the determining factors in the dynamical and chemical evolution of galaxies.

Our group has been involved in massive-star surveys for the last two decades, starting with the Galactic O-Star Spectroscopic Survey 
(GOSSS, \citealt{Maizetal11}). In parallel, OWN began as a multi-epoch high-resolution spectroscopic survey of massive stars in 
the southern hemisphere \citep{Barbetal10,Barbetal17} specifically oriented towards spectroscopic multiplicity. In the northern
hemisphere several spectroscopic (NoMaDS, \citealt{Maizetal12}; CAF\'E-BEANS, \citealt{Neguetal15a}; IACOB, \citealt{SimDetal11a})
and high-resolution imaging \citep{Maiz10a} surveys converged into an OWN-equivalent project, MONOS (Multiplicity Of Northern O-type
Stars, \citealt{Maizetal19b}). The large amount of epochs generated by those projects coupled with the availability of
spectroscopic databases led to the creation of \lili\ (Library of Libraries of Massive-star high-Resolution spectra, 
\citealt{Maizetal19a}) in order to uniformly process the currently $\sim 10^5$ such epochs of hot (mostly massive) stellar 
high-resolution spectroscopic data. Both OWN and MONOS have been highly successful. In 2026 the first paper on 23 SB1 orbits has 
been accepted (Barb\'a et al. 2026, OWN~I) and a second paper on SB2 orbits will be submitted soon (Gamen et al. in prep., OWN~II). 
Yet, OWN results on individual stars had already appeared as separate papers 
\citep{Gameetal06a,Gameetal08b,Gameetal15a,Gameetal15b,Ariaetal10,Campetal19,Barbetal20,Putketal21,Putketal22,Putketal23,Putketal26,Ansietal23}.
MONOS has produced three papers: MONOS~I on the multiplicity statistics of northern O-type systems \citep{Maizetal19b}, MONOS~II with
35 SB1 orbits \citep{Trigetal21}, and MONOS~III with 10 SB2E orbits \citep{Holgetal25a}. 

As OWN and MONOS have run their courses in the last years, we decided to continue them with the new project presented in this
paper, Multiplicity of Massive stars in the Milky Way (\MTW). The main reason for a follow-up project is that, despite the advances
provided by the two projects (and by those of other groups, see next section), we have only scratched the surface of the problem of 
massive-star multiplicity and its main objectives are not yet solved. \MTW\ gets rid of the northern-southern hemisphere separation 
of the previous projects (which made sense previously, when spectroscopic archives were scarce in data, but not so much currently), 
the limitations of the OWN sample, and of the restriction to
O (and WN) stars by including other types of massive stars. It is a timely project because of the incoming availability of epoch 
astrometry and spectroscopy with \Gaia~DR4 in late 2026 \citep{Prusetal16} and of ground-based multi-fibre spectroscopic 
surveys such as WEAVE \citep{Jinetal24} and 4MOST \citep{DeJoetal19}. It will also make use of the tools we have developed for OWN 
and MONOS, such as UNWIND (presented in this paper but already used in OWN~I and MONOS~III), that will allow us for a more productive
exploitation of the data.

This paper is organised as follows. First, we describe the project, second we present the UNWIND tool, third we apply it to a
benchmark SB2 system, GLS~\num{11448}, and we finish the main section of the paper with a summary and an outlook of the incoming papers
of the series. The appendices present a glossary of terms, the DIB catalog generated for UNWIND, and an additional table. 

\section{Project description}

\subsection{Objectives}

$\,\!$\indent The immediate goal of \MTW\ is to determine the kinematical and dynamical properties (systemic or centre-of-mass 
RV, orbital parameters, mass, and $k_2$ internal structure constant, \citealt{Rosu25}) 
of as many Galactic massive stars as possible. Such individual analyses are a necessary step before any subsequent ones on the 
collective properties because systemic RVs of massive stars are frequently biased (an issue that
will be explored in a paper of the \MTW\ series) and some published orbits are either contaminated by third-light effects or even 
false orbits generated by e.g. poor temporal sampling of pulsating stars (MONOS~II, \citealt{SimDetal24}). Furthermore, 
massive stars are scarce so any additional information about one or several objects may be crucial to solve problems relevant to a 
class.

Beyond determining individual characteristics, the next goal of \MTW\ is to determine the multiplicity of massive stars
in the solar neighbourhood (within a few kpc of our Galactic position). There are analyses on the spectroscopic multiplicity of 
massive stars in Galactic clusters \citep{Sanaetal08b,Sanaetal09,Sanaetal11a,Kobuetal14,Banyetal22} but those are limited in sample 
size and scope. Furthermore, for a complete understanding of massive-star multiplicity, one needs to include the effect of 
longer-period systems \citep{Maiz10a,Sanaetal14}. Spectroscopic studies with larger samples have been done in the Magellanic Clouds
\citep{Sanaetal13b,Dunsetal15,Almeetal17,Shenetal24a,Britetal23,Britetal25,Villetal25,Sanaetal25}
which inform us of the metallicity dependence. 
Yet, at those distances our knowledge of multiplicity is limited by spatial resolution, hence limiting our understanding of its 
behaviour at large separations. Those limitations are the reason for expanding the sample of over 100 O-type orbits in the solar 
neighbourhood carried out by OWN and MONOS into a larger volume-limited sample containing several hundreds of massive stars with 
\MTW, leading to a thorough study of massive-star multiplicity in our nearby Galactic environment.

Those individual and collective analyses of the motion of massive stars will yield subsequent studies with different objectives. 
The first one will be the analysis of massive binaries with compact companions (neutron stars or black holes), expected to become 
gravitational-wave sources, for which we will combine our spectroscopic data with \Gaia~DR4 epoch astrometry to derive accurate masses. 
Another direct application will be the use of long time baselines to determine the apsidal motion of eccentric binaries and in that way 
probe their stellar interiors \citep{Rosu22,Rosuetal20a,Rosuetal20b,Rosuetal22a,Rosuetal22b}. 
The determination of accurate Keplerian stellar masses, solving 
the inclination problem with eclipses or astrometry, should address the discrepancies between evolutionary and spectroscopic masses,
still present after all these years \citep{Herretal92,Herr07,Holgetal20,Holgetal25b}. This, in turn, should lead to the
necessary corrections in the massive-star mass function due to hidden multiplicity \citep{Maiz08a} and other issues related to binary
interactions, such as the stars expelled from clusters as runaways \citep{Maizetal22b}. Runaway stars are another field that will be
impacted by \MTW, for example by the derivation of more accurate systemic RVs that should allow a better backtracing of
Galactic trajectories and by the detection of hidden companions to elucidate their formation mechanisms 
\citep{Maizetal18b,Carretal26}.

These objectives constitute the primary motivation for this project. Additional goals are expected to emerge during its execution, 
including the study of specific systems of interest that may exhibit unique characteristics.

\subsection{Data}

\subsubsection{Sample} 

$\,\!$\indent The samples for OWN and MONOS were built from the Galactic O-Star Catalog (GOSC,
\citealt{Maizetal04b,Sotaetal08}) which started as a compilation of high-quality spectral classifications of O stars from 
\citet{Walb72,Walb73a,Walb82a} and was later expanded to include the results of GOSSS and other types of massive stars (massive B
stars, WRs\ldots). For \MTW\ we
use the more extensive Alma Luminous Star (ALS) catalog, which also started as a literature compilation \citep{Reed03} but was 
later cross-matched with \Gaia\ and expanded with additional sources \citep{Pantetal21,Pantetal25b}. ALS gives us access to a
significantly larger sample of Galactic massive stars ($\sim\num{15000}$) that will soon be expanded by an order of magnitude 
using a systematic search with \Gaia\ data (Ma\'{\i}z Apell\'aniz et al. in prep.) and an extension into the Magellanic system 
\citep{MoliMaiz25}.

The stated primary goal of \MTW\ is to study the multiplicity of massive stars in the solar neighbourhood. Our previous 
experience with OWN and MONOS indicates that such an endeavour carries three complications. First, one needs to determine not only the
orbits of the multiple systems but also to establish which ones are truly single stars. This is more easily said than done, as completely
excluding the presence of a companion requires multiple epochs and a combination of techniques that only keeps eliminating possible
companions of lower masses and inclinations and different separations (longer for spectroscopic orbits, shorter for visual ones). In that
way, excluding that a 10~\Msun\ star at 1~kpc has a 0.5~\Msun\ companion in a circular 10~au orbit with an inclination of 10\degr\ is
impossible with our current means (see Fig.~7 in \citealt{SanaVran26}). Therefore, identifying a system as a ``single star'' always 
carries an asterisk that excludes extreme cases of binarity. Second, the availability of new data, instruments, surveys, and techniques as 
time goes by makes the use of a fixed sample counterproductive if one intends to do an analysis as complete as possible, as new massive 
stars are still being discovered in a fixed volume around the Sun (unless such volume is smaller than $\sim$300~pc, which leads to a very 
small sample). This is a lesson we learned well with the fixed sample of OWN. Third, the total sample of massive stars within a reasonable
distance of the Sun is quite large (e.g. there are 9016 likely massive stars within 2.5~kpc in \citealt{Pantetal25b} and that is not a
complete sample), some of them are too faint to be analysed in the optical region due to their high extinctions, and they are scattered
over the sky (with a strong preference for Galactic latitudes close to zero).

Those circumstances constrain the level of completeness of any multiplicity study but we should keep in mind that they only get
worse with increasing distance (with the only exception of extinction in the case of the Magellanic Clouds), so the solar neighbourhood is
the place to start. This prompts us to divide our sample into two: a core of known bright O stars within 2.5~kpc of the Sun that can be 
studied with \lili\ from day one and a larger growing sample based on ALS that extends to longer distances, higher extinctions, and lower
masses. Limiting the core sample to $B = 12$~mag leaves a current total of 515 O stars based on GOSSS spectral classifications. The number 
is expected to grow over the lifetime of the project but not by much, as our experience with GOSSS over the last two decades shows that the
discovery rate for new bright Galactic O stars has slowed significantly and the majority of them are already known. The much larger 
extended sample, on the other hand, is expected to grow by an order of magnitude in the near future mostly thanks to \textit{Gaia}, WEAVE, 
and 4MOST. Even though it will be studied in lesser detail than the core sample, it will serve as the basis for the statistical study of
multiplicity and its impact on the mass functions (initial and present day). The extended sample is also the one expected to include the 
majority of the more interesting objects, such as most of the massive binaries with compact objects and most of the systems with 
detectable apsidal motions.

\subsubsection{\lili} 

\begin{table}
\caption{\lili\ spectrograph configurations used in \MTW\ divided by hemisphere, note that the GLS~\num[detect-all]{11448} uses
         all of the northern ones except those from OHP and SONG.}
\addtolength{\tabcolsep}{-1mm}
\begin{tabular}{cccrr@{-}rr}
Code   & Telescope & Spectrograph & \mci{$R$}        & \mcii{Wav. range}      & \mci{Epochs}        \\
       &           &              &                  & \mcii{(\AA)}           & \mci{$\times 10^3$} \\
\midrule
HET-B  & HET       & HRS          & \num{30000}      & \num{3820}&\num{4705}  &  0.1                \\
       &           &              & \num{30000}      & \num{4765}&\num{5725}  &  0.1                \\
HET-R  & HET       & HRS          & \num{30000}      & \num{5320}&\num{6250}  &  0.2                \\
       &           &              & \num{30000}      & \num{6395}&\num{7320}  &  0.2                \\
CAR    & CAHA3.5   & CARMENES     & \num{95000}      & \num{5240}&\num{9850}  &  1.9                \\
       &           &              & \num{80000}      & \num{9610}&\num{17100} &  1.9                \\
HAR-N  & TNG       & HARPS-N      & \num{115000}     & \num{3875}&\num{6905}  &  1.0                \\
FIES-L & NOT       & FIES         & \num{25000}      & \num{3700}&\num{9000}  &  1.7                \\
FIES-M & NOT       & FIES         & \num{46000}      & \num{3700}&\num{9000}  &  2.5                \\
FIES-H & NOT       & FIES         & \num{67000}      & \num{3700}&\num{9000}  &  0.4                \\
CAF\'E & CAHA2.2   & CAF\'E       & \num{95000}      & \num{3925}&\num{9225}  &  0.8                \\
OHP    & OHP1.9    & ELODIE       & \num{42000}      & \num{4000}&\num{6800}  &  2.0                \\
       &           & SOPHIE       & \num{40000}      & \num{3875}&\num{6930}  &  0.9                \\
       &           &              & \num{75000}      & \num{3875}&\num{6930}  &  3.0                \\
Merc   & Mercator  & HERMES       & \num{85000}      & \num{3785}&\num{8995}  &  9.0                \\
SONG   & SONG      & SONG         & \num{100000}     & \num{4400}&\num{6900}  & 18.0                \\
\midrule
SALT   & SALT      & HRS          & \num{65000}      & \num{3850}&\num{5560}  &  1.0                \\
       &           &              & \num{65000}      & \num{5450}&\num{8800}  &  1.0                \\
UVES-U & VLT       & UVES         & \num{59000}      & \num{3030}&\num{3880}  &  5.4                \\
UVES-u & VLT       & UVES         & \num{59000}      & \num{3260}&\num{4540}  &  2.8                \\
UVES-B & VLT       & UVES         & \num{59000}      & \num{3760}&\num{4975}  &  5.4                \\
UVES-y & VLT       & UVES         & \num{66000}      & \num{4580}&\num{6680}  &  3.0                \\
UVES-V & VLT       & UVES         & \num{66000}      & \num{4760}&\num{6840}  &  3.8                \\
UVES-I & VLT       & UVES         & \num{66000}      & \num{5650}&\num{9460}  &  2.1                \\
UVES-z & VLT       & UVES         & \num{66000}      & \num{6600}&\num{10600} &  4.1                \\
HAR-S  & OLS3.6    & HARPS        & \num{120000}     & \num{3785}&\num{6905}  &  4.5                \\
FEROS  & OLS2.2    & FEROS        & \num{48000}      & \num{3570}&\num{9210}  & 17.4                \\
\midrule
\end{tabular}
\addtolength{\tabcolsep}{1mm}
\label{spectrographs}   
\end{table}

$\,\!$\indent The bulk of the \MTW\ spectroscopic data, at least in the first years of the project, will be the \lili\ 
high-resolution spectroscopy \citep{Maizetal19a}, which is uniformly processed and already adapted to the UNWIND needs (see Sect.~3).
The characteristics of the spectrographs used are listed in Table~\ref{spectrographs}. We have already collected $\sim10^5$ epochs from 
database searches (such as ESO and OHP) and our own programs at ORM, CAHA, and OLS (see Appendix~A for a list of acronyms used in
this paper). Table~\ref{spectrographs} includes the current number of epochs in \lili\ for each configuration 
but in incoming years we will continue adding high-resolution spectra both from archival searches and our own programs at an expected
pace of $\sim$5000 spectra per year.

The diversity of stellar magnitudes and spectrographs in our sample inevitably leads to a variety in S/N of the data. This is
aggravated by the sometimes intense extinction experienced by the stars, which can lead to simultaneously high values of S/N at the red 
end and low S/N values at the blue end of the same spectrograph. These complications require that the orbits of different stars are often
measured using different wavelength regions: faint stars are usually more extinguished, which constrains their analysis to longer 
wavelengths, while for bright stars we have more choices. One of the advantages of our UNWIND code, described in the next section, is its 
ability to clean wavelength regions with telluric lines and DIBs (more common at longer wavelengths in the optical spectrum), thus allowing
us to determine spectroscopic orbits under more circumstances. In any case, all the epochs listed in Fig.~\ref{spectrographs} have 
wavelength regions with S/N $> 30$, the minimum we need for orbit determinations, and most have regions with S/N $> 100$.

For the 515 stars in the core sample we currently have $\sim$\num[detect-all]{26000} \lili\ epochs, with a median of 8 per star. In
order to have a minimum of 10 epochs for each star we require 2000 additional epochs. The number increases to 5200 if the minimum is 
established at 20 epochs. The first one is the value we will use to establish the spectroscopic binarity of a system and the second one the
one to determine a spectroscopic orbit. Currently 237 systems satisfy the 10-epoch criterion and 156 the 20-epoch one but, at the rate of 
growth of the \lili\ database, all of the core sample should satisfy the first one in 5 years.

\subsubsection{Additional ground-based spectroscopy} 

$\,\!$\indent High-resolution spectroscopy ($R > \num{10000}$) is the best adapted for the analysis 
of spectroscopic binaries. However, intermediate-resolution ($R = 2000 - \num{10000}$) spectroscopy can suffice for
large RV amplitudes\footnote{As an example, the SB2 nature of the main target in this paper, GLS~\num{11448}, was
discovered in an $R\sim 2500$ spectrum from CAHA3.5.}. Furthermore, high-resolution spectra are usually obtained with a single
circular aperture that (for the most part) erases the spatial information from visual binaries
(e.g. \citealt{SimDetal15a}), something that can be overcome at lower spectral resolutions with 
long slits or IFUs \citep{Maizetal18a,Maizetal21b,MaizBarb20}. We expect to include in \MTW\ intermediate-resolution spectra from 
GOSSS, WEAVE \citep{Jinetal24}, 4MOST \citep{DeJoetal19}, and possibly other ground-based sources. They will be used mostly to
expand and analyse the extended sample.

\subsubsection{Gaia astrometry and spectroscopy} 

$\,\!$\indent The most useful new contribution of \Gaia~DR4 to \MTW\ will be the epoch astrometry, which
will allow for the full orbital determination of massive stars with a compact companion, where the photocentre corresponds to the
visible stars, and the placement of constraints for systems with a significant magnitude difference between the two components. We 
note, however, the special characteristics of the epoch photometry, with errors much larger in one direction than in another 
\citep{LindBast22}. The combined astrometry will also be useful, with significantly improved parallaxes and (especially) proper
motions. We caution, however, against rushing to publish parameters for massive binaries from \Gaia~DR4 data for two reasons: the 
need to evaluate the
precision and accuracy of the astrometry against calibrators \citep{Maizetal21c,Maiz22} and the likely possibility that the
originally derived parallaxes and (to a lesser degree) proper motions will be affected by the orbital motion of the system. When
searching for a solution in a many-dimensional space, one is likely to find local minima that are not the desired one and, in such 
cases, using external information (e.g. orbital periods or eccentricities) can significantly improve the solution, hence the usefulness 
of having previous spectroscopic orbits available. Astrometry will not be the only \Gaia\ data that we will use, as DR4 will also 
include RVS time series for all sources brighter than $G_{\rm RVS} = 14$~mag that will complement \lili, low-resolution epoch
spectrophotometry, and epoch photometry (see next).

\subsubsection{Epoch photometry} 

$\,\!$\indent To model the eclipsing binaries and ellipsoidal variables (\citealt{MarRetal24} and 
MONOS~III), \MTW\ will also use epoch photometry from different sources. \textit{TESS} \citep{Ricketal15} will be the most useful,
given its all-sky coverage, dynamic range, and high cadence, but its poor spatial resolution, low-frequency photometric calibration
issues, and uneven long-term coverage require complementary data. On the one hand, \Gaia~DR4 will be even more useful than DR3
to study eclipsing binaries \citep{Mowletal23}, with a more uniform long-term coverage and better spatial resolution and photometric 
stability than TESS (at the cost of worse cadences). On the other hand, MUDEHaR \citep{Holgetal24} will complement \Gaia\ epoch 
photometry, targeting areas rich in massive stars to increase the time baseline and include H$\alpha$ variability.

\subsubsection{High-resolution imaging and interferometry} 

$\,\!$\indent Finally, for some systems we will complement all of the above information with lucky imaging and interferometry 
\citep{Maiz10a,SanBetal13,Sanaetal14,Aldoetal15,Lantetal23}. The first is useful to determine or constrain long orbits with 
separations of 0\farcs1-1\farcs0 and the second for shorter orbits, in some cases by itself and in others in concert with 
high-resolution spectroscopy.  We will also use the information available in the Washington Double Star catalog (WDS, 
\citealt{Masoetal01}).

\section{UNWIND}

\subsection{Package description}

$\,\!$\indent Within OWN, MONOS, and other multiplicity projects we have developed several codes related to orbit fitting written
in IDL. The main novelty of the software package developed for the project is the UNWIND\footnote{To be pronounced as 
\textipa{/2n"w2Ind/}, not as \textipa{/2n"wInd/}, as the purpose of the code is not to eliminate the stellar wind features in 
the stellar spectra.} disentangling code, which will be used within \MTW\ for the main purpose of separating two or 
more components in a spectroscopic binary but also for other tasks such as the fitting and subtraction of the ISM component and the 
generation of high S/N spectra of SB1 systems (e.g. OWN~I).

In addition to UNWIND, the package includes routines for single- and multi-component line fitting, cross correlation, and 
spectroscopic and visual orbit fitting, among others. Some of those will be used in this paper. Depending on the circumstances, the
function fitting routines use (a) the non-linear least squares package MPFIT \citep{Mark09} or (b) an analysis of the full relevant
parameter space in the style of CHORIZOS for photometry \citep{Maiz04c}. As the number of dimensions can make the latter too
computationally expensive, a search through the parameter space has to be performed in the case of visual orbits to look for
different local minima \citep{Maizetal17a,Maiz19}. For spectroscopic orbits we use MPFIT instead but we add a search with different 
seeds around the primary solution to check for the existence of alternative solutions. 

UNWIND is at this point a beta software. We have been able to run it in different computers under Mac OS and Linux but it currently
lacks a manual or an installation package, so its dependencies have to be solved one by one. Eventually, we expect to address those issues
but currently it can only be shared without support.

\subsection{UNWIND preprocessing and modules}

\begin{figure*}
 \centerline{\includegraphics*[width=\linewidth]{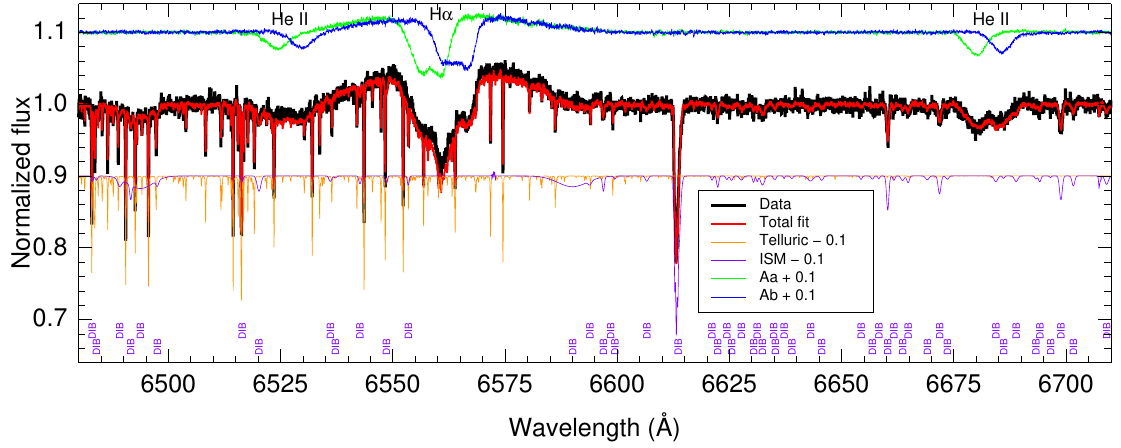}}
 \caption{Example of UNWIND output for the H$\alpha$ region of a GLS~\num{11448}~Aa,Ab spectroscopic epoch and the result of combining 
          77 epochs. The top spectra (green and blue) show the normalized output for the two components, shifted horizontally to 
          the RVs of the epoch and upwards 0.1 continuum units, and diluted by their flux fractions. The bottom (orange and 
          purple) spectra show the fitted telluric lines, specific to this epoch, and ISM spectrum, mostly DIBs and common to all 
          epochs, in the two cases shifted downwards 0.1 units. The central spectra show the data (black) and fit (red) for this 
          epoch, the fit being the sum of the green and blue spectra (minus one to leave the continuum fixed) multiplied by the orange
          and purple spectra.}
 \label{disentangling}      
\end{figure*}

$\,\!$\indent Here we describe the data flow though the different modules (each with a different task) that make up UNWIND. In the
next subsection we describe some of the issues that can be addressed with UNWIND.

As a preliminary step, all spectra have to be preprocessed using the \lili\ procedures for header and data uniformisation for 
different spectrographs, pre-rectification (to be revised later, see below), and telluric-absorption calculation. The latter is done 
applying the telluric model of \citet{Gardetal13} with an extension into the near infrared that was added when CARMENES data were 
included in \lili. Alternative telluric corrections can be used instead. By default, UNWIND 
requires that both the telluric-absorption corrected spectrum and the correction itself be passed along to the code in order to 
compute the weights used to combine spectra. However, it is possible to deactivate the requirement and to not apply telluric 
weights, for example for blue-violet spectra.

UNWIND itself uses the following modules in sequence: 

\textit{Data reading, resampling, and rectification.} The spectra are first read and resampled into a spectral grid in 
the barycentric frame with equal
spacing in $\Delta\lambda/\lambda$ (logarithmic binning) to simplify RV calculations. Rectification is done using wavelength
points provided by the user after a careful examination of the spectra. This is done because of the high sensitivity of disentangling
to rectification: it may be time consuming but it reduces systematic errors.

\textit{Telluric emission, standard ISM, and DIB fitting.} A median spectrum from all epochs is used to eliminate the
non-stellar contributions to the spectra. Telluric emission lines (and possibly residuals from broad telluric O$_2$ absorption lines)
are removed by fitting (narrow) Gaussian profiles. The standard ISM
(ionic and molecular) lines in absorption and emission are then directly subtracted but not fitted, as their profiles are usually 
non-Gaussian. In a future version we will implement the possibility of the subtraction of ISM emission lines such as H$\alpha$ or
[\ion{O}{iii}]~$\lambda$\,5007 being refined on an
epoch-by-epoch basis, as different spectrograph apertures and seeing conditions provide different degrees of contamination.
Finally, DIBs are fitted with Gaussian profiles, as described below and in Appendix~B. The default use is to fit only the overall
RV of the DIBs and the EW of each one, leaving the central wavelengths and FWHM fixed, but it is possible
to fit more parameters. The UNWIND result on the median spectrum can be analysed at this point with an interactive spectrum viewer 
to fine-tune the rectification points and the fitted ISM (standard or DIBs) lines.

\textit{Spectral Energy Distribution, flux fraction, and RV guesses.} The code then reads the input guesses for the 
disentangling process. The rectified SED guesses can be derived from the data or from an input SED. In the current version, UNWIND 
includes a library of TLUSTY SEDs available from \citet{LanzHube03,LanzHube07} but users can include their own preferred SEDs. 
Flux fraction and RV values are fixed and provided by the user, the latter either as an orbital solution or as epoch 
RVs\footnote{The current version of the code assumes that the two stars are not too different in \Teff, so that the flux 
fraction does not change much over the wavelength range of interest. For systems composed of e.g. a blue and a red star, a
solution is to divide the disentangling process in several wavelength ranges and apply different flux fractions in each.}. 
Note that those are not fitted by UNWIND itself but require an outer iteration (see next subsection). Also, in the current version 
flux fractions are constant for each component throughout all epochs but variable ones may be implemented in the future for eclipsing
binaries.

\textit{Disentangling.} UNWIND uses the iterative shift and subtract disentangling technique of \citet{GonzLeva06}, see
\citet{Hadr09} for a comparison between different disentangling methods. The
essence of the technique is that rectified SED guesses are provided for each component, shifted in RV for each epoch, and 
subtracted from the observed data. The residuals are then calculated and incorporated into new guesses, with the outcome iterated until 
convergence. Each iteration is displayed on the screen for the user to assess convergence and possibly increase the number of
cycles if needed or decrease it if convergence is achieved faster. As it is common in iterative fitting procedures, the result may alternate 
between two solutions in consecutive iterations. For such a case, UNWIND offers the possibility of accelerating convergence by using
the average of two iterations at a point in the process. It is possible in principle to include RV changes into the next
iteration but we do not do it, as UNWIND leaves that to an outer iteration (see next subsection and examples in subsequent 
papers). The algorithm was originally designed to include only two components but with UNWIND it has been expanded into three or more. 
However, if an unconstrained component is added to the iterative procedure, the most likely outcomes are divergent non-physical solutions. 
For that reason, UNWIND allows only for fixed third lights, with the possibility of using the outer iteration to alternate among which 
components to fit, two at a time.

\textit{Output.} During the disentangling process, the result of each iteration is displayed on the screen and, optionally, the 
final result is shown on an interactive spectrum viewer. In addition to the final result, the residuals for each epoch are also
produced. An example of the final products for a spectroscopic GLS~\num{11448}~Aa,Ab epoch (from the analysis in the next section) is 
shown in Fig.~\ref{disentangling}.

\subsection{Disentangling issues}

$\,\!$\indent A typical use of UNWIND is iterative. For example, a combination of guesses for the orbit and flux fractions 
of the components may be tried, the output analysed, and a new orbit and/or flux 
fractions calculated. Under that scenario, the new orbit can be calculated by cross-correlating the UNWIND output with the epoch data 
(for simple SB1 systems) or by profile-fitting the epoch data with the UNWIND output (when two or more lights are present) and the new 
flux fractions obtained by comparing with single-star spectra or model SEDs\footnote{Using the wrong flux fractions for a system leads to
components with too strong or too weak lines but, otherwise, has only a small influence in the disentangling process itself through the
initial guesses. Hence, it is something that can be adjusted at the end of the process.}. Those new values may then be used for a new 
UNWIND run. Therefore, one can consider the process as containing an inner loop (the disentangling itself) and an outer loop over the 
orbital parameters and flux fractions. A simple case, such as one with two components with large RV amplitudes and similar fluxes 
may require relatively little work but more complicated ones may require more effort and it is possible that the available epochs may 
not be enough to reach a satisfactory solution. 

At this point we have tested UNWIND under quite different situations and in this subsection we
discuss some of the issues that can complicate the disentangling of spectroscopic binaries. This should be taken as an
introduction to such issues, as real examples provide a better guide to the process and its limitations. One such example,
GLS~\num{11448}, is analysed in Sect.~4 and future papers of the series will study others. Of course, there are other
issues that can lead to disentangling problems besides the ones mentioned here. Two obvious ones are the lack of adequate orbital 
phase coverage for the system and the addition of variability to one of both stars in the form of pulsations, winds, or disks.
Another, perhaps less obvious one, is the need for an adequate rectification of the spectra, already mentioned above.

\textit{ISM subtraction.} The ISM signature during the disentangling process could be approximated as a third light with fixed 
RV and intensity\footnote{This would be an approximation, not an exact fit, for two reasons: [a] the variability of the ISM
among epochs due to (rare) intrinsic changes or (common) aperture/seeing changes for emission lines and [b] the fact that the ISM
imprint is a multiplicative signature in the sum of the observed fluxes.}. However, as noted above, an unconstrained disentangling
process usually diverges from the physical solution. For those reasons, UNWIND calculates the ISM signature as a preliminary step
and subtracts it from the observed spectra. Doing that is already a considerable task already for the standard (ionic or molecular) ISM
lines as it requires compiling a library of the existing lines. For the case of GLS~\num{11448} below we detected almost 200 such
lines of \ion{He}{i}, \ion{Li}{i}, \ion{Na}{i}, \ion{Ca}{i}, \ion{Ca}{ii}, \ion{K}{i}, \ion{Rb}{i}, C$_2$, CH, CH+, and CN. The
problem becomes more important for DIBs, as they are more common, have broader profiles (in some cases similar to those
of OB stellar lines), and can dominate the observed spectra of OB stars in some wavelength regions. That is one of the reasons why
some previous disentangling efforts have been restricted to small wavelength regions instead of to the full spectral region
observed. As obtaining such a broad wavelength range is one of the goals of UNWIND, we determined the most complete DIB library ever
built, with a total of 631 in the 4000-\num{17100}~\AA\ range (Appendix~B), 
and incorporated its subtraction into UNWIND. In Sect.~4 we 
show the result of the process for GLS~\num{11448}, the first time an OB-type SB2 system has been disentangled over the entire
3820-\num{11000}~\AA\ range. The extracted DIB spectrum is of astrophysical interest by itself but we defer its study to a paper of 
the CollDIBs series \citep{Maiz15a,Maizetal19a}.

\textit{Telluric-line subtraction.} A similar issue is the elimination of telluric lines in disentangled spectra. This is a minor or
non-existent problem at short wavelengths but becomes more important at longer ones, where it is compounded with the DIB subtraction
issue mentioned above. An accurate telluric subtraction requires not only a good atmospheric model but also a spectrograph with good
stability and a pipeline that delivers an accurate calibration. Among the spectrographs used in \lili, CARMENES 
\citep{Quiretal14,Cabaetal25} 
excels in those two aspects and thanks to its data we have been able to disentangle some \ion{He}{ii} lines in the spectra of
GLS~\num{11448} that are located in regions of high telluric contamination (see Sect.~4). In addition, the stability of CARMENES has
allowed us to apply UNWIND to extract the spectra of some fullerene DIBs (see Appendix B) in regions with high telluric 
contamination by making the Earth play the role of a spectroscopic binary: the telluric lines move back and forth along with our
orbital motion around the Sun while the DIBs remain fixed in RV. 

\begin{table}
\caption{GLS~\num{11448} orbit results.}
\begin{tabular}{l@{\hspace{1mm}}c@{\hspace{1.5mm}}r@{.}l@{$\pm$}r@{.}l@{\hspace{0mm}}r@{.}l@{$\pm$}r@{.}l}
Quantity                          & Units   & \mciv{}                                           & \mciv{}                               \\
\midrule
$T_0$                             & BRJD    &       \num{56003}&299        &       0&092        & \mciv{}                               \\
                                  &         & \cre{\num{56002}}&\cre{80}   & \cre{0}&\cre{25}   & \mciv{}                               \\
$P$                               & d       &                97&1696       &       0&0023       & \mciv{}                               \\
                                  &         & \cre{         97}&\cre{168}  & \cre{0}&\cre{025}  & \mciv{}                               \\
$e$                               &         &                 0&5831       &       0&0030       & \mciv{}                               \\
                                  &         & \cre{          0}&\cre{5627} & \cre{0}&\cre{0061} & \mciv{}                               \\
$\omega$                          & degrees &               130&67         &       0&52         & \mciv{}                               \\
                                  &         & \cre{        126}&\cre{1}    & \cre{1}&\cre{2}    & \mciv{}                               \\
$(M_{\rm Aa}+M_{\rm Ab})\sin^3 i$ & \Msun   &                69&37         &       0&55         & \mciv{}                               \\
                                  &         & \cre{         74}&\cre{9}    & \cre{1}&\cre{0}    & \mciv{}                               \\
$q$                               &         &                 0&9175       &       0&0095       & \mciv{}                               \\
                                  &         & \cre{          0}&\cre{9622} & \cre{0}&\cre{0051} & \mciv{}                               \\
$(a_{\rm Aa}+a_{\rm Ab})\sin i$   & au      &                 1&6997       &       0&0045       & \mciv{}                               \\
                                  &         & \cre{          1}&\cre{7438} & \cre{0}&\cre{0077} & \mciv{}                               \\
\midrule
                                  &         & \mciv{Aa}                                   & \mciv{Ab}                                   \\
\midrule
$K$                               & km/s    &          114&86        &       0&56         &         119&38         &       0&45         \\
                                  &         & \cre{   112}&\cre{7}   & \cre{1}&\cre{2}    & \cre{  122}&\cre{9}    & \cre{1}&\cre{3}    \\
$\gamma$                          & km/s    &         $-$6&8         &       4&7          &       $-$11&7          &       3&4          \\
                                  &         & \cre{ $-$17}&\cre{68}  & \cre{0}&\cre{98}   & \cre{$-$20}&\cre{88}   & \cre{1}&\cre{04}   \\
$M\sin^3 i$                       & \Msun   &           35&36        &       0&49         &          34&01         &       0&51         \\
                                  &         & \cre{    38}&\cre{80}  & \cre{0}&\cre{83}   & \cre{   35}&\cre{60}   & \cre{0}&\cre{77}   \\
$a\sin i$                         & au      &            0&8334      &       0&0035       &           0&8662       &       0&0029       \\
                                  &         & \cre{     0}&\cre{8325}& \cre{0}&\cre{0078} & \cre{    0}&\cre{9073} & \cre{0}&\cre{0081} \\
\midrule
\multicolumn{10}{l}{Values in black correspond to this paper and values in red to} \\
\multicolumn{10}{l}{$\;\;$\citet{Maizetal15a}}
\vspace{-5mm}
\end{tabular}
\label{orbitalparameters}   
\end{table}

\textit{Third (and additional) lights.} Given the high-order multiplicity typical of OB-type systems (MONOS~I), it is common
to find systems where three or more stars contribute to the observed spectra. In those cases, UNWIND requires that the flux(es) of
the third (or higher) light(s) is/are fixed to an initial guess (but allowed to move by an orbital motion) and then an
outer iteration is applied to the process, fixing one of the initial two stars and letting the other and the third one vary in a new
UNWIND run. The process may be repeated as needed until convergence. An example of this UNWIND application is shown for
BD~$-$13~4929 in MONOS~III. The system consists of two early B stars in a 2.9302~d orbit, each contributing 18\% of the total light,
in a long orbit around an O star that contributes the remaining 64\% (and, which, for the purposes of UNWIND, can be considered as
stationary). The UNWIND result in Fig.~A.6 of MONOS~III shows a clear separation among the three components. For some stars 
in OWN~I ($\tau$~CMa~Aa,Ab, HD~\num{101190}~Aa,Ab, HD~\num{101205}~A,B, and HD~\num{152723}~Aa,Ab) we also applied third-light
analyses with UNWIND, some of which will be presented in more detail in future papers of this series.

\textit{Initial guesses.} Any non-linear fitting process can be affected by the choice of initial guesses due to the possible
existence of multiple local minima in the parameter space and this is especially so for a many-dimensional iterative process such as
disentangling. This issue did not receive much attention when disentangling techniques were first applied \citep{GonzLeva06} but
subsequent analyses \citep{Quinetal20,Rosuetal23} showed that a wrong choice of initial guesses can introduce spurious structures 
such as ``horns'' (emission peaks on both sides) around absorption lines. As in UNWIND it is possible to use 
different initial guesses, in subsequent papers we will explore the dependency of the final outcome on them.

\textit{Search for hidden components.} A currently hot topic is the search for the signature of a companion in SB1 systems in order 
to detect whether it is a lower-mass normal star or a compact object (\citealt{Shenetal20,Shenetal22b,Shenetal22a}, MONOS~II, 
\citealt{Mahyetal22a}). In OWN~I we already used UNWIND for this task and we note one aspect of the code that makes it especially
well suited for it: Using the full spectral window for a given spectrograph gives us more chances to discover the hypothetical 
signature of the lower-mass companion. We will also explore this in future papers of the series.

\section{Sample application: GLS~\num[detect-all]{11448} (= LS~III~+46~11)}

$\,\!$\indent \citet{Maizetal15a} analysed GLS~\num{11448} using multi-epoch spectroscopy and discovered it is a near-twin,
eccentric, very massive system with minimum masses of 38.8~\Msun\ and 35.6~\Msun, respectively. In a follow-up paper, 
\citet{Maizetal15c} analysed the foreground ISM affecting both GLS~\num{11448} and the nearby GLS~\num{11449} (= LS~III~+46~12), and
detected two distinct kinematic components for both stars, indicating the presence of two clouds at different RVs. Despite 
their proximity in the plane of the sky, GLS~\num{11448} is significantly more extinguished than GLS~\num{11449} and different ISM
indicators led \citet{Maizetal15c} to suggest that the sightline of the former goes through the UV-shielded core of the progenitor
cloud ($\zeta$ sightline, see Appendix~B) while the sightline of the latter goes through less UV-shielded regions
($\sigma$~sightline, see Appendix~B). Within the Villafranca project \citep{Maizetal20b} that is analysing Galactic stellar clusters 
with massive stars, \citet{Maizetal22b} used \Gaia~DR3 data to measure a distance of $2741^{+86}_{-81}$~pc to Berkeley~90, the
cluster where both stars reside, which is consistent with the spectrophotometric distance to GLS~\num{11448} previously determined by 
\citet{Maizetal15a} and within 2 sigmas of the value from \citet{MarcNegu17}.

In this section we use UNWIND to [a] derive a new SB2 orbit for GLS~\num{11448}, [b] disentangle the two components (Aa and Ab) in 
the 3820-\num{11000}~\AA\ range, [c] determine new spectral classifications and stellar parameters, and [d] analyse the signature of 
the ISM in the disentangled spectra. In principle, this could be done by first using a small wavelength range with one or several 
prominent lines to derive the new orbit using as a seed the old one, applying the new orbit to the whole wavelength 
range to obtain the disentangled Aa and Ab plus the ISM spectra, and then proceeding to [c] and [d] above. However, that leaves us 
with the issue of partially undetermined systemic RVs, $\gamma_{\rm Aa}$ and $\gamma_{\rm Ab}$, because for O stars in general
(and even more so for early-type supergiants such as GLS~\num{11448}) wind infilling shifts the centroid of absorption 
lines, thus altering their values in different ways for different lines. This is a general problem for O stars that is especially
relevant to the derivation of the peculiar 3-D velocities of runaway stars and will be treated in detail in a future paper of the \MTW\
series. For GLS~\num{11448} we address it adding an outer iteration to the procedure above. First, we calculate the orbit (with
approximate systemic RVs), disentangle the spectra according to it, and fit FASTWIND models to each of the two
components. Those preliminary FASTWIND models are used to (a) check the flux fractions and (b) calculate the RV shifts for 
\ion{H}{i}, \ion{He}{i}, and \ion{He}{ii} lines; select those lines with a good S/N in the spectra and small enough shifts; and use 
the difference between the disentangled spectra and the FASTWIND predictions to recompute new values of 
$\gamma_{\rm Aa}$ and $\gamma_{\rm Ab}$ to obtain the final orbit (using the dispersions as uncertainties). The final disentangled 
spectra can be calculated by running UNWIND one last time for the whole spectrum with the new orbit (if the flux fractions have changed)
or simply by shifting in RV the original UNWIND output (if they have not).

We use spectra from six spectrograph+telescope combinations, the northern ones in Table~\ref{spectrographs} except for OHP and
SONG. Four of the spectrographs (HARPS-N@TNG, FIES@NOT, CAF\'E@CAHA2.2, and HERMES@Mercator) have a single arm and the other two (HRS@HET 
and CARMENES@CAHA3.5) two simultaneous arms. CARMENES has a fixed configuration that yields a large coverage between 5240~\AA\
and \num{17100}~\AA\ (with some gaps) between the two arms. HRS@HET covers a smaller wavelength range and can be set up in
either a blue (HET-B) or a red (HET-R) configuration, but it has the advantage of the much larger effective aperture of the 
telescope (9.2~m at the times the data were obtained). We used a total of 80 epochs, listed in Table~\ref{epochs}. We note that
different wavelengths were observed on a different number of epochs due to the coverage of each spectrograph and to the fact that the
high extinction experienced by GLS~\num{11448} yields a poor S/N in the blue-violet region for small-aperture telescopes.

\subsection{The new orbit}

\subsubsection{The orbital parameters}

$\,\!$\indent As explained above, our approach is to determine an initial orbit with UNWIND without paying attention to the systemic 
RVs (as those only shift the disentangled spectra in wavelength), obtain the stellar parameters from the result in
subsection~4.2, and run UNWIND again applying the new $\gamma$ values determined from them in order to obtain a self-consistent
result. For the initial UNWIND run we start from the \citet{Maizetal15a} orbit, disentangle the spectra, and use the results to fit 
the RVs to each epoch in the original data as a composite of the profiles of the two components, Aa and Ab. We do 
this in four wavelength ranges: 4305-4705~\AA\ (blue), 5390-5430~\AA\ (\ion{He}{ii}~$\lambda$5411.53), 6505-6705~\AA\ (red), and 
8208-8267~\AA\ (\ion{He}{ii}~$\lambda$8236.79). Each one of them has advantages and disadvantages, such as the larger number of lines in 
the blue range and the increasing S/N for the longer-wavelength regions. As it turns out, the solutions for the four wavelength ranges 
yield parameters that are consistent within two sigmas, so we selected the \ion{He}{ii}~$\lambda$5411.53 range as our primary 
source for three reasons: it has one of the strongest isolated lines in the whole optical spectrum, the S/N of the data there is 
considerably better than in the blue, and its central location in the spectrum allows us to use the largest number of epochs of any 
range. The \ion{He}{ii}~$\lambda$5411.53 RVs are listed in Table~\ref{epochs}.

\begin{figure}
 \centerline{\includegraphics*[width=\linewidth]{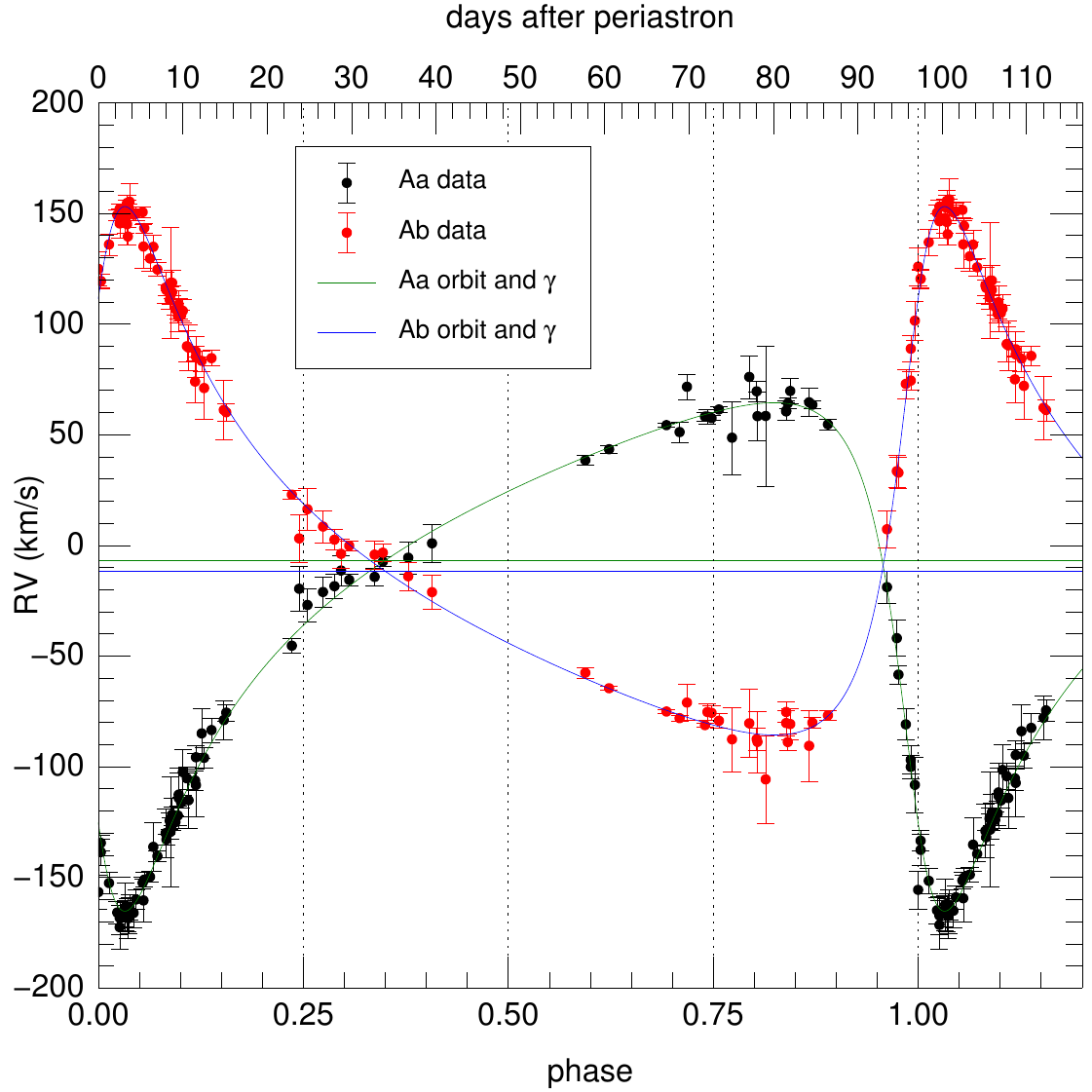}}
 \caption{Phased RV curves for GLS~\num{11448}~Aa~and~Ab.}
 \label{orbit}      
\end{figure}

The orbital parameters (including the systemic RV resulting from the outer iteration derived in the next subsubsection) 
are fitted using an IDL code we already tested and applied in previous papers, such as MONOS III and OWN I. They
are listed in Table~\ref{orbitalparameters} and the orbit is plotted in Fig.~\ref{orbit}. Not including the $\gamma$ values, the
parameters are very similar to those of \citet{Maizetal15a}, in many cases within two sigmas. The RV amplitudes are slightly
lower, leading to lower minimum (Keplerian) masses, and the mass ratio is now even closer to 1.0 than before. The masses are analysed
in Sect.~4.2 in conjunction with the values determined from the disentangled spectra. The most significant change for the parameters
is the reduction of the uncertainties, with the one for the period now an order of magnitude lower. This is an expected effect
of adding epochs with better measurements and, for the period, of extending the time coverage.

\subsubsection{The systemic RVs}

\begin{figure}
 \centerline{\includegraphics*[width=0.49\linewidth]{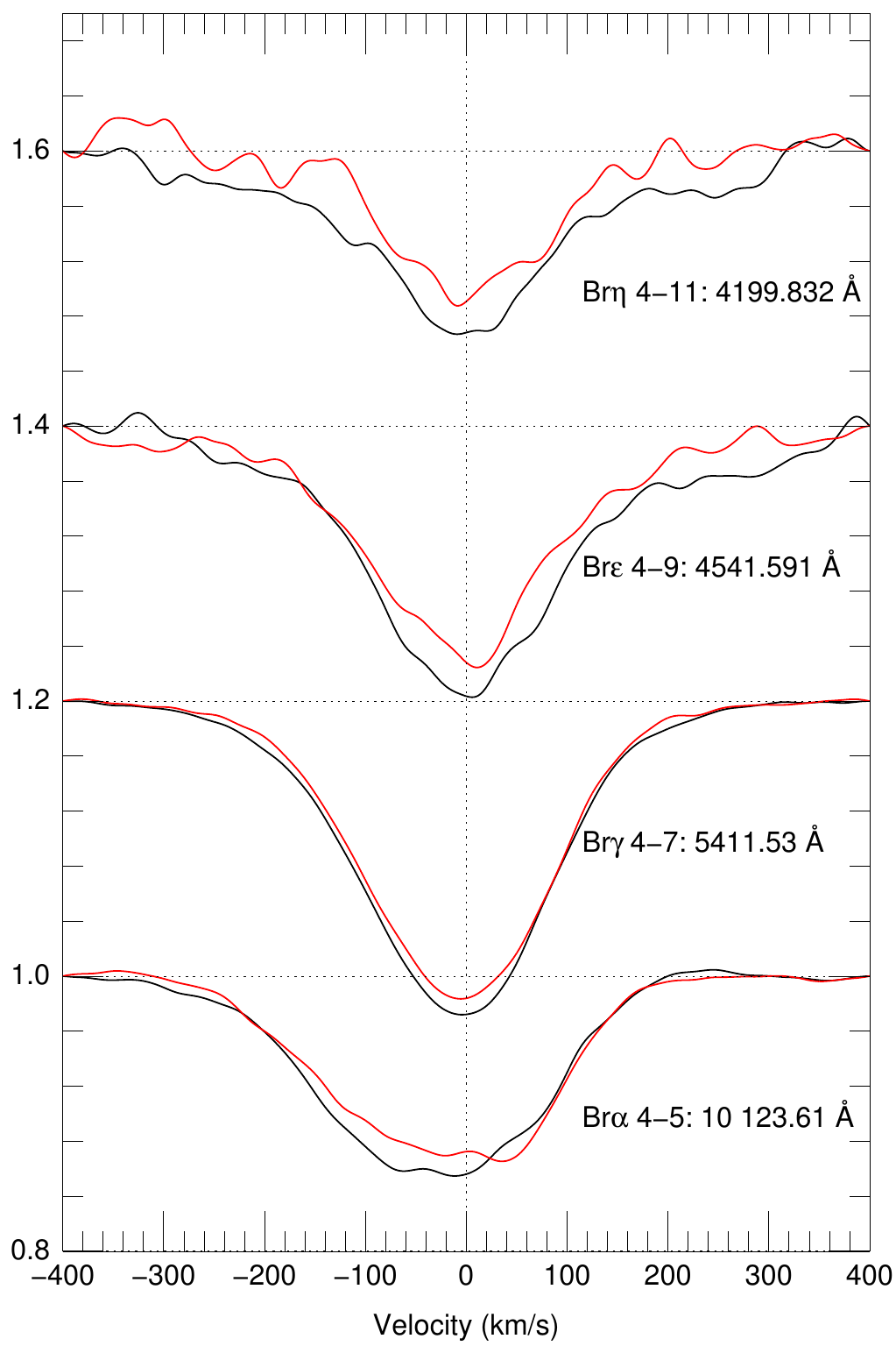} \
             \includegraphics*[width=0.49\linewidth]{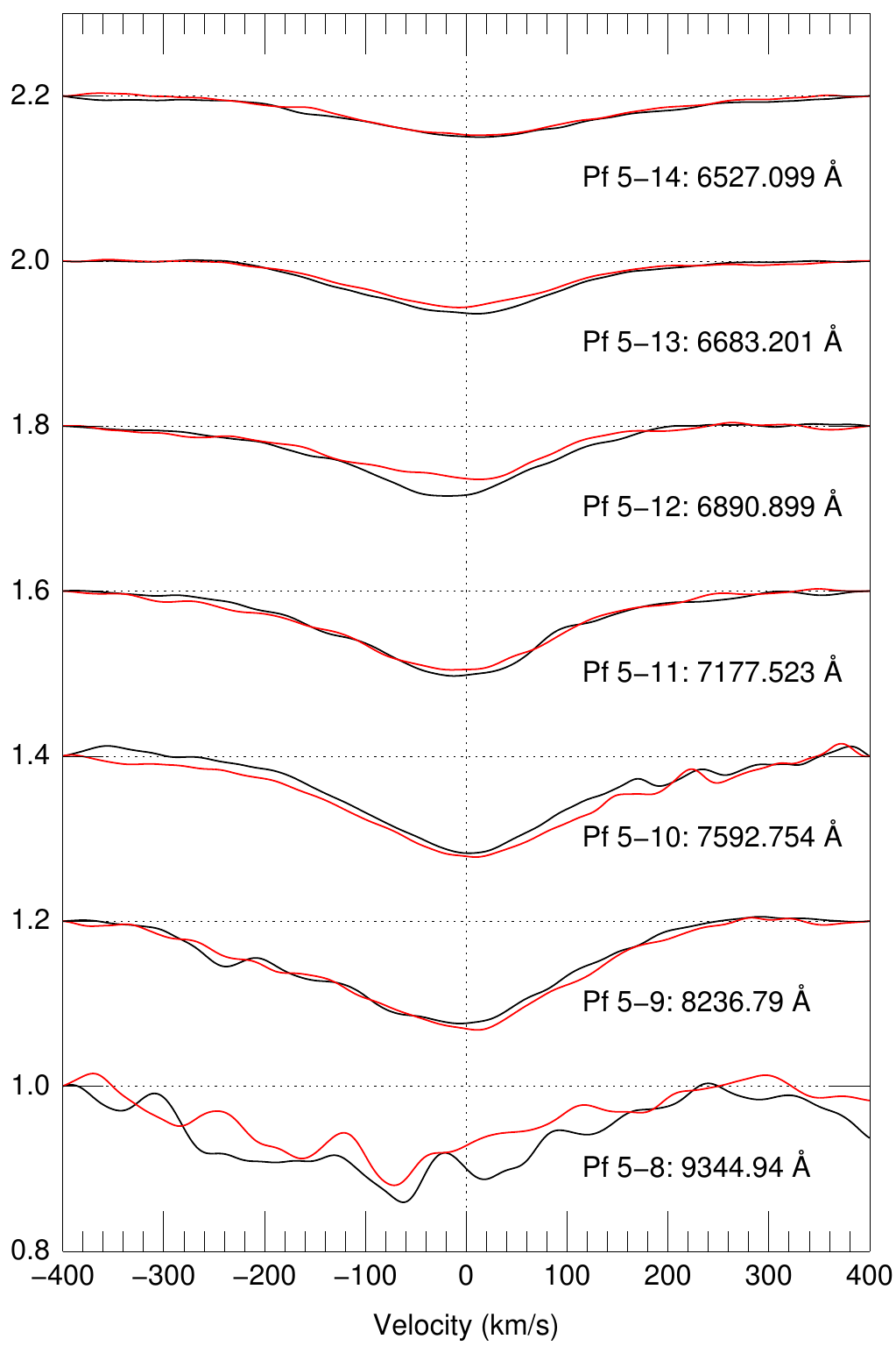}}
 \caption{Disentangled Brackett (left) and Pfund (right) \ion{He}{ii} series of GLS~\num{11448}~Aa (black) and Ab (red). The spectra
          are normalised, displaced in continuum units, and placed in the rest velocity frame of each component systemic RV
          without taking into account wind infilling effects (hence, the small displacement in centroids towards the blue). 
          Differences in S/N are caused by extinction increasing towards the blue and by the residuals of telluric-line subtraction 
          (the most obvious ones being for Pf 5-8 and 5-10). The even-numbered Brackett series is not shown because of the proximity of
          the \ion{H}{i} Balmer series. The spectra have been degraded to $R=\num{10000}$ for display purposes.}
 \label{He_II}
\end{figure}

$\,\!$\indent In Sect.~4.2 we fit FASTWIND models to the disentangled spectra to derive synthetic H+He spectra for 
GLS~\num{11448}~Aa~and~Ab that include the effect of unresolved multiplets and lines from different species and of infilling from
stellar winds. Here we use them to adjust the systemic RVs, for which we analyse the different choices:

\FloatBarrier

\ion{\textit{H}}{i}. Hydrogen lines are intrinsically broader than He lines; they include 
contributions from partially unresolved \ion{He}{ii} lines, which are especially important for stars of very early subtypes such as
GLS~\num{11448}; and some (especially H$\alpha$) are strongly affected by wind infilling. For H$\alpha$, the inclusion of the 
\ion{He}{ii} component in the FASTWIND models leads to a displacement towards the blue of 8-15 km/s depending on the method used to 
determine the central value of such an asymmetric profile\footnote{In addition to the effect of the two species, each line is itself 
a multiplet with a small RV spacing due to fine structure that is usually erased by the effects of rotation and turbulence.}. 
The inclusion of wind infilling leads to an even more complicated effect in H$\alpha$, with an absorption in the core and emission in
the wings (Fig.~\ref{disentangling}), with the absorption core being displaced towards the blue by an additional 25-80 km/s (with the 
exact values depending again on the method used to determine them). However, even for a line less affected by wind infilling such as 
H$\beta$, the effect is $\sim$15~km/s for GLS~\num{11448}~Aa~and~Ab. For those reasons, we did not use \ion{H}{i} 
lines to determine the systemic RVs.

\ion{\textit{He}}{i}. Neutral helium lines have the advantage of being intrinsically narrower than either \ion{H}{i} or \ion{He}{ii} 
lines, which would make them better candidates for measuring the systemic RVs, but for GLS~\num{11448} they have the 
disadvantage of their weakness. Indeed, only three \ion{He}{i} lines are detected in our data in the 3820-\num{11000}~\AA\ range 
($\lambda$4471.55, $\lambda$5875.70, and $\lambda$7065.3, all weak) and, furthermore, all are multiplets. Hence, we decided against 
using \ion{He}{i} lines to determine the systemic RVs.

\ion{\textit{He}}{ii}. 
Ionised helium lines\footnote{For \ion{He}{ii} we use the Atomic Line List v.2.04,
\url{https://linelist.pa.uky.edu/atomic/}, for the central wavelengths.}
have an intrinsic width that is intermediate between those of \ion{H}{i} and \ion{He}{i} lines but for GLS~\num{11448} they have two 
important advantages: they are singlets (other than for small fine-structure splittings) and they are very strong. For those reasons, we
selected the best five \ion{He}{ii} lines in terms of strength, isolation, and absence of telluric contamination 
($\lambda$4199.832, $\lambda$4541.591, $\lambda$5411.531, $\lambda$7177.523, and $\lambda$8236.79, Fig.~\ref{He_II})
and averaged their results for our primary determination of the systemic RVs. We exclude the \ion{He}{ii}~$\lambda$4685.71 
Paschen line because it is strongly affected by winds, which was the reason why \citet{Walb71a} used it as the primary luminosity 
criterion for stars earlier than O9 (see subsection~4.2 for its effect on spectral classification).

\textit{Metallic lines}. There is an \ion{O}{iii} line ($\lambda$5592.252), two \ion{N}{iv} lines ($\lambda$5200.41 and 
$\lambda$6380.7), and two \ion{N}{v} lines ($\lambda$4603.74 and $\lambda$4619.97), all in absorption, that, though weak, can be used to
determine the systemic RVs, as they are even intrinsically narrower than \ion{He}{i} lines. However, our current FASTWIND 
models do not include them, so we cannot calculate their wind infilling. However, we retain them to compute their averages as an 
alternative determination of the systemic RVs.

Our results for the systemic RVs based on the five \ion{He}{ii} lines are given in Table~\ref{orbitalparameters}. They agree 
within one sigma for the two components, a good sign of the absence of systematic effects\footnote{This is a weak argument because
both components have the same spectral type. In a subsequent paper of the \MTW\ series we will explore this issue in more detail for a
variety of O spectral subtypes.}. The correction of the wind
infilling is significant, between 7~km/s and 11~km/s depending on the line, in all cases in the sense of the wind moving the
centroid of the line towards the blue. The alternative values derived from the five metallic lines are $-7.2\pm3.1$~km/s for Aa and 
$-11.5\pm6.3$~km/s for Ab. Those are consistent with the ones from \ion{He}{i}, with the scatter likely arising from the
weakness of the lines and the uncorrected effect of wind infilling affecting them to different degrees. This issue could be verified
in the future with an atmospheric modelling of those metallic lines. The $\gamma$ values for both components are higher 
than the ones from \citet{Maizetal15a} by about $\sim$10~km/s, which is consistent with the previous paper not taking into account  
wind infilling effects. As we will see in future papers of the series, such systematic effects are commonly lurking among 
published RV results for O stars in the literature.

\begin{figure*}
 \centerline{\includegraphics*[width=1.1\linewidth]{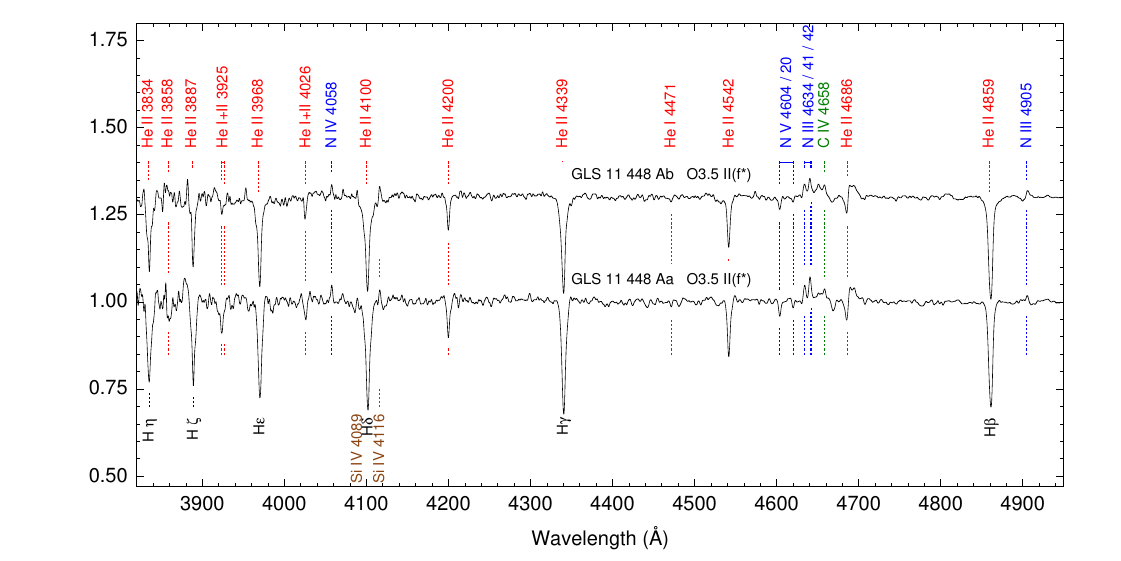}}
 \caption{Disentangled blue-violet spectra of GLS~\num{11448}~Aa,Ab shifted to the stellar reference frame and degraded to the 
          spectral classification $R = 2500$. The two spectra are almost identical and there is no sign of the subtracted 
          ISM. This is an update of Fig.~2 in \citet{Maizetal15a}, where the S/N is higher but the spectra are not disentangled and the 
          ISM is not subtracted.}
 \label{blue-violet}      
\end{figure*}

\begin{table}
\caption{GLS~\num{11448} parameter results.}
\begin{tabular}{lcr@{.}l@{$\pm$}r@{.}lr@{.}l@{$\pm$}r@{.}l}
Quantity          & Units & \mciv{Aa}          & \mciv{Ab}          \\
\midrule
Spectral type     &       & \mciv{O3.5 II(f*)} & \mciv{O3.5 II(f*)} \\
\Teff\            & kK    &    43&8  &  3&0    &    45&3  &  3&0    \\
\logg\            & cgs   &     3&86 &  0&27   &     3&93 &  0&21   \\
$R$               & \Rsun &    15&6  &  0&8    &    15&6  &  0&6    \\
$\logL$           &       &     5&91 &  0&12   &     5&97 &  0&11   \\
$\logLL$          &       &     4&09 &  0&16   &     4&05 &  0&12   \\
\logQ\            & cgs   & $-$12&33 &  0&95   & $-$12&31 &  0&95   \\
\vsini\           & km/s  &    96&0  &  9&6    &    87&0  &  8&7    \\
$\Theta_{\rm RT}$ & km/s  &    64&0  &  6&4    &    60&0  &  6&0    \\
\Mspec            & \Msun &    65&0  & 36&0    &    77&0  & 34&0    \\
\Mevol            & \Msun &    70&0  & 10&0    &    76&0  & 11&0    \\
age               & Ma    &     1&59 &  0&56   &     1&32 &  0&56   \\
\midrule
\end{tabular}
\label{stellarparameters}   
\end{table}

\subsubsection{On the existence of an apsidal motion}

$\,\!$\indent Given the high eccentricity of the GLS~\num{11448}~Aa,Ab orbit and our 14-year spectroscopic coverage, we investigate
whether our data are good enough to determine a possible apsidal motion ($d\omega/dt$) in the system. For the three wavelength ranges 
for which we have the best coverage (blue, \ion{He}{ii}~$\lambda$5411.53, and red) we fitted orbits leaving $d\omega/dt$ as a free 
parameter \citep{Rosuetal20a}. 
In the three cases we obtained very low values, with the highest one being $0.22\pm0.13$~degrees/a, which corresponds to 
$3.0\pm1.8$~degrees in 14 years, so our results are consistent with a non-detection of apsidal motion. How does this compare with 
what we expect from theory? For the general relativistic effect, assuming two 75~\Msun\ stars (see subsection~4.2), the predicted 
result \citep{Rosu25} is $0.004$~degrees/a, which is more than one order of magnitude smaller than the uncertainty in the measurement
above. For the Newtonian effect caused by tides and rotation, the exact value depends on the $k_2$ internal stellar structure constant,
for which there is a possible range of values, but the final result is a few times lower ($0.0005$ to $0.002$~degrees/a) than the 
general relativistic effect. Therefore, we conclude that our non-detection of an apsidal motion is consistent with theory and that 
GLS~\num{11448} is a system where apsidal motion should be very hard to detect, as even a century of data would yield a total apsidal 
motion of a fraction of a degree after almost 400 orbits.

\subsection{The disentangled spectra}

$\,\!$\indent After deriving the new orbital solution in the previous subsection, we used it to disentangle the full 
3820-\num{11000}~\AA\ range. To our knowledge, this is the first time an O-type SB2 system has been disentangled in such a large
wavelength range, including regions of significant telluric absorption and hundreds of DIBs. Such broad wavelength coverage allows 
for the analysis of a large number of lines of the same species (Fig.~\ref{He_II}, which does not include several Paschen and Humphreys
\ion{He}{ii} lines also detected). For similar efforts over a range about one half of that (but avoiding regions of strong telluric 
absorption), see e.g. \citet{Rosuetal20a,Rosuetal22a}.

As part of the outer iteration described above and, given the previous results of \citet{Maizetal15a}, we initially assumed a flux 
fraction for Aa of 0.50 (i.e. a flux ratio of 1:1 between the two components). Such an assumption leads to the two components having
almost identical spectra (Figs.~\ref{He_II}~and~\ref{blue-violet}), with EWs that conform to what is expected from observed stars of 
similar spectral types and to FASTWIND models for the corresponding \Teff\ and luminosities (see below). 
Therefore, we left the value of the flux 
fractions as 0.50 and we estimate that the magnitude difference between the two components is at most 0.1~mag. This should be verified 
in the near future by resolving the system with interferometry, something that may be possible with CHARA given that the maximum 
separation in the eccentric orbit is expected to be around 1~mas.

The disentangled blue-violet spectra were degraded to $R = 2500$ for their spectral classification. \citet{Maizetal15a} classified both
components as O3.5~If*, with the spectral subtype derived from the ratios of nitrogen lines \citep{Walbetal02b} and the luminosity
class derived from the \ion{He}{ii}~$\lambda$4685.71 line. At that time the spectral classification grid only had a II luminosity class
down to O6 \citep{Maizetal16} but since then it has been extended to earlier O subtypes when \ion{He}{ii}~$\lambda$4685.71 has a 
P-Cygni profile, following the same criterion as for mid-O subtypes (Ma\'{\i}z Apell\'aniz et al. in prep.). As that is the case for 
both components of GLS~\num{11448}, we change their spectral classification to O3.5~II(f*).

\FloatBarrier

After spectral disentangling, the spectrum of each component was analysed independently following the methodology described in 
\citet{Holgetal25b}. The effective temperature (\Teff) and surface gravity (\logg) were derived through quantitative spectroscopic 
analysis using a grid of FASTWIND stellar atmosphere models \citep{Santetal97,Pulsetal05} and the IACOB-GBAT fitting tool 
\citep{SimDetal11d}, while projected rotational velocities (\vsini) were determined with IACOB-BROAD following the Fourier-transform 
and goodness-of-fit approach described by \citet{SimDHerr07,SimDHerr14}. In both cases, the analysis was based on the 
\ion{N}{v}~$\lambda 4603.74$ line. We used the Berkeley~90 distance of \citet{Maizetal22b} and the extinction parameters of 
\citet{MaizBarb18}, determined by fitting the optical-NIR photometry of GLS~\num[detect-all]{11448} to the SED grid of 
\citet{Maiz13a} with the CHORIZOS code \citep{Maiz04c} and the extinction-law family of \citet{Maizetal14a}.
Stellar radii ($R$), luminosities ($L$), and spectroscopic luminosities ($\mathcal{L}=\Teff^4/g$,
\citealt{LangKudr14,deBuetal24a}) and masses (\Mspec) were derived using standard relations involving the synthetic flux of the 
best-fitting FASTWIND model \citep{Kudr80,Holg19}. The surface gravity was corrected for centrifugal acceleration following 
\citet{Herretal92} and \citet{Repoetal04}. Uncertainties in the final fundamental parameters were 
estimated through Monte Carlo propagation of the errors in the spectroscopic and photometric quantities. To calculate the 
evolutionary initial masses (\Mevol) and ages, we used the 3-D grid of stellar models of \citet{Maiz13a} which for massive
stars is based on the Geneva evolutionary tracks without rotation \citep{LejeScha01}. The masses were computed from the real luminosity
instead of from the spectroscopic luminosity, as the distance and extinction are more precisely determined than the gravities, which is
also the reason why the \Mspec\ uncertainties are larger than those of \Mevol. Our results include the effect of the correlated
nature of the derived quantities and were compared with the Bonnsai tool \citep{Schnetal16}, which uses the alternative Bonn 
evolutionary tracks, with an agreement better than one sigma. 

The parameters listed in Table~\ref{stellarparameters} confirm the results of the spectral classification in the sense that the two
components are very similar, to the point that all results are within half a sigma of either uncertainty. The \Teff\ uncertainties are
relatively large because for very early spectral subtypes the \ion{He}{i} lines used in our procedure become very weak 
\citep{Walbetal02b}. In order to improve them a future analysis should use nitrogen lines instead.

The most important result is the extremely high evolutionary masses derived for both components of 70$\pm$10~\Msun\ and 
76$\pm$11~\Msun, respectively\footnote{The derived evolutionary mass for Ab is higher than for Aa but the error bars are large
enough for the mass ratio to be compatible with the more precise one derived in Table~\ref{orbitalparameters}.}. 
They are the highest masses ever derived for any Galactic O star (excluding ``early-slash'' stars, see 
below).  The closest contenders are Cyg~OB2-9~A \citep{Nazeetal10}, Cyg~OB2-B17~A \citep{Stroetal10}, HD~\num{93129}~Aa 
(\citealt{Maizetal17a}, an improved measurement will be the subject of an incoming \MTW\ paper), and WR~21a~B \citep{Barbetal22}, all 
of them with values 10-20~\Msun\ lower than the ones for GLS~\num{11448} (in different degrees of certainty). Stars more massive than 
GLS~\num{11448}~Aa,Ab are expected to be born with winds strong enough to appear of type WNh even in the ZAMS, with WR20a being a twin 
system of such stars with masses of 83$\pm$5~\Msun\ and 82$\pm$5~\Msun, respectively \citep{Bonaetal04}. Similarly, GLS~\num{22159}~A
(= WR21a~A), an ``early-slash'' or early Of/WN star \citep{Sotaetal14} classified as O2~Ifc*/WN5 (OWN~II) has a mass of 
$93.2^{+2.2}_{-1.9}$~\Msun\ \citep{Barbetal22}. Therefore, if our results are confirmed, the limit between O and ``early-slash''+ WNh 
stars appears to be around 80~\Msun\ but it is possible that once rotation, metallicity, and evolutionary effects are included, an 
overlap in masses may reveal itself (for example, Cyg~OB2-9~A and Cyg~OB2-B17~A are more evolved than GLS~\num{11448}~Aa,Ab and 
HD~\num{93129}~Aa). The main obstacle to elucidate this topic is the rarity of these objects, which enhances the importance of our 
result. Given that our high masses are calculated from evolutionary tracks, they should be confirmed with a measurement of the $\sin i$
of the orbit from interferometry. To reconcile the values of the Keplerian minimum masses with the evolutionary masses the inclination 
should be around 50~degrees.

\begin{figure}
 \centerline{\includegraphics*[width=\linewidth]{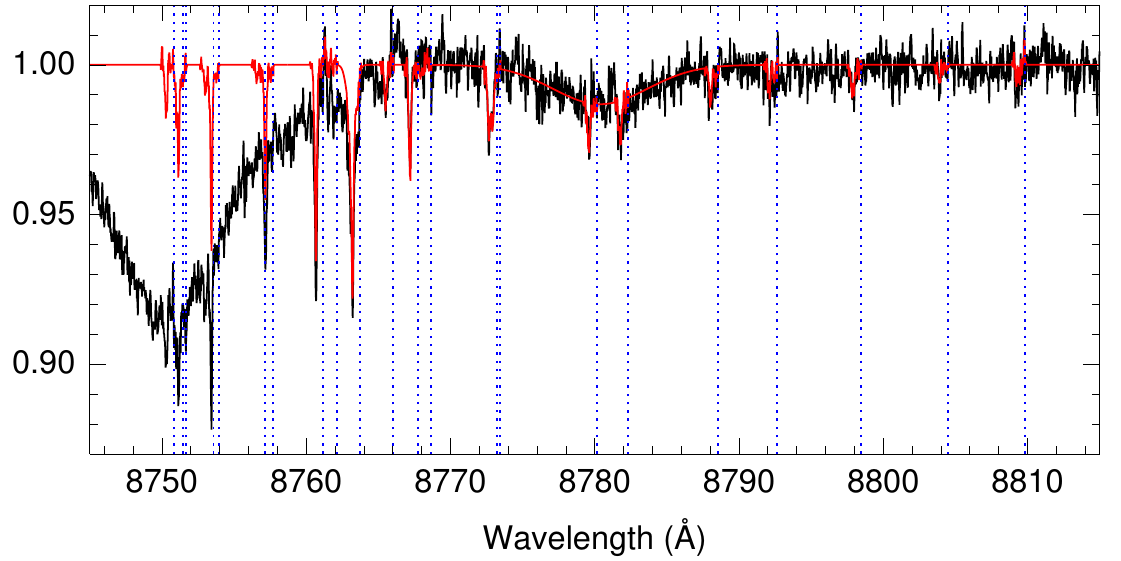}}
 \caption{Section of the UNWIND extraction for GLS~\num{11448}~Aa (in the rest frame of the star determined from
         Table~\ref{orbitalparameters}) with a significant portion of the Phillips C$_2$ (2,0) band. The black line is the star+ISM 
         line (with the most prominent feature being the \ion{H}{i} Pa$\iota$ line) and the red line the fitted ISM, including the 
         C$_2$ lines, DIBN8764, and DIBI8781. The vertical lines mark the expected position of the C$_2$ lines in the rest frame of the 
         star, note the RV displacement.}
 \label{C2}      
\end{figure}

\begin{figure}
 \centerline{\includegraphics*[width=\linewidth]{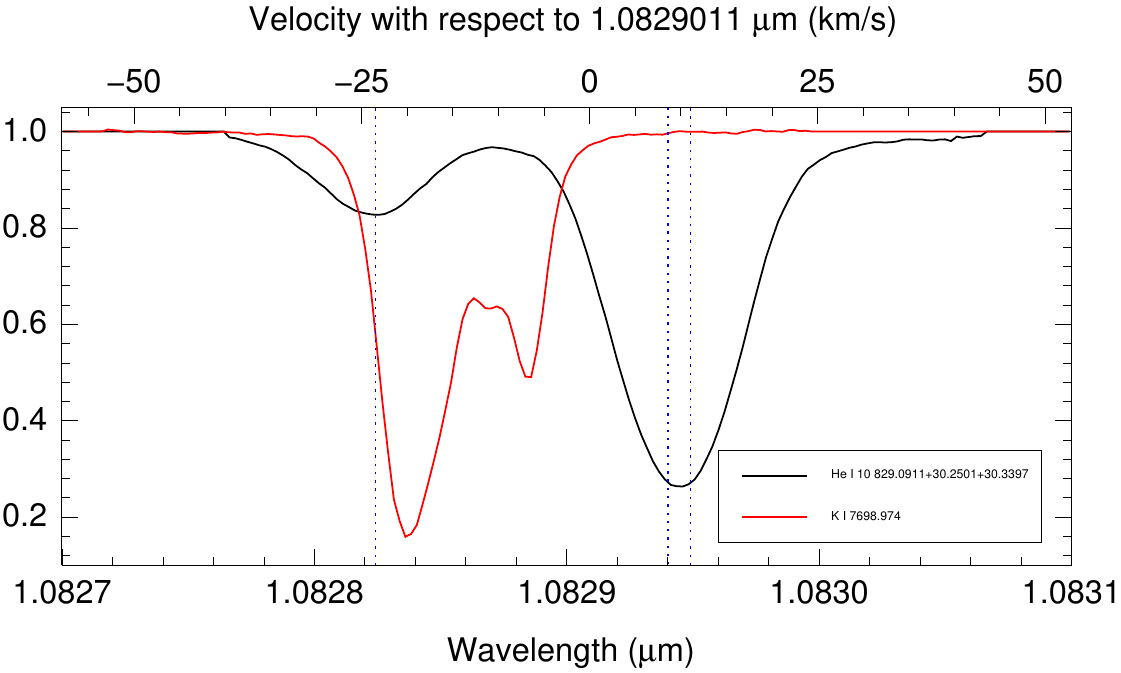}}
 \caption{A comparison of the profiles of the interstellar \ion{He}{i}~$\lambda$\num{10830} triplet and \ion{K}{i}~$\lambda$7698.974 
          line in the GLS~\num{11448} sightline, with the latter placed in the RV reference frame of the first component of the 
          \ion{He}{i} triplet (\num{10829.0911}~\AA), see also Fig.~\ref{ISM_RV}. The dashed blue vertical lines mark the tentative
          \ion{He}{i} RV of $-23.5$~km/s in our barycentric reference frame, which is blueshifted with respect to either 
          \ion{K}{i} cloud. Stellar lines are broader in RV than the plotted range (see e.g. Fig.~\ref{He_II}).}
 \label{He_I}      
\end{figure}

\FloatBarrier

\subsection{The foreground ISM}

$\,\!$\indent As described in Appendix~B, we used GLS~\num{11448} as the primary standard to build the DIB library for UNWIND. 
Table~\ref{dibs_all} list the characteristics of the 631 DIBs in the library, including the EWs for GLS~\num{11448}, and 
Table~\ref{dibs_mult} gives more details about the 37 for which we used multiple Gaussians. Those tables can be considered expanded
versions of Table~2 in \citet{Maizetal15c}. In this subsection we use the UNWIND results to briefly expand the results from that paper.

\citet{Maizetal15c} proposed that the reason why the GLS~\num{11448} sightline is more extinguished than that of GLS~\num{11449} but 
with lower EWs for most DIBs is that the additional extinction originates in a UV-shielded ($\zeta$ sightline) cloud core that 
contributes little to the DIB EWs. One consequence of that would be the presence of C$_2$ absorption on the GLS~\num{11448} 
sightline, something that is readily apparent in the UNWIND ISM output, where we detect strong signals in the Phillips (1,0), (2,0), and
(3,0) bands around \num{10150}~\AA, 8765~\AA, and 7720~\AA, respectively (Figs.~\ref{C2}~and~\ref{ISM_RV}). We even detect the seldom 
seen Phillips (0,0) band around \num{12100}~\AA\ \citep{Hamaetal19b} in a range of our CARMENES spectra that we did not process with 
UNWIND, though for that band we lose some lines that fall in an order gap. The RV of the C$_2$ lines indicates that it is 
associated with the cloud with the most negative RV of the two along the sightline (Fig.~\ref{ISM_RV}). On the other hand, we 
also analysed equivalent spectra (with shorter total exposure time and, hence, S/N) of GLS~\num{11449} that we also obtained with 
CARMENES and we were only able to detect a weak signal, with EWs roughly an order of magnitude lower than for GLS~\num{11448}. 
Therefore, the existence of such a molecular-gas rich, UV-shielded core along the GLS~\num{11448} sightline (that has little or no
effect on the GLS~\num{11449} sightline) is confirmed.

An aspect of the GLS~\num{11448} sightline that could not be studied by \citet{Maizetal15c} is the presence of \ion{He}{i} interstellar
absorption. Under normal conditions in the ISM, most \ion{He}{i} atoms in the ISM are in the singlet 1s$^2$~$^1$S$_0$ ground state,
which requires EUV photons to excite and, for that reason, makes \ion{He}{i} not detectable in absorption at longer wavelengths (the 
same reason why \ion{H}{i} is
not detected, either, even though in that case the required energies are lower). In an \ion{H}{ii} region, however, some recombinations
take place through the triplet configuration and those end up in the metastable 1s2s~2~$^3$S$_1$ level. There, they may stay long 
enough to absorb radiation from a background star and produce interstellar absorption lines \citep{Drai26}. Three such lines have 
wavelengths longer than 3000~\AA: the hard-to access from the ground $\lambda$3187.7 plus $\lambda$3888.6 and $\lambda$\num{10830}. All
three are triplets but the separation of the third one is considerable larger (1.25~\AA) than that of the other two (0.16~\AA\ for the 
second and even less for the first). $\lambda$3888.6 was first detected in absorption in an O star immersed in an \ion{H}{ii} region by
\citet{Wils37}. Some authors have attempted or suggested the detection of $\lambda$\num{10830} in absorption in OB-star sightlines
\citep{Indretal09,GalaKrel12} but, to our knowledge, without success.

In the UNWIND output we detect a clear \ion{He}{i}~$\lambda$3888.6 interstellar absorption with an EW of $\sim$60~m\AA\ but with a
large uncertainty due to the low S/N in our data at short wavelengths. More importantly, we detect for the first time the 
$\lambda$\num{10830} triplet in absorption in an OB-star (Fig.~\ref{He_I}), with a very strong EW of 568~m\AA\ including the three 
components of the triplet. Both $\lambda$\num{10829.0911} and the unresolved $\lambda\lambda$\num{10830.2501},{30.3397} doublet are 
consistent with a (solar) barycentric RV of $-23.5$~km/s, blueshifted with respect to the two components detected in the rest of 
the absorption lines (Fig.~\ref{ISM_RV}). This could be caused by an expansion of the \ion{H}{ii} region or possibly by the complicated 
radiation transport involved \citep{Drai26}. The lack of stellar absorption in such an early O-spectrum makes the detection of the
interstellar line easier but we note that the stellar component would be much broader (Fig.~\ref{He_II}).

\section{Summary and future work}

$\,\!$\indent In this first \MTW\ paper we have established the objectives of the project, which revolve around the determination of
the individual and collective multiplicity properties of the largest sample of massive stars in the solar neighbourhood ever
assembled. The sample will be built from the ALS catalog and data will consist of a combination of spectroscopy, astrometry, and 
photometry obtained from the ground and from space. We have presented the software package that will be used for the analysis, with 
an emphasis on the UNWIND spectral disentangling tool. We have produced the most complete DIB library to date and for the massive 
near-twin binary GLS~\num{11448} we have (a) computed a new SB2 orbit, (b) determined that it is composed of the two non-early-slash 
O stars with the highest evolutionary masses known, and (c) detected the interstellar \ion{He}{i}~$\lambda$\num{10830} triplet in 
absorption in an OB-star for the first time ever.

\MTW\ is a long-term project for which we have already planned or started writing several papers, including two that are near completion.
The first one will be an analysis of the systemic RVs and variability of single O stars to better establish the detection limits of 
spectroscopic binaries and determine the systematic and random uncertainties in the systemic RVs caused by pulsations and line infilling. 
The second paper will be on the complex hierarchical system $\tau$~CMa \citep{Rosuetal25}.

\begin{acknowledgements}
J.~M.~A., S.~R., E.~M.~F., and J.~A.~M.-C acknowledge support from the Spanish Government Ministerio de Ciencia e 
Innovaci\'on and Agencia Estatal de Investigaci\'on (\num{10.13039}/\num{501100011033}) through grant PID2022-\num{136640}~NB-C22.
R.~C.~G. acknowledges support in the form of travel funds from the Red Espa\~nola de \Gaia.
G.~H. and S.~S.-D. acknowledge support from the Spanish Government Ministerio de Ciencia e Innovaci\'on, Agencia Estatal de 
Investigaci\'on (\num{10.13039}/\num{501100011033}) and the European Regional Development Fund, FEDER under grants 
PID2021-\num{122397}~NB-C21 and PID2024-\num{159329}~NB-C21
S.~R. acknowledges support from the Swiss National Foundation (SNF) project No \num{212143}.
M. A.-M. acknowledges support from the ``La Caixa'' Foundation (ID \num{100010434}) under the fellowship code 
LCF/BQ/PI23/\num{11970035}.
This paper includes data obtained with:
the 1.2~m Mercator Telescope (MT),
the 2.6~m Nordic Optical Telescope (NOT), 
the 3.6~m Telescopio Nazionale Galileo (TNG),
the 4.2~m William Herschel Telescope (WHT),
the 10.4~m Gran Telescopio Canarias (GTC) at the Observatorio del Roque de los Muchachos, La Palma, Spain;
the 2.2~m and 3.5~m Telescopes (CAHA2.2 and CAHA3.5) at the Centro Astron\'omico Hispano-Alem\'an/Andaluz, Almer{\'\i}a, Spain;
the 2.2~m Telescope (OLS2.2) and 
the 3.6~m Telescope (OLS3.6) at the Observatorio~de~La~Silla, Chile; 
the 8.2~m Very Large Telescope (VLT) at Cerro Paranal, Chile; and
the 9.2~m Hobby-Eberly Telescope (HET) at MacDonald Observatory, U.S.A.
We thank the staff at those observatories for their support.
We also thank Alexander Ebenbichler for corrections on a previous version of this paper.
The paper is dedicated to Rodolfo H. Barb\'a, who spent a considerable amount of time in laying the original work for this project
but unfortunately passed away before seeing the fruits of his work in the form of this first paper of the series.
\end{acknowledgements}

\bibliographystyle{aa} 
\bibliography{general}   

$\,\!$

\vfill

\eject

\begin{appendix}

\section{Glossary}

$\,\!$\indent We provide a list of acronyms and terms used in this paper. 

\begin{itemize}
 \item $a$: (Orbital) semi-major axis.
 \item ALS: Alma Luminous Star catalog.
 \item BRJD: Barycentric Julian Date $-2.4\cdot 10^6$~d.
 \item CAHA: Centro Astron\'omico Hispano-Alem\'an/Andaluz.
 \item CAHA2.2: CAHA 2.2~m telescope.
 \item CAHA3.5: CAHA 3.5~m telescope.
 \item CollDIBs: Collection of DIBs \citep{Maiz15a}.
 \item DIB: Diffuse Interstellar Band.
 \item DR3: (\Gaia) third data release.
 \item DR4: (\Gaia) fourth data release.
 \item $e$: (Orbital) eccentricity.
 \item ESO: European Southern Observatory.
 \item EW: Equivalent Width.
 \item FWHM: Full-Width at Half Maximum.
 \item GLS: Galactic Luminous Star.
 \item GOSC: Galactic O-Star Catalog.
 \item GOSSS: Galactic O-Star Spectroscopic Survey.
 \item GTC: Gran Telescopio Canarias.
 \item $\gamma$: Systemic radial velocity.
 \item HET: Hobby-Eberly Telescope.
 \item $i$: (Orbital or rotational) inclination.
 \item IFU: Integral Field Unit.
 \item ISM: Interstellar Medium.
 \item $K$: Semi-amplitude of the RV curve.
 \item $\mathcal{L}$: Spectroscopic luminosity ($=\Teff^4/g$).
 \item LiLiMaRlin: \textbf{Li}brary of \textbf{Li}braries of \textbf{Ma}ssive-Star 
       High-\textbf{R}eso\textbf{l}ut\textbf{i}o\textbf{n} Spectra.
 \item \logQ: Wind-strength parameter \citep{Holgetal18}.
 \item \Mspec: Spectroscopic mass.
 \item \Mevol: Initial evolutionary mass.
 \item MT: Mercator Telescope.
 \item \MTW: Multiplicity of Massive stars in the Milky Way.
 \item NOT: Nordic Optical telescope.
 \item $\omega$: Argument of periastron.
 \item OHP: Observatoire de Haute-Provence.
 \item OHP1.9: Observatoire de Haute-Provence 1.9~m telescope.
 \item OLS: Observatorio de La Silla.
 \item OLS2.2: Observatorio de La Silla 2.2~m telescope.
 \item OLS3.6: Observatorio de La Silla 3.6~m telescope.
 \item OP: Observatorio Paranal.
 \item ORM: Observatorio del Roque de los Muchachos.
 \item $P$: (Orbital) period.
 \item $q$: Mass ratio.
 \item RV(S): Radial Velocity (Spectrometer).
 \item SAAO: South African Astronomical Observatory.
 \item SALT: Southern African Large Telescope.
 \item SB1(E): Single-lined spectroscopic (Eclipsing) Binary.
 \item SB2(E): Double-lined spectroscopic (Eclipsing) Binary.
 \item SED: Spectral Energy Distribution.
 \item SONG: Stellar Observations Network Group.
 \item $T_0$: Time of periastron passage.
 \item $\Theta_{\rm RT}$: Macroturbulent velocity.
 \item TNG: Telescopio Nazionale Galileo.
 \item VLT: Very Large Telescope.         
 \item WDS: Washington Double Star catalog \citep{Masoetal01}.
 \item WHT: William Herschel Telescope. 
 \item (ZA)MS: Zero-Age Main Sequence. 
\end{itemize}

\vfill

\eject

\section{The DIB catalog}

\subsection{Building the catalog}

$\,\!$\indent We present the details of the DIB catalog associated with UNWIND and this paper. To build the
 catalog we used six ISM standard stars (one primary and five secondary):

\begin{itemize}
 \item \textit{GLS~\num[detect-all]{11448}} is the object of study of section~4 in this paper and is our primary 
       ISM standard for three reasons: high extinction, very scarce presence of stellar features due to the early-O spectral types
       of its two components, and its SB2 nature (which allows for an additional differentiation between stellar and ISM features 
       using multi-epoch spectroscopy).
 \item \textit{Cyg~OB2-12} is selected due to its even higher extinction than GLS~\num{11448} and its contrast in spectral
       classification to the primary standard, as it is a late-B hypergiant with few stellar features in common with it other than
       H and some \ion{He}{i} lines. We note its higher number of stellar features and variability.
 \item \textit{HD~\num[detect-all]{183143}} is a late-B supergiant that has been previously used as an ISM standard. Its presence 
       allows for the verification of some DIBs and a comparison with previous results.
 \item \textit{V747~Cep} is another obscured early-O spectroscopic binary that we use to verify some of the DIBs
       seen only in the primary standard (avoiding contamination by stellar lines in the two B stars above) that has the
       advantage of being in an \ion{H}{ii} region (Berkeley~59), thus probing a different environment.
 \item \textit{$\zeta$~Oph} is a fast-rotating late-O runaway star that has a significantly lower extinction than the first four
       standards. It is included here because it is the prototype for (quasi-)single-kinematic-component low-UV-exposure ISM
       sightlines (so-called $\zeta$ sightlines after it, \citealt{Kreletal97}).
 \item \textit{$\sigma$~Sco~Aa,Ab} is the counterpart of $\zeta$~Oph for (quasi-)single-kinematic-component ISM 
       sightlines for the high-UV exposure case ($\sigma$ sightlines, \citealt{Kreletal97}). It is a visual double with a
       Hipparcos separation of 0\farcs428 and with Aa being itself an SB2 system \citep{Maizetal21b}. 
       Our data includes the combined light of the three stars.
\end{itemize}

\begin{figure*}
 \vspace{-2mm}
 \centerline{\includegraphics*[width=0.49\linewidth]{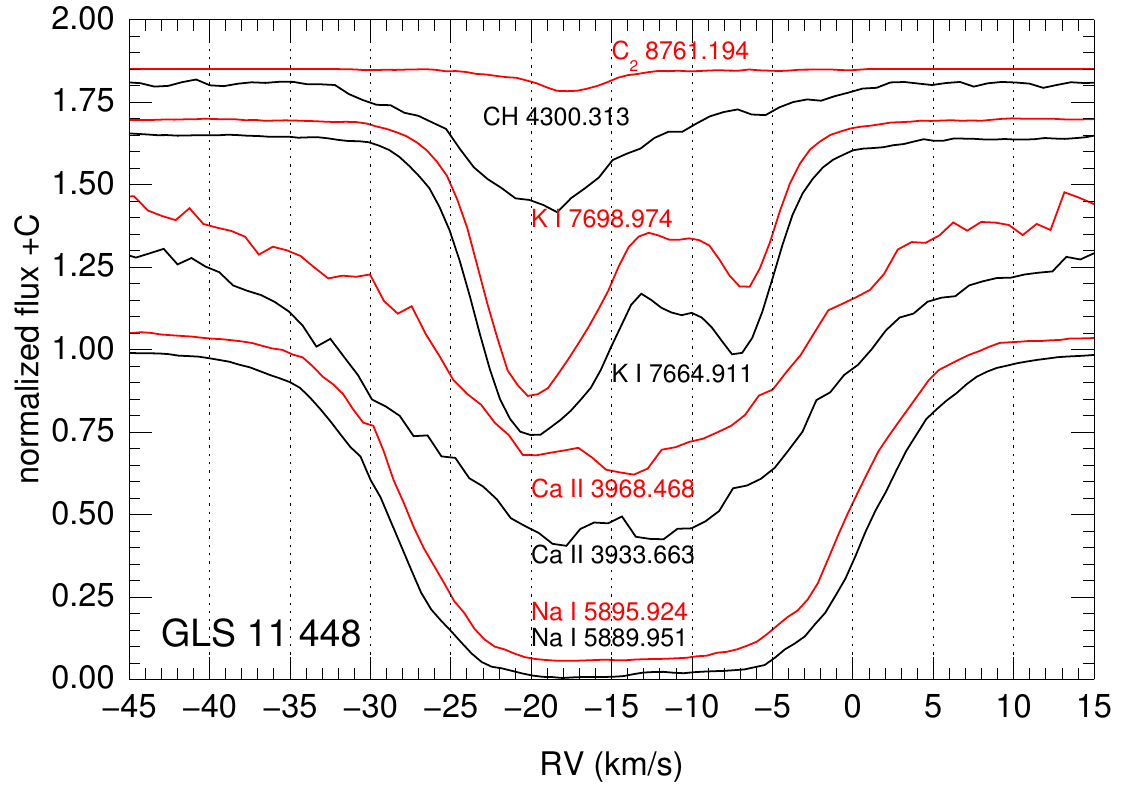} \
             \includegraphics*[width=0.49\linewidth]{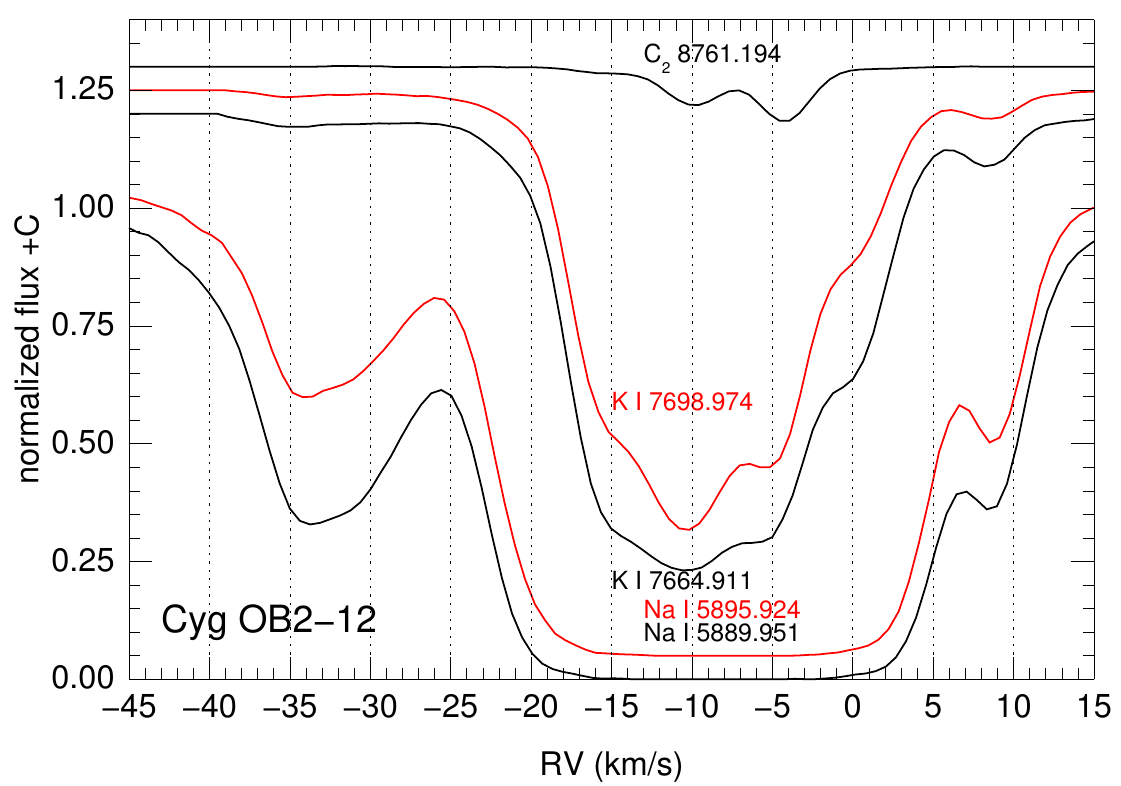}}
 \centerline{\includegraphics*[width=0.49\linewidth]{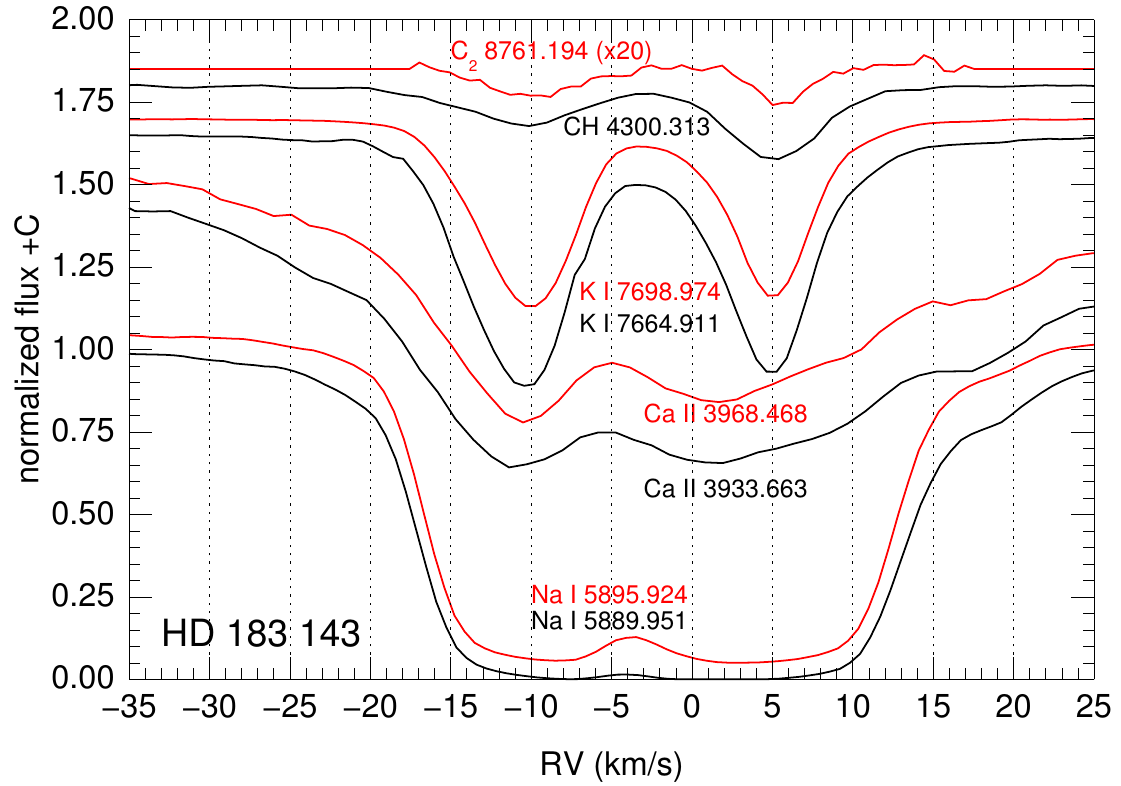} \
             \includegraphics*[width=0.49\linewidth]{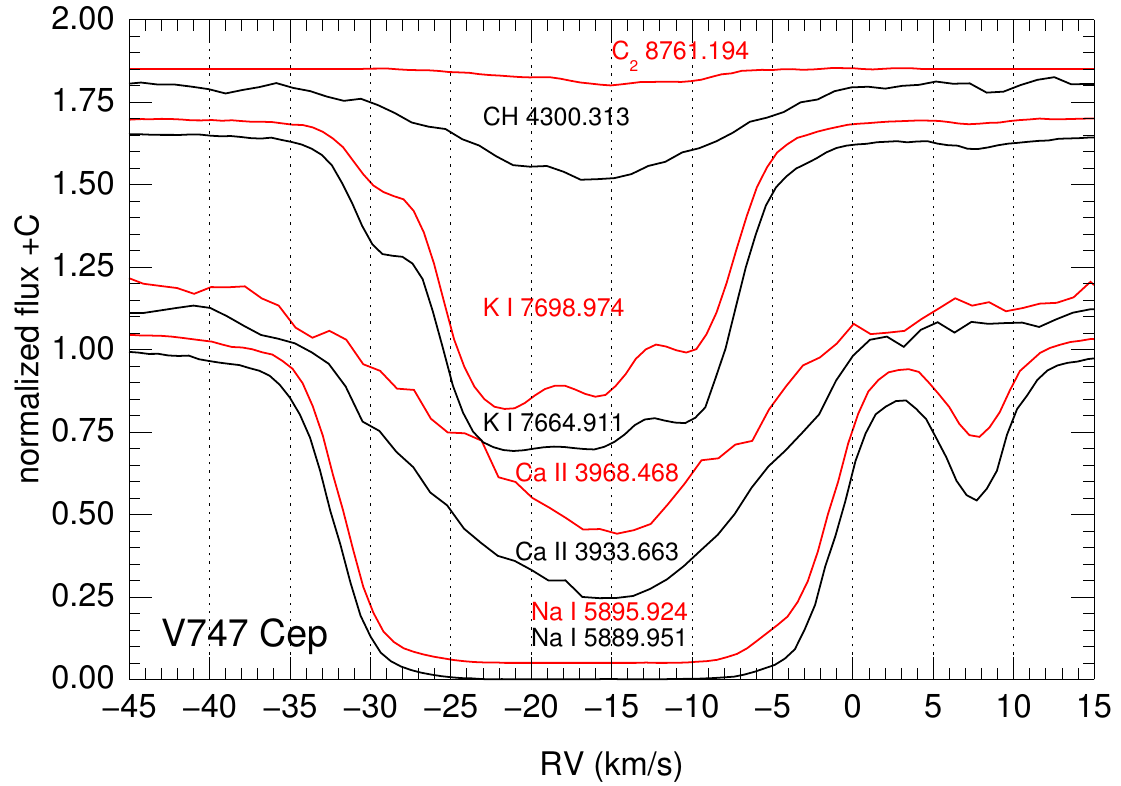}}
 \centerline{\includegraphics*[width=0.49\linewidth]{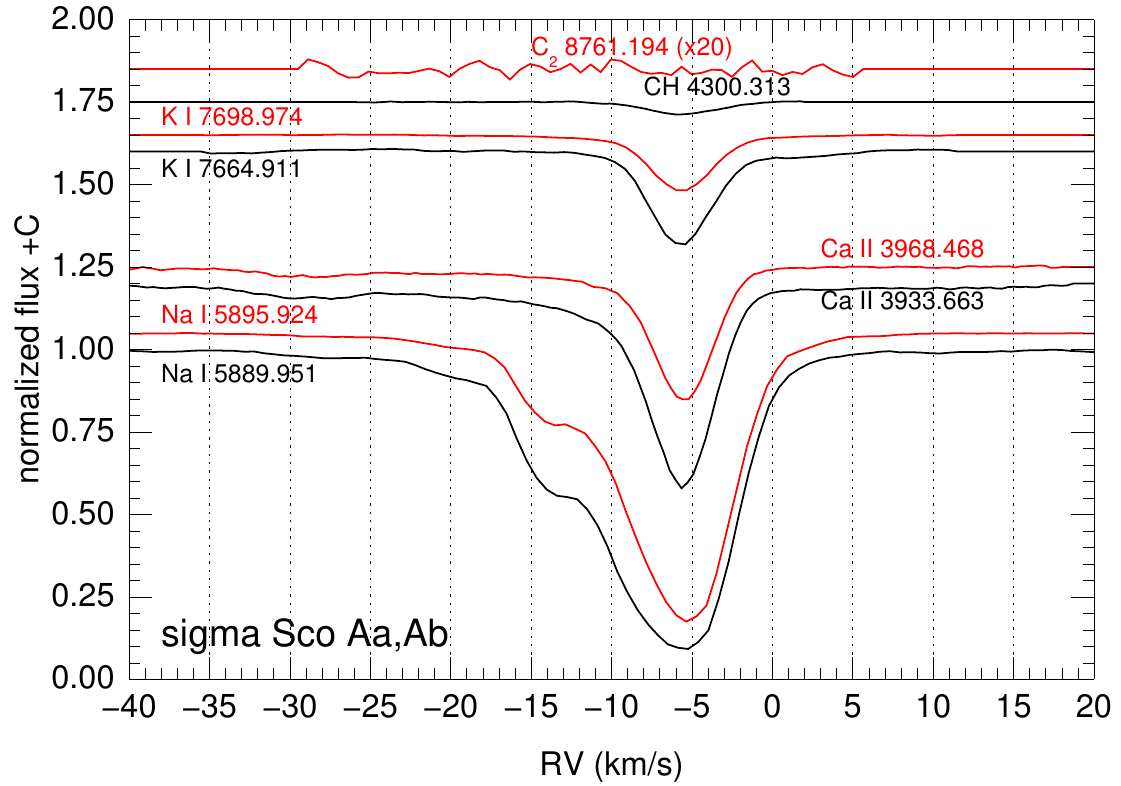} \
             \includegraphics*[width=0.49\linewidth]{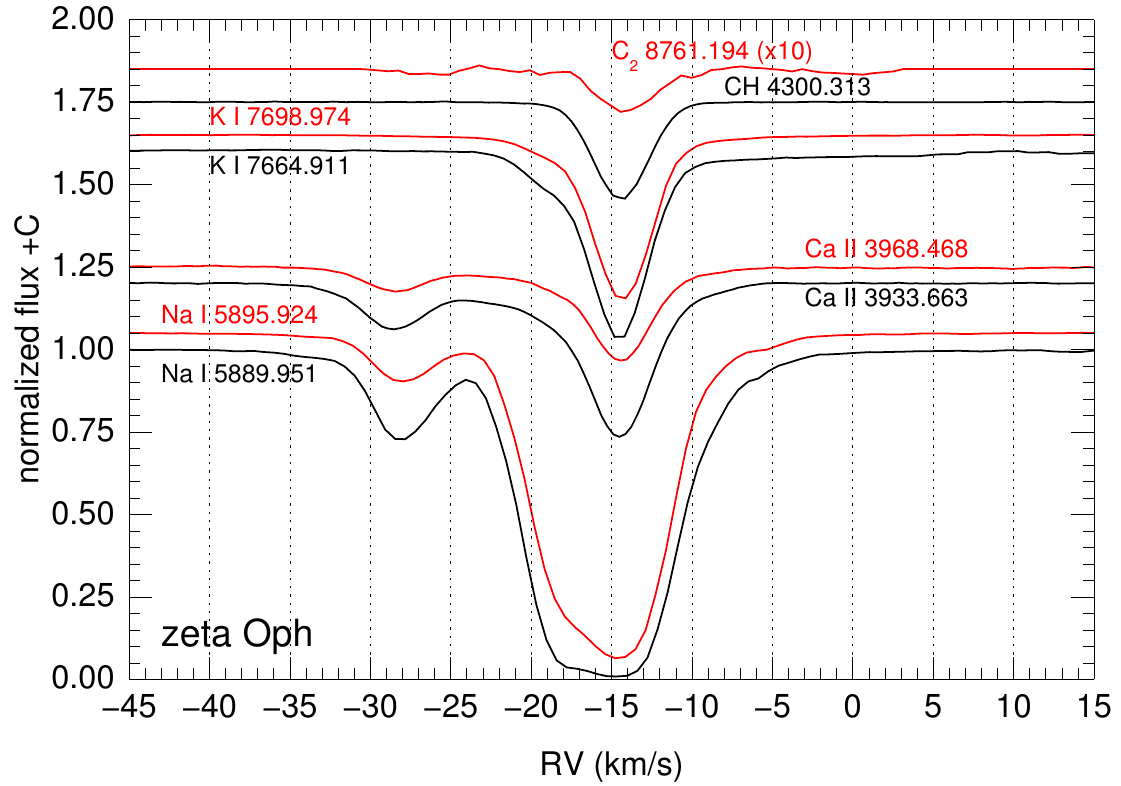}}
 \caption{Absorption lines used to determine the ISM rest frame for the six standard stars. The \ion{Ca}{ii} lines are not available
          for Cyg~OB2-12 in our data. The vertical scale of the C$_2$~$\lambda$\,8761.194 line was expanded in those cases where 
          it is weak.}
 \label{ISM_RV}   
 \vspace{-3mm}   
\end{figure*}

We used the multi-epoch high-resolution \lili\ spectroscopy to produce high-S/N spectra of the six standards. 
Before analysing the DIBs, we studied ISM lines with known central wavelengths
(\ion{Na}{i}~$\lambda\lambda$\,5889.951,5895.924; \ion{Ca}{ii}~$\lambda\lambda$\,3933.633,3968.468; 
\ion{K}{i}~$\lambda\lambda$\,7664.911,7698.974; CH~$\lambda$\,4300.313; and C$_2$~$\lambda$\,8761.194) to determine the average 
RV of each ISM sightline (Fig.~\ref{ISM_RV} and Table~\ref{standards}) and, in that way, measure the DIB central wavelengths 
$\lambda_0$ with respect to the ISM rest frame. While the $\zeta$~Oph and $\sigma$~Sco~Aa,Ab sightlines are dominated by a single 
cloud (with weak secondary components that are unlikely to contribute much for DIBs), the other four are more complex. 
GLS~\num{11448}, Cyg~OB2-12, and HD~\num{183143} have (at least) two main kinematic components and V747~Cep has a broad main 
component and a weak secondary one. This is important for narrow DIBs with possible intrinsic level structure 
\citep{Galaetal02,Bernetal18}, where the fitted DIB profiles should appear kinematically broadened for those four stars.

\begin{table}
\caption{Information for the six ISM standard stars}
\begin{tabular}{lclr}
\midrule
Star               & ID & Spectral                  & \mci{RV}        \\
                   &    & $\;$type                  & \mci{(km/s)}    \\
\midrule
GLS~\num{11448}    &  1 & O3.5 II(f*) + O3.5 II(f*) & $-$14.9$\pm$2.0 \\
Cyg~OB2-12         &  2 & B5 Ia+                    &  $-$9.2$\pm$1.6 \\
HD~\num{183143}    &  3 & B8 Iae                    &  $-$0.6$\pm$3.8 \\
V747~Cep           &  4 & O5.5~V(n)((f))            & $-$15.6$\pm$1.7 \\
$\zeta$~Oph        &  5 & O9.2 IVnn(e)              & $-$14.8$\pm$2.0 \\
$\sigma$~Sco~Aa,Ab &  6 & B0.7 IV + B1: V           &  $-$7.3$\pm$1.8 \\
\midrule
\end{tabular}
\vspace{-5mm}
\label{standards}
\end{table}

\begin{table*}
\caption{DIBs used for the UNWIND database, with their average $\lambda_0$ and FWHM, the EW measured for GLS~\num{11448}, and the number of Gaussians ($n_{\rm G}$) and stars (IDs in Table~\ref{standards}) used to determine the profile.}
\addtolength{\tabcolsep}{-1mm}
\hspace{-10mm}\begin{tabular}{lr@{.}lr@{.}l@{}rc@{}c@{\hspace{5mm}}lr@{.}lr@{.}l@{}rc@{}c@{\hspace{5mm}}lr@{.}lr@{.}l@{}rc@{}c}
\midrule
Name      & \mcii{$\lambda_0$} & \mcii{$\!\!$FWHM}  & \mci{$\!\!\!\!$EW}     & $n_{\rm G}$ & $\!\!$Stars & Name      & \mcii{$\lambda_0$} & \mcii{$\!\!$FWHM}  & \mci{$\!\!\!\!$EW}     & $n_{\rm G}$ & $\!\!$Stars & Name      & \mcii{$\lambda_0$} & \mcii{$\!\!$FWHM}  & \mci{$\!\!\!\!$EW}     & $n_{\rm G}$ & $\!\!$Stars \\
          & \mcii{(\AA)}       & \mcii{$\!\!$(\AA)} & \mci{$\!\!\!\!$(m\AA)} &             & $\!\!$used  &           & \mcii{(\AA)}       & \mcii{$\!\!$(\AA)} & \mci{$\!\!\!\!$(m\AA)} &             & $\!\!$used  &           & \mcii{(\AA)}       & \mcii{$\!\!$(\AA)} & \mci{$\!\!\!\!$(m\AA)} &             & $\!\!$used  \\
\midrule
\cgr DIBB4075  & \num{ 4074}&7  &  10&5  &  156 &     1 &     123    & \cre DIBI5671  & \num{ 5671}&1  &   9&9  &   75 &     1 &     14     &      DIBN5986  & \num{ 5986}&44 &   0&66 &    2 &     1 &     1234   \\
\cgr DIBB4179  & \num{ 4178}&7  &  20&8  &  230 &     1 &     123    & \cgr DIBB5705  & \num{ 5704}&8  &  12&3  &  150 &     1 &     124    &      DIBN5988  & \num{ 5988}&08 &   0&77 &    8 &     1 &     12345  \\
     DIBN4259  & \num{ 4258}&95 &   1&29 &    5 &     1 &     134    &      DIBN5705  & \num{ 5705}&16 &   2&39 &   98 &     1 &     123456 &      DIBN5989  & \num{ 5989}&42 &   0&53 &    5 &     1 &     1234   \\
     DIBN4364  & \num{ 4363}&82 &   0&47 &    8 &     1 &     14     &      DIBN5708  & \num{ 5707}&67 &   0&47 &    3 &     1 &     125    &      DIBN5996  & \num{ 5995}&79 &   0&73 &    6 &     1 &     1234   \\
     DIBN4372  & \num{ 4371}&68 &   2&77 &   34 &     1 &     1234   &      DIBN5710  & \num{ 5710}&46 &   0&66 &    6 &     1 &     1245   &      DIBN6000  & \num{ 5999}&80 &   0&84 &    5 &     1 &     1234   \\
\cgr DIBB4429  & \num{ 4428}&8  &  17&7  & 1962 & \cog2 &     134    &      DIBN5711  & \num{ 5711}&39 &   0&68 &   10 &     1 &     1245   &      DIBN6005  & \num{ 6005}&08 &   2&65 &   43 &     1 &     123456 \\
     DIBN4495  & \num{ 4494}&74 &   2&20 &   31 &     1 &     1356   &      DIBN5716  & \num{ 5716}&23 &   0&38 &    2 &     1 &     145    &      DIBN6010  & \num{ 6010}&48 &   4&18 &  147 &     1 &     123456 \\
     DIBN4502  & \num{ 4501}&93 &   2&99 &  170 & \cog3 &     156    &      DIBN5720  & \num{ 5719}&60 &   0&66 &   17 & \cog2 &     14     &      DIBN6015  & \num{ 6014}&98 &   1&01 &    3 &     1 &     123456 \\
\cgr DIBB4590  & \num{ 4590}&3  &  23&1  &  283 &     1 &     13     &      DIBN5741  & \num{ 5741}&36 &   2&33 &   29 &     1 &     14     &      DIBN6019  & \num{ 6019}&41 &   1&01 &   15 &     1 &     12346  \\
     DIBN4727  & \num{ 4726}&85 &   2&42 &  234 & \cog2 &     14     &      DIBN5745  & \num{ 5744}&56 &   1&34 &   16 &     1 &     14     &      DIBN6024  & \num{ 6023}&59 &   1&28 &    3 &     1 &     1234   \\
\cgr DIBB4761  & \num{ 4761}&2  &  19&1  &  444 &     1 &     13     &      DIBN5747  & \num{ 5746}&89 &   1&86 &   30 &     1 &     14     &      DIBN6027  & \num{ 6027}&38 &   1&63 &   24 &     1 &     123456 \\
     DIBN4763  & \num{ 4762}&57 &   2&10 &  112 & \cog2 &     34     &      DIBN5749  & \num{ 5748}&94 &   1&99 &   19 &     1 &     14     & \cre DIBI6037  & \num{ 6037}&1  &   5&8  &  118 &     1 &     1234   \\
     DIBN4780  & \num{ 4780}&44 &   1&56 &  115 & \cog2 &     1      &      DIBN5763  & \num{ 5762}&65 &   0&58 &    5 &     1 &     123456 &      DIBN6037  & \num{ 6037}&34 &   1&34 &   13 &     1 &     123456 \\
\cgr DIBB4884  & \num{ 4883}&7  &  16&7  &  861 & \cog2 &     1234   &      DIBN5766  & \num{ 5765}&91 &   0&74 &   21 & \cog2 &     1345   & \cre DIBI6045  & \num{ 6045}&4  &   7&5  &  107 &     1 &     1234   \\
     DIBN4947  & \num{ 4947}&21 &   0&53 &    3 &     1 &     23     &      DIBN5769  & \num{ 5769}&05 &   0&43 &    7 &     1 &     1234   &      DIBN6051  & \num{ 6051}&37 &   0&62 &    2 &     1 &     1234   \\
     DIBN4951  & \num{ 4951}&19 &   0&87 &    5 &     1 &     12345  &      DIBN5773  & \num{ 5772}&58 &   1&50 &   20 &     1 &     1234   &      DIBN6057  & \num{ 6057}&33 &   0&55 &    3 &     1 &     12345  \\
     DIBN4957  & \num{ 4956}&60 &   1&10 &    1 &     1 &     3456   &      DIBN5776  & \num{ 5775}&86 &   0&83 &    7 &     1 &     12346  &      DIBN6059  & \num{ 6059}&21 &   0&56 &    5 &     1 &     1234   \\
     DIBN4964  & \num{ 4963}&87 &   0&62 &   37 &     1 &     12345  & \cre DIBI5778  & \num{ 5778}&4  &   8&2  &  245 &     1 &     123456 &      DIBN6060  & \num{ 6060}&13 &   0&69 &    7 &     1 &     12346  \\
     DIBN4969  & \num{ 4969}&06 &   0&76 &    8 &     1 &     12345  &      DIBN5781  & \num{ 5780}&69 &   2&06 &  522 & \cog2 &     123456 &      DIBN6065  & \num{ 6065}&19 &   0&65 &   11 &     1 &     123456 \\
     DIBN4980  & \num{ 4979}&55 &   0&64 &    9 &     1 &     14     &      DIBN5785  & \num{ 5784}&98 &   1&19 &   19 &     1 &     1234   &      DIBN6068  & \num{ 6068}&07 &   1&35 &    8 &     1 &     1234   \\
     DIBN4982  & \num{ 4982}&13 &   0&32 &    2 &     1 &     123    &      DIBN5788  & \num{ 5788}&05 &   3&47 &   58 &     1 &     13456  &      DIBN6071  & \num{ 6071}&18 &   0&99 &    6 &     1 &     1234   \\
     DIBN4985  & \num{ 4984}&77 &   0&50 &   16 &     1 &     12346  &      DIBN5793  & \num{ 5793}&18 &   1&27 &   17 &     1 &     134    &      DIBN6081  & \num{ 6081}&02 &   0&44 &    2 &     1 &     12345  \\
     DIBN4988  & \num{ 4987}&65 &   0&97 &    4 &     1 &     12346  &      DIBN5795  & \num{ 5795}&38 &   1&82 &   42 &     1 &     123456 &      DIBN6085  & \num{ 6084}&81 &   0&85 &    6 &     1 &     12345  \\
     DIBN5004  & \num{ 5003}&57 &   0&94 &   13 &     1 &     14     &      DIBN5797  & \num{ 5797}&23 &   0&70 &  189 & \cog4 &     156    &      DIBN6090  & \num{ 6089}&80 &   0&60 &   16 &     1 &     123456 \\
     DIBN5028  & \num{ 5027}&53 &   0&63 &    5 &     1 &     134    &      DIBN5801  & \num{ 5800}&96 &   2&74 &   20 &     1 &     1235   &      DIBN6099  & \num{ 6098}&54 &   0&62 &    4 &     1 & \cma12345  \\
     DIBN5055  & \num{ 5054}&80 &   0&48 &    4 &     1 &     124    &      DIBN5807  & \num{ 5806}&50 &   1&21 &    5 &     1 &     136    &      DIBN6103  & \num{ 6102}&65 &   0&87 &    2 &     1 &     12345  \\
     DIBN5062  & \num{ 5061}&52 &   0&49 &   10 &     1 &     12345  &      DIBN5809  & \num{ 5809}&28 &   1&45 &   35 &     1 &     12356  &      DIBN6106  & \num{ 6106}&33 &   0&70 &    3 &     1 &     12345  \\
     DIBN5074  & \num{ 5074}&47 &   0&55 &    7 &     1 &     145    &      DIBN5812  & \num{ 5811}&91 &   1&65 &   19 &     1 &     12356  &      DIBN6107  & \num{ 6107}&25 &   0&61 &    4 &     1 &     12345  \\
     DIBN5092  & \num{ 5092}&05 &   0&79 &    3 &     1 &     2345   &      DIBN5814  & \num{ 5814}&21 &   0&49 &    3 &     1 &     123456 &      DIBN6108  & \num{ 6108}&04 &   0&46 &   10 &     1 &     123456 \\
\cre DIBI5111  & \num{ 5111}&1  &   5&7  &   46 &     1 &     12345  &      DIBN5816  & \num{ 5815}&77 &   0&53 &    2 &     1 &     123456 &      DIBN6110  & \num{ 6109}&88 &   0&60 &    4 &     1 &     12345  \\
\cre DIBI5154  & \num{ 5154}&1  &   6&0  &   41 &     1 & \cma134    &      DIBN5819  & \num{ 5818}&66 &   0&63 &    5 &     1 &     123456 &      DIBN6113  & \num{ 6113}&14 &   0&83 &   23 &     1 &     123456 \\
     DIBN5170  & \num{ 5170}&47 &   0&35 &    4 &     1 &     245    &      DIBN5821  & \num{ 5821}&10 &   0&30 &    1 &     1 &     12456  &      DIBN6117  & \num{ 6116}&79 &   0&92 &   14 &     1 &     123456 \\
     DIBN5176  & \num{ 5175}&98 &   0&61 &   14 &     1 &     123456 &      DIBN5828  & \num{ 5828}&48 &   0&73 &   11 &     1 &     123456 &      DIBN6119  & \num{ 6118}&53 &   0&62 &    5 &     1 &     12345  \\
     DIBN5236  & \num{ 5236}&16 &   1&49 &   33 &     1 &     12456  &      DIBN5838  & \num{ 5838}&01 &   0&33 &    3 &     1 &     123456 &      DIBN6128  & \num{ 6128}&43 &   2&18 &   10 &     1 &     1234   \\
\cre DIBI5246  & \num{ 5245}&8  &   5&7  &   74 &     1 &     12345  &      DIBN5841  & \num{ 5840}&61 &   0&58 &    2 &     1 &     123456 &      DIBN6133  & \num{ 6133}&39 &   0&76 &    3 &     1 &     1346   \\
     DIBN5299  & \num{ 5299}&45 &   4&24 &   42 &     1 &     1245   &      DIBN5844  & \num{ 5843}&73 &   4&26 &   58 &     1 &     123456 &      DIBN6136  & \num{ 6135}&78 &   1&24 &    5 &     1 &     1234   \\
     DIBN5340  & \num{ 5340}&24 &   0&63 &    5 &     1 &     1234   &      DIBN5845  & \num{ 5844}&89 &   0&65 &    9 &     1 &     123456 &      DIBN6140  & \num{ 6139}&92 &   0&66 &   12 &     1 &     123456 \\
     DIBN5359  & \num{ 5358}&80 &   0&28 &    1 &     1 &     14     &      DIBN5850  & \num{ 5849}&75 &   0&93 &   70 &     1 &     123456 &      DIBN6142  & \num{ 6142}&31 &   0&99 &    5 &     1 &     124    \\
     DIBN5364  & \num{ 5363}&72 &   2&10 &   36 &     1 &     123456 &      DIBN5854  & \num{ 5854}&38 &   0&54 &    4 &     1 &     13456  &      DIBN6146  & \num{ 6145}&67 &   0&58 &    3 &     1 &     12456  \\
     DIBN5404  & \num{ 5404}&49 &   0&90 &   20 &     1 &     12346  &      DIBN5856  & \num{ 5855}&54 &   0&49 &    3 &     1 &     123456 &      DIBN6148  & \num{ 6148}&23 &   1&04 &    6 &     1 &     123456 \\
     DIBN5419  & \num{ 5418}&84 &   0&72 &   24 &     1 &     12345  &      DIBN5859  & \num{ 5859}&11 &   2&03 &    4 &     1 &     123    &      DIBN6151  & \num{ 6150}&83 &   1&26 &    8 &     1 &     123456 \\
     DIBN5424  & \num{ 5424}&17 &   0&94 &    6 &     1 &     123456 &      DIBN5885  & \num{ 5885}&32 &   0&48 &    2 &     1 &     14     &      DIBN6157  & \num{ 6157}&18 &   2&70 &    1 &     1 &     236    \\
\cgr DIBB5449  & \num{ 5449}&0  &  12&1  &  253 &     1 &     1234   &      DIBN5894  & \num{ 5893}&50 &   0&47 &    4 &     1 &     124    &      DIBN6159  & \num{ 6158}&52 &   0&74 &    5 &     1 &     2345   \\
     DIBN5487  & \num{ 5487}&40 &   4&57 &  129 &     1 &     123456 &      DIBN5900  & \num{ 5900}&43 &   0&76 &   10 &     1 &     1234   &      DIBN6162  & \num{ 6161}&83 &   0&39 &    6 &     1 &     123456 \\
     DIBN5494  & \num{ 5494}&05 &   0&57 &   27 &     1 &     123456 &      DIBN5905  & \num{ 5904}&58 &   0&69 &    2 &     1 &     124    &      DIBN6164  & \num{ 6163}&56 &   0&66 &    3 &     1 &     123456 \\
     DIBN5497  & \num{ 5497}&02 &   2&13 &   10 &     1 &     1235   &      DIBN5908  & \num{ 5907}&81 &   0&98 &    4 &     1 &     1246   &      DIBN6166  & \num{ 6165}&61 &   1&33 &    7 &     1 &     1234   \\
     DIBN5503  & \num{ 5503}&02 &   0&72 &    3 &     1 &     123    &      DIBN5911  & \num{ 5910}&51 &   0&84 &   15 &     1 &     123456 &      DIBN6168  & \num{ 6167}&79 &   0&47 &    1 &     1 &     123    \\
     DIBN5504  & \num{ 5504}&26 &   0&50 &    4 &     1 &     123    &      DIBN5914  & \num{ 5913}&73 &   0&47 &    2 &     1 &     1345   &      DIBN6171  & \num{ 6170}&59 &   1&02 &    2 &     1 &     1234   \\
     DIBN5506  & \num{ 5506}&09 &   0&97 &   19 &     1 &     1245   &      DIBN5915  & \num{ 5914}&69 &   0&42 &    1 &     1 &     145    &      DIBN6175  & \num{ 6174}&59 &   0&43 &    1 &     1 &     1245   \\
     DIBN5508  & \num{ 5508}&15 &   2&08 &   63 &     1 &     12456  &      DIBN5917  & \num{ 5916}&99 &   1&33 &   10 &     1 &     12345  & \cgr DIBB6176  & \num{ 6176}&2  &  23&9  & 1007 & \cog2 &     123456 \\
     DIBN5513  & \num{ 5512}&65 &   0&57 &   13 &     1 &     123456 &      DIBN5922  & \num{ 5922}&23 &   0&35 &    3 &     1 &     1234   &      DIBN6178  & \num{ 6177}&80 &   1&49 &    8 &     1 &     124    \\
     DIBN5516  & \num{ 5515}&93 &   0&69 &    3 &     1 &     12346  &      DIBN5923  & \num{ 5923}&16 &   0&64 &   18 &     1 &     12345  &      DIBN6183  & \num{ 6182}&61 &   0&74 &    1 &     1 &     12346  \\
\cre DIBI5525  & \num{ 5524}&9  &   6&7  &   54 &     1 &     12346  &      DIBN5924  & \num{ 5924}&14 &   0&95 &   10 &     1 &     12345  &      DIBN6186  & \num{ 6185}&80 &   0&61 &    6 &     1 &     12346  \\
     DIBN5535  & \num{ 5535}&28 &   1&78 &   42 &     1 &     123456 &      DIBN5926  & \num{ 5925}&83 &   0&85 &   16 &     1 &     12345  &      DIBN6187  & \num{ 6187}&20 &   1&04 &   10 &     1 &     12346  \\
     DIBN5538  & \num{ 5537}&70 &   1&09 &    8 &     1 &     134    &      DIBN5928  & \num{ 5927}&64 &   0&51 &    5 &     1 &     12345  &      DIBN6189  & \num{ 6189}&38 &   0&58 &    4 &     1 &     12345  \\
     DIBN5542  & \num{ 5541}&80 &   0&69 &   10 &     1 &     12345  &      DIBN5945  & \num{ 5945}&44 &   0&62 &    2 &     1 &     123456 &      DIBN6195  & \num{ 6194}&68 &   0&54 &   10 &     1 &     123456 \\
     DIBN5545  & \num{ 5544}&99 &   0&77 &   27 &     1 &     123456 &      DIBN5947  & \num{ 5947}&22 &   0&58 &    7 &     1 &     123456 &      DIBN6196  & \num{ 6195}&94 &   0&52 &   68 &     1 &     123456 \\
     DIBN5546  & \num{ 5546}&38 &   0&80 &   10 &     1 &     12345  &      DIBN5949  & \num{ 5948}&87 &   0&39 &    2 &     1 &     123456 &      DIBN6199  & \num{ 6198}&90 &   0&70 &    7 &     1 &     1234   \\
     DIBN5547  & \num{ 5547}&37 &   0&35 &    2 &     1 &     124    &      DIBN5952  & \num{ 5952}&27 &   0&55 &    1 &     1 &     12456  &      DIBN6203  & \num{ 6202}&83 &   1&31 &  103 & \cog2 &     123456 \\
     DIBN5601  & \num{ 5600}&54 &   1&26 &    6 &     1 &     12346  &      DIBN5954  & \num{ 5954}&20 &   0&26 &    1 &     1 &     123456 & \cre DIBI6204  & \num{ 6204}&0  &   5&5  &  211 &     1 &     123456 \\
     DIBN5610  & \num{ 5609}&75 &   2&26 &   33 &     1 &     12346  &      DIBN5958  & \num{ 5958}&48 &   0&47 &   10 &     1 &     12456  &      DIBN6204  & \num{ 6204}&38 &   0&97 &    2 &     1 &     123456 \\
     DIBN5635  & \num{ 5634}&77 &   1&05 &    8 &     1 &     1234   &      DIBN5959  & \num{ 5959}&12 &   0&57 &   12 &     1 &     12456  &      DIBN6212  & \num{ 6211}&63 &   0&53 &    6 &     1 &     123456 \\
     DIBN5640  & \num{ 5640}&31 &   0&63 &    5 &     1 & \cma14     &      DIBN5963  & \num{ 5962}&53 &   1&46 &    8 &     1 &     123456 &      DIBN6213  & \num{ 6212}&91 &   0&52 &    2 &     1 &     12345  \\
     DIBN5644  & \num{ 5643}&51 &   3&60 &   19 &     1 &     14     &      DIBN5974  & \num{ 5973}&78 &   0&47 &    3 &     1 &     123456 &      DIBN6223  & \num{ 6223}&49 &   0&46 &    4 &     1 &     123456 \\
     DIBN5645  & \num{ 5645}&43 &   0&41 &    2 &     1 &     14     &      DIBN5976  & \num{ 5975}&69 &   0&36 &    3 &     1 &     123456 &      DIBN6226  & \num{ 6226}&18 &   0&49 &    4 &     1 &     12346  \\
     DIBN5652  & \num{ 5652}&01 &   1&09 &    6 &     1 &     124    &      DIBN5983  & \num{ 5982}&86 &   1&24 &   10 &     1 &     12345  &      DIBN6234  & \num{ 6233}&98 &   0&60 &   15 &     1 &     123456 \\
\midrule
\end{tabular}
\addtolength{\tabcolsep}{1mm}
\label{dibs_all}
\end{table*}

\addtocounter{table}{-1}
\begin{table*}
\caption{(Continued).}
\addtolength{\tabcolsep}{-1mm}
\hspace{-10mm}\begin{tabular}{lr@{.}lr@{.}l@{}rc@{}c@{\hspace{5mm}}lr@{.}lr@{.}l@{}rc@{}c@{\hspace{5mm}}lr@{.}lr@{.}l@{}rc@{}c}
\midrule
Name      & \mcii{$\lambda_0$} & \mcii{$\!\!$FWHM}  & \mci{$\!\!\!\!$EW}     & $n_{\rm G}$ & $\!\!$Stars & Name      & \mcii{$\lambda_0$} & \mcii{$\!\!$FWHM}  & \mci{$\!\!\!\!$EW}     & $n_{\rm G}$ & $\!\!$Stars & Name      & \mcii{$\lambda_0$} & \mcii{$\!\!$FWHM}  & \mci{$\!\!\!\!$EW}     & $n_{\rm G}$ & $\!\!$Stars \\
          & \mcii{(\AA)}       & \mcii{$\!\!$(\AA)} & \mci{$\!\!\!\!$(m\AA)} &             & $\!\!$used  &           & \mcii{(\AA)}       & \mcii{$\!\!$(\AA)} & \mci{$\!\!\!\!$(m\AA)} &             & $\!\!$used  &           & \mcii{(\AA)}       & \mcii{$\!\!$(\AA)} & \mci{$\!\!\!\!$(m\AA)} &             & $\!\!$used  \\
\midrule
     DIBN6237  & \num{ 6236}&71 &   0&55 &    6 &     1 &     123456 &      DIBN6600  & \num{ 6599}&84 &   0&51 &    2 &     1 &     123    &      DIBN6822  & \num{ 6821}&57 &   1&45 &    8 &     1 &     1234   \\
     DIBN6244  & \num{ 6244}&48 &   0&80 &    8 &     1 &     123456 &      DIBN6607  & \num{ 6606}&95 &   0&79 &    6 &     1 &     123    &      DIBN6823  & \num{ 6823}&38 &   0&41 &    4 &     1 &     1234   \\
     DIBN6245  & \num{ 6245}&28 &   0&59 &    6 &     1 &     12346  &      DIBN6614  & \num{ 6613}&71 &   0&96 &  226 & \cog4 &     56     &      DIBN6827  & \num{ 6827}&28 &   0&81 &   10 &     1 &     123456 \\
     DIBN6251  & \num{ 6250}&85 &   0&62 &    5 &     1 &     123456 &      DIBN6621  & \num{ 6621}&48 &   0&53 &    3 &     1 &     12345  &      DIBN6834  & \num{ 6834}&31 &   0&68 &    3 &     1 &     12345  \\
     DIBN6252  & \num{ 6252}&35 &   0&38 &    1 &     1 &     1234   &      DIBN6623  & \num{ 6622}&75 &   0&56 &   10 &     1 &     12345  &      DIBN6838  & \num{ 6837}&61 &   0&84 &    9 &     1 &     1234   \\
     DIBN6260  & \num{ 6259}&62 &   0&56 &    2 &     1 &     12345  &      DIBN6625  & \num{ 6624}&79 &   0&60 &    3 &     1 &     12345  &      DIBN6839  & \num{ 6839}&41 &   0&55 &    3 &     1 &     12345  \\
     DIBN6266  & \num{ 6266}&44 &   1&42 &    7 &     1 &     123456 &      DIBN6626  & \num{ 6625}&83 &   0&66 &    3 &     1 &     12345  &      DIBN6842  & \num{ 6841}&56 &   0&69 &   12 &     1 &     1235   \\
     DIBN6270  & \num{ 6269}&83 &   1&27 &  140 & \cog2 &     123456 &      DIBN6628  & \num{ 6628}&02 &   0&95 &    4 &     1 &     12345  &      DIBN6844  & \num{ 6843}&51 &   1&09 &   33 &     1 &     123456 \\
     DIBN6278  & \num{ 6277}&99 &   3&02 &  174 &     1 &     1234   &      DIBN6631  & \num{ 6630}&72 &   0&51 &    4 &     1 &     12345  &      DIBN6845  & \num{ 6845}&31 &   0&55 &    4 &     1 &     123456 \\
     DIBN6280  & \num{ 6280}&31 &   1&46 &   73 &     1 &     1234   &      DIBN6632  & \num{ 6631}&55 &   0&45 &    3 &     1 &     12345  &      DIBN6846  & \num{ 6846}&40 &   0&40 &    2 &     1 &     1234   \\
     DIBN6284  & \num{ 6284}&04 &   3&31 &  974 & \cog3 &     1234   &      DIBN6633  & \num{ 6632}&73 &   1&01 &   13 &     1 &     12346  &      DIBN6848  & \num{ 6847}&59 &   0&56 &    3 &     1 &     123    \\
     DIBN6288  & \num{ 6287}&52 &   0&74 &   14 &     1 &     12346  &      DIBN6635  & \num{ 6635}&19 &   0&34 &    1 &     1 & \cma1234   &      DIBN6852  & \num{ 6852}&43 &   0&73 &   11 &     1 &     123456 \\
     DIBN6289  & \num{ 6289}&19 &   2&70 &  117 &     1 &     12346  &      DIBN6636  & \num{ 6635}&59 &   0&46 &    2 &     1 &     12345  &      DIBN6860  & \num{ 6860}&00 &   0&68 &   12 &     1 &     123456 \\
     DIBN6302  & \num{ 6302}&40 &   1&45 &   27 &     1 &     1456   &      DIBN6638  & \num{ 6637}&53 &   0&36 &    1 &     1 &     12345  &      DIBN6862  & \num{ 6862}&47 &   0&47 &    5 &     1 &     123456 \\
\cre DIBI6309  & \num{ 6309}&1  &   7&2  &  281 &     1 &     145    &      DIBN6639  & \num{ 6639}&27 &   0&61 &    3 &     1 &     12345  &      DIBN6878  & \num{ 6877}&57 &   0&91 &   16 &     1 &     1234   \\
     DIBN6309  & \num{ 6309}&42 &   1&29 &   16 &     1 &     1246   &      DIBN6643  & \num{ 6643}&42 &   1&45 &   11 &     1 &     12345  &      DIBN6887  & \num{ 6886}&89 &   0&89 &   12 &     1 &     1234   \\
\cre DIBI6317  & \num{ 6317}&4  &   7&4  &  256 &     1 &     145    &      DIBN6646  & \num{ 6645}&92 &   0&81 &    6 &     1 &     12345  &      DIBN6900  & \num{ 6900}&30 &   0&96 &    7 &     1 & \cma13     \\
     DIBN6318  & \num{ 6318}&22 &   0&53 &    7 &     1 &     146    &      DIBN6655  & \num{ 6654}&68 &   0&58 &    3 &     1 &     12345  &      DIBN6904  & \num{ 6903}&86 &   0&49 &    4 &     1 &     1234   \\
     DIBN6325  & \num{ 6324}&75 &   0&86 &   22 &     1 &     12346  &      DIBN6657  & \num{ 6657}&21 &   0&67 &    4 &     1 &     1234   &      DIBN6905  & \num{ 6905}&01 &   1&41 &   14 &     1 & \cma1234   \\
     DIBN6330  & \num{ 6330}&01 &   0&80 &   10 &     1 &     123456 &      DIBN6659  & \num{ 6658}&63 &   0&51 &    2 &     1 &     12345  &      DIBN6914  & \num{ 6913}&69 &   2&06 &   30 &     1 & \cma1234   \\
     DIBN6346  & \num{ 6346}&08 &   1&28 &    9 &     1 & \cma146    &      DIBN6661  & \num{ 6660}&63 &   0&61 &   31 &     1 &     123456 &      DIBN6919  & \num{ 6919}&09 &   0&93 &   38 &     1 &     1234   \\
     DIBN6349  & \num{ 6349}&39 &   0&42 &    3 &     1 &     14     &      DIBN6662  & \num{ 6662}&00 &   0&56 &    4 &     1 &     12345  &      DIBN6927  & \num{ 6926}&67 &   1&07 &    9 &     1 &     1234   \\
     DIBN6353  & \num{ 6353}&28 &   1&66 &   27 &     1 &     12346  &      DIBN6664  & \num{ 6663}&94 &   0&83 &    6 &     1 &     12345  &      DIBN6930  & \num{ 6929}&87 &   2&11 &   29 &     1 & \cma124    \\
     DIBN6358  & \num{ 6358}&31 &   0&60 &    4 &     1 &     12346  &      DIBN6665  & \num{ 6665}&17 &   0&66 &    7 &     1 &     12345  & \cgr DIBB6937  & \num{ 6936}&5  &  12&1  &  204 &     1 & \cma123456 \\
     DIBN6362  & \num{ 6362}&27 &   1&74 &   25 &     1 &     12346  &      DIBN6669  & \num{ 6669}&43 &   0&92 &    6 &     1 &     1234   &      DIBN6945  & \num{ 6944}&56 &   0&96 &   23 &     1 &     1234   \\
     DIBN6367  & \num{ 6367}&26 &   0&47 &   13 &     1 &     123456 &      DIBN6672  & \num{ 6672}&17 &   0&73 &   20 &     1 &     12346  &      DIBN6947  & \num{ 6946}&72 &   1&88 &    9 &     1 & \cma1234   \\
     DIBN6369  & \num{ 6369}&11 &   1&45 &   14 &     1 &     1456   &      DIBN6674  & \num{ 6673}&92 &   0&75 &    4 &     1 & \cma1234   &      DIBN6951  & \num{ 6950}&51 &   0&65 &    8 &     1 &     1236   \\
     DIBN6376  & \num{ 6375}&95 &   0&56 &   25 &     1 &     123456 &      DIBN6685  & \num{ 6684}&71 &   0&97 &    9 &     1 &     1234   &      DIBN6952  & \num{ 6951}&71 &   0&47 &    6 &     1 &     1234   \\
     DIBN6377  & \num{ 6376}&84 &   1&91 &   28 &     1 &     123456 &      DIBN6686  & \num{ 6686}&44 &   0&50 &    2 &     1 & \cma1234   &      DIBN6971  & \num{ 6971}&43 &   1&16 &   10 &     1 & \cma123456 \\
     DIBN6379  & \num{ 6379}&23 &   0&68 &   86 &     1 &     123456 &      DIBN6689  & \num{ 6689}&25 &   0&92 &    8 &     1 &     1234   &      DIBN6974  & \num{ 6973}&59 &   0&55 &    8 &     1 &     1234   \\
     DIBN6397  & \num{ 6396}&89 &   1&14 &   19 &     1 &     12456  &      DIBN6694  & \num{ 6693}&65 &   0&57 &    3 &     1 &     12345  &      DIBN6978  & \num{ 6978}&33 &   0&79 &    7 &     1 &     12345  \\
     DIBN6400  & \num{ 6400}&38 &   0&70 &    5 &     1 &     145    &      DIBN6695  & \num{ 6694}&50 &   0&62 &    6 &     1 &     1234   &      DIBN6982  & \num{ 6982}&24 &   0&91 &    8 &     1 &     123456 \\
\cgr DIBB6407  & \num{ 6406}&6  &  10&1  &  294 &     1 & \cma145    &      DIBN6697  & \num{ 6696}&90 &   0&50 &    2 &     1 &     12346  &      DIBN6993  & \num{ 6993}&14 &   0&80 &   91 & \cog2 &     123456 \\
     DIBN6410  & \num{ 6410}&09 &   1&05 &   12 &     1 &     145    &      DIBN6699  & \num{ 6699}&22 &   0&72 &   25 &     1 &     12346  &      DIBN6997  & \num{ 6996}&79 &   0&60 &    4 &     1 &     1234   \\
     DIBN6414  & \num{ 6413}&52 &   3&06 &   34 &     1 &     12345  &      DIBN6702  & \num{ 6701}&99 &   0&71 &   12 &     1 &     12346  &      DIBN6999  & \num{ 6998}&66 &   0&98 &   13 &     1 &     1234   \\
     DIBN6419  & \num{ 6418}&53 &   0&53 &    4 &     1 &     123456 &      DIBN6709  & \num{ 6709}&40 &   0&92 &   11 &     1 &     1234   &      DIBN7002  & \num{ 7002}&25 &   1&07 &   13 &     1 &     1234   \\
     DIBN6426  & \num{ 6425}&64 &   0&64 &   15 &     1 &     12345  &      DIBN6713  & \num{ 6713}&47 &   0&76 &    6 &     1 &     1234   &      DIBN7004  & \num{ 7004}&31 &   0&39 &    2 &     1 &     1234   \\
     DIBN6438  & \num{ 6438}&24 &   0&75 &    5 &     1 &     12346  &      DIBN6729  & \num{ 6729}&16 &   0&90 &   15 &     1 &     12345  &      DIBN7020  & \num{ 7020}&41 &   2&05 &   16 &     1 & \cma123456 \\
     DIBN6439  & \num{ 6439}&44 &   0&70 &   20 &     1 &     123456 &      DIBN6733  & \num{ 6733}&26 &   0&93 &    8 &     1 &     12345  &      DIBN7025  & \num{ 7025}&14 &   1&14 &   14 &     1 &     1234   \\
     DIBN6443  & \num{ 6442}&58 &   1&37 &    7 &     1 &     12356  &      DIBN6737  & \num{ 6737}&16 &   0&64 &    6 &     1 &     123456 &      DIBN7030  & \num{ 7030}&18 &   0&73 &    7 &     1 &     12346  \\
     DIBN6445  & \num{ 6445}&24 &   0&77 &   34 &     1 &     12346  &      DIBN6741  & \num{ 6740}&91 &   1&09 &   12 &     1 &     12345  &      DIBN7031  & \num{ 7031}&47 &   0&56 &    7 &     1 &     12346  \\
     DIBN6449  & \num{ 6449}&18 &   0&85 &   23 &     1 &     123456 &      DIBN6742  & \num{ 6742}&49 &   0&90 &    8 &     1 &     1234   &      DIBN7046  & \num{ 7045}&81 &   0&66 &    9 &     1 &     1234   \\
     DIBN6456  & \num{ 6455}&97 &   1&21 &   34 &     1 &     123456 &      DIBN6744  & \num{ 6743}&56 &   1&22 &    8 &     1 &     12345  &      DIBN7061  & \num{ 7060}&98 &   0&60 &   19 &     1 &     12346  \\
     DIBN6460  & \num{ 6460}&30 &   0&65 &    3 &     1 &     123456 &      DIBN6748  & \num{ 6747}&72 &   0&87 &    7 &     1 &     12346  &      DIBN7063  & \num{ 7062}&65 &   0&57 &   20 &     1 &     12345  \\
     DIBN6464  & \num{ 6463}&59 &   1&00 &   13 &     1 &     123456 &      DIBN6751  & \num{ 6750}&69 &   1&27 &   10 &     1 &     12345  &      DIBN7069  & \num{ 7069}&45 &   1&13 &   28 &     1 &     1234   \\
     DIBN6465  & \num{ 6465}&41 &   0&28 &    1 &     1 &     124    &      DIBN6752  & \num{ 6752}&33 &   1&09 &    8 &     1 &     12345  &      DIBN7078  & \num{ 7077}&99 &   1&07 &   14 &     1 &     1234   \\
     DIBN6467  & \num{ 6466}&83 &   0&54 &    5 &     1 &     12346  &      DIBN6756  & \num{ 6755}&73 &   1&23 &    3 &     1 &     1234   &      DIBN7080  & \num{ 7080}&03 &   1&01 &    6 &     1 & \cma1234   \\
     DIBN6469  & \num{ 6468}&66 &   0&85 &    7 &     1 &     123456 &      DIBN6757  & \num{ 6757}&18 &   0&91 &    3 &     1 &     1234   &      DIBN7084  & \num{ 7083}&65 &   0&92 &    7 &     1 &     1234   \\
     DIBN6474  & \num{ 6474}&18 &   0&73 &    7 &     1 &     1234   &      DIBN6759  & \num{ 6759}&12 &   1&11 &    2 &     1 &     12345  &      DIBN7085  & \num{ 7085}&25 &   1&45 &   17 &     1 &     1234   \\
     DIBN6477  & \num{ 6476}&95 &   0&55 &    3 &     1 &     1234   &      DIBN6762  & \num{ 6762}&04 &   0&60 &    2 &     1 &     123    &      DIBN7087  & \num{ 7086}&69 &   0&45 &    5 &     1 &     1234   \\
     DIBN6483  & \num{ 6483}&38 &   0&54 &    4 &     1 & \cma145    &      DIBN6765  & \num{ 6765}&24 &   0&67 &    2 &     1 &     12345  &      DIBN7093  & \num{ 7092}&51 &   0&62 &    3 &     1 &     1234   \\
     DIBN6484  & \num{ 6484}&33 &   0&61 &    4 &     1 &     15     &      DIBN6768  & \num{ 6767}&66 &   0&57 &    2 &     1 &     12346  &      DIBN7096  & \num{ 7096}&33 &   0&56 &    4 &     1 &     1234   \\
     DIBN6489  & \num{ 6489}&49 &   1&06 &    9 &     1 &     123456 &      DIBN6770  & \num{ 6770}&11 &   0&57 &    8 &     1 &     12345  &      DIBN7101  & \num{ 7101}&26 &   0&79 &    2 &     1 &     124    \\
     DIBN6492  & \num{ 6491}&96 &   0&53 &   12 &     1 &     12346  &      DIBN6779  & \num{ 6778}&94 &   0&59 &    5 &     1 &     12345  &      DIBN7105  & \num{ 7105}&18 &   0&59 &    4 &     1 &     1234   \\
\cre DIBI6494  & \num{ 6494}&1  &   6&5  &  123 &     1 &     12346  &      DIBN6781  & \num{ 6780}&63 &   0&76 &    4 &     1 &     12345  &      DIBN7116  & \num{ 7116}&32 &   0&89 &   13 &     1 &     123456 \\
     DIBN6498  & \num{ 6497}&88 &   0&69 &    6 &     1 &     12345  &      DIBN6786  & \num{ 6786}&48 &   0&54 &    2 &     1 &     12346  &      DIBN7120  & \num{ 7119}&77 &   1&32 &   25 &     1 &     123456 \\
     DIBN6517  & \num{ 6516}&77 &   0&57 &    5 &     1 & \cma1234   &      DIBN6789  & \num{ 6788}&75 &   0&71 &    7 &     1 &     12345  &      DIBN7124  & \num{ 7124}&08 &   1&25 &    4 &     1 &     1234   \\
     DIBN6521  & \num{ 6520}&54 &   0&98 &   23 &     1 &     12345  &      DIBN6792  & \num{ 6792}&46 &   0&72 &    7 &     1 &     123456 &      DIBN7137  & \num{ 7136}&78 &   1&29 &    8 &     1 &     1234   \\
     DIBN6536  & \num{ 6536}&43 &   0&73 &    5 &     1 &     123456 &      DIBN6795  & \num{ 6795}&18 &   0&63 &    9 &     1 &     123456 &      DIBN7138  & \num{ 7138}&36 &   0&75 &    5 &     1 &     1234   \\
     DIBN6538  & \num{ 6537}&57 &   0&64 &    1 &     1 & \cma12346  &      DIBN6801  & \num{ 6801}&35 &   0&75 &   10 &     1 &     1234   &      DIBN7139  & \num{ 7139}&09 &   3&72 &   31 &     1 &     124    \\
     DIBN6543  & \num{ 6543}&04 &   0&60 &    6 &     1 &     12345  &      DIBN6803  & \num{ 6803}&20 &   0&67 &    6 &     1 &     12345  &      DIBN7143  & \num{ 7142}&94 &   0&60 &    3 &     1 &     1234   \\
     DIBN6549  & \num{ 6548}&94 &   0&48 &    4 &     1 &     1234   &      DIBN6804  & \num{ 6804}&30 &   1&29 &    7 &     1 &     12345  &      DIBN7154  & \num{ 7153}&97 &   1&01 &   10 &     1 &     1234   \\
     DIBN6554  & \num{ 6553}&80 &   0&48 &    7 &     1 &     1234   &      DIBN6809  & \num{ 6809}&37 &   0&40 &    1 &     1 &     1234   &      DIBN7159  & \num{ 7159}&32 &   0&57 &    8 &     1 &     1234   \\
\cre DIBI6590  & \num{ 6590}&4  &   7&7  &  162 &     1 &     1234   &      DIBN6811  & \num{ 6811}&06 &   0&60 &    8 &     1 &     123    &      DIBN7160  & \num{ 7160}&04 &   0&84 &    5 &     1 &     1234   \\
     DIBN6594  & \num{ 6594}&25 &   0&86 &    5 &     1 &     1234   &      DIBN6812  & \num{ 6811}&74 &   1&11 &   10 &     1 &     123    &      DIBN7163  & \num{ 7162}&85 &   0&52 &    5 &     1 &     1234   \\
     DIBN6597  & \num{ 6597}&28 &   0&52 &   11 &     1 &     123456 &      DIBN6813  & \num{ 6812}&82 &   0&81 &    8 &     1 &     1235   &      DIBN7166  & \num{ 7166}&38 &   0&82 &    4 &     1 &     12     \\
     DIBN6599  & \num{ 6598}&87 &   0&80 &    1 &     1 & \cma23     &      DIBN6818  & \num{ 6818}&36 &   1&12 &    5 &     1 &     1234   &      DIBN7198  & \num{ 7198}&46 &   0&75 &    5 &     1 &     123456 \\
\midrule
\end{tabular}
\addtolength{\tabcolsep}{1mm}
\end{table*}

\addtocounter{table}{-1}
\begin{table*}
\caption{(Continued).}
\addtolength{\tabcolsep}{-1mm}
\hspace{-10mm}\begin{tabular}{lr@{.}lr@{.}l@{}rc@{}c@{\hspace{5mm}}lr@{.}lr@{.}l@{}rc@{}c@{\hspace{5mm}}lr@{.}lr@{.}l@{}rc@{}c}
\midrule
Name      & \mcii{$\lambda_0$} & \mcii{$\!\!$FWHM}  & \mci{$\!\!\!\!$EW}     & $n_{\rm G}$ & $\!\!$Stars & Name      & \mcii{$\lambda_0$} & \mcii{$\!\!$FWHM}  & \mci{$\!\!\!\!$EW}     & $n_{\rm G}$ & $\!\!$Stars & Name      & \mcii{$\lambda_0$} & \mcii{$\!\!$FWHM}  & \mci{$\!\!\!\!$EW}     & $n_{\rm G}$ & $\!\!$Stars \\
          & \mcii{(\AA)}       & \mcii{$\!\!$(\AA)} & \mci{$\!\!\!\!$(m\AA)} &             & $\!\!$used  &           & \mcii{(\AA)}       & \mcii{$\!\!$(\AA)} & \mci{$\!\!\!\!$(m\AA)} &             & $\!\!$used  &           & \mcii{(\AA)}       & \mcii{$\!\!$(\AA)} & \mci{$\!\!\!\!$(m\AA)} &             & $\!\!$used  \\
\midrule
     DIBN7204  & \num{ 7203}&54 &   0&61 &   14 &     1 &     1234   &      DIBN7707  & \num{ 7706}&69 &   0&76 &    7 &     1 &     123456 &      DIBN9988  & \num{ 9987}&99 &   1&26 &   10 &     1 & \cma1234   \\
     DIBN7224  & \num{ 7224}&06 &   1&24 &  239 & \cog2 &     12356  &      DIBN7708  & \num{ 7708}&06 &   0&69 &    9 &     1 &     12346  &      DIBN9990  & \num{ 9989}&56 &   4&39 &   19 &     1 & \cma123456 \\
     DIBN7228  & \num{ 7228}&30 &   1&24 &   17 &     1 &     123456 &      DIBN7722  & \num{ 7721}&90 &   0&73 &   23 &     1 &     123456 &      DIBN10004 & \num{10004}&29 &   1&27 &    7 &     1 & \cma12346  \\
     DIBN7232  & \num{ 7231}&59 &   1&30 &   18 &     1 & \cma1234   & \cre DIBI7748  & \num{ 7748}&0  &   7&3  &   81 &     1 &     123456 &      DIBN10007 & \num{10006}&78 &   1&35 &   10 &     1 &     12346  \\
     DIBN7249  & \num{ 7249}&18 &   1&64 &   21 &     1 &     1234   &      DIBN7758  & \num{ 7758}&33 &   2&79 &   14 &     1 & \cma123    &      DIBN10009 & \num{10008}&57 &   3&20 &   16 &     1 & \cma1236   \\
     DIBN7258  & \num{ 7257}&50 &   1&01 &   12 &     1 &     1234   &      DIBN7761  & \num{ 7761}&40 &   2&84 &   20 &     1 & \cma123    &      DIBN10080 & \num{10080}&30 &   2&31 &   11 &     1 & \cma12456  \\
     DIBN7268  & \num{ 7268}&19 &   1&00 &    6 &     1 &     1234   & \cgr DIBB7779  & \num{ 7779}&4  &  12&2  &  111 &     1 & \cma12345  &      DIBN10162 & \num{10162}&14 &   1&30 &   12 &     1 & \cma123    \\
     DIBN7274  & \num{ 7273}&71 &   0&74 &    7 &     1 & \cma1234   &      DIBN7780  & \num{ 7780}&37 &   0&76 &    1 &     1 & \cma12345  &      DIBN10208 & \num{10208}&03 &   0&71 &    4 &     1 & \cma123    \\
     DIBN7275  & \num{ 7274}&93 &   0&68 &   12 &     1 &     1234   &      DIBN7828  & \num{ 7827}&64 &   0&77 &    6 &     1 &     1234   &      DIBN10233 & \num{10233}&12 &   2&99 &   16 &     1 & \cma123456 \\
     DIBN7277  & \num{ 7276}&55 &   0&68 &   18 &     1 &     1234   &      DIBN7829  & \num{ 7829}&02 &   0&75 &    2 &     1 &     1234   &      DIBN10259 & \num{10258}&80 &   1&43 &    6 &     1 & \cma12     \\
     DIBN7301  & \num{ 7300}&95 &   0&81 &    8 &     1 & \cma1234   &      DIBN7833  & \num{ 7832}&79 &   0&70 &   19 &     1 &     123456 &      DIBN10263 & \num{10262}&60 &   1&23 &   11 &     1 &     1234   \\
     DIBN7303  & \num{ 7302}&93 &   0&85 &    8 &     1 &     1234   &      DIBN7841  & \num{ 7840}&80 &   0&68 &    5 &     1 &     1234   &      DIBN10280 & \num{10280}&46 &   1&05 &    9 &     1 & \cma123    \\
     DIBN7322  & \num{ 7321}&88 &   0&61 &    4 &     1 &     1234   &      DIBN7844  & \num{ 7843}&54 &   1&74 &    4 &     1 &     14     &      DIBN10283 & \num{10283}&15 &   0&85 &    5 &     1 & \cma1234   \\
     DIBN7330  & \num{ 7329}&91 &   1&57 &   11 &     1 &     123    & \cre DIBI7844  & \num{ 7844}&3  &   8&0  &   42 &     1 & \cma12346  &      DIBN10284 & \num{10284}&35 &   0&55 &    5 &     1 & \cma124    \\
     DIBN7334  & \num{ 7334}&39 &   0&97 &   39 &     1 &     12346  &      DIBN7846  & \num{ 7845}&64 &   0&82 &    3 &     1 & \cma1234   &      DIBN10288 & \num{10288}&35 &   1&19 &   12 &     1 &     1234   \\
     DIBN7340  & \num{ 7340}&02 &   1&29 &    5 &     1 &     1234   &      DIBN7856  & \num{ 7855}&91 &   0&77 &    2 &     1 & \cma1234   &      DIBN10327 & \num{10327}&31 &   1&46 &    9 &     1 & \cma12346  \\
     DIBN7343  & \num{ 7342}&58 &   1&91 &   11 &     1 &     1234   &      DIBN7862  & \num{ 7862}&33 &   0&72 &    8 &     1 &     1234   &      DIBN10391 & \num{10391}&30 &   0&56 &    2 &     1 & \cma124    \\
     DIBN7347  & \num{ 7347}&20 &   2&05 &   25 &     1 &     12346  &      DIBN7866  & \num{ 7866}&39 &   1&09 &    5 &     1 & \cma1234   &      DIBN10393 & \num{10393}&31 &   1&07 &   16 & \cog2 &     124    \\
     DIBN7350  & \num{ 7349}&75 &   1&06 &   25 &     1 &     123456 & \cre DIBI7874  & \num{ 7874}&3  &   5&0  &   32 &     1 & \cma1234   &      DIBN10409 & \num{10409}&03 &   0&91 &    5 &     1 & \cma1234   \\
     DIBN7355  & \num{ 7354}&78 &   0&85 &    7 &     1 &     12346  &      DIBN7905  & \num{ 7904}&78 &   1&63 &    7 &     1 &     1234   &      DIBN10439 & \num{10438}&57 &   3&46 &   45 &     1 &     12345  \\
     DIBN7357  & \num{ 7357}&45 &   1&31 &   44 &     1 &     12346  &      DIBN7915  & \num{ 7915}&08 &   1&08 &    3 &     1 &     1234   &      DIBN10459 & \num{10458}&71 &   2&67 &   18 &     1 & \cma12345  \\
     DIBN7361  & \num{ 7360}&56 &   0&76 &   12 &     1 &     12346  & \cgr DIBB7928  & \num{ 7927}&6  &  10&0  &  157 &     1 &     1234   &      DIBN10476 & \num{10476}&45 &   2&20 &    8 &     1 & \cma1234   \\
     DIBN7364  & \num{ 7363}&83 &   0&47 &    4 &     1 &     1234   &      DIBN7935  & \num{ 7934}&68 &   0&47 &    4 &     1 & \cma123    &      DIBN10505 & \num{10504}&51 &   1&60 &   59 &     1 &     12345  \\
     DIBN7366  & \num{ 7366}&10 &   1&11 &   23 &     1 &     123456 &      DIBN7936  & \num{ 7935}&50 &   0&73 &    4 &     1 &     123    &      DIBN10507 & \num{10507}&18 &   1&58 &   49 &     1 &     12345  \\
     DIBN7367  & \num{ 7367}&07 &   0&55 &   27 &     1 &     123456 &      DIBN7943  & \num{ 7943}&07 &   1&20 &    3 &     1 & \cma1234   &      DIBN10522 & \num{10521}&54 &   3&29 &    8 &     1 & \cma1234   \\
     DIBN7376  & \num{ 7375}&82 &   0&68 &   13 &     1 &     12346  &      DIBN7948  & \num{ 7947}&67 &   1&78 &    9 &     1 & \cma1234   &      DIBN10528 & \num{10528}&47 &   1&17 &    5 &     1 & \cma1234   \\
     DIBN7382  & \num{ 7382}&43 &   1&11 &    8 &     1 &     1234   &      DIBN7951  & \num{ 7951}&23 &   1&51 &    9 &     1 &     1234   &      DIBN10563 & \num{10562}&72 &   2&77 &    8 &     1 & \cma1234   \\
     DIBN7386  & \num{ 7385}&83 &   0&53 &    6 &     1 &     123456 &      DIBN7968  & \num{ 7967}&59 &   2&41 &    5 &     1 &     1234   &      DIBN10568 & \num{10568}&07 &   2&18 &    7 &     1 & \cma1234   \\
\cre DIBI7398  & \num{ 7398}&1  &   5&3  &   22 &     1 &     12346  &      DIBN7971  & \num{ 7971}&08 &   1&10 &    5 &     1 &     1234   &      DIBN10574 & \num{10573}&60 &   4&17 &   27 &     1 & \cma12346  \\
     DIBN7402  & \num{ 7402}&34 &   1&72 &   15 &     1 &     123456 &      DIBN8026  & \num{ 8026}&17 &   0&71 &   27 &     1 &     12346  &      DIBN10579 & \num{10578}&95 &   1&09 &    7 &     1 & \cma12346  \\
     DIBN7406  & \num{ 7405}&53 &   0&68 &    7 &     1 &     123    &      DIBN8038  & \num{ 8037}&52 &   2&30 &   38 &     1 &     1234   &      DIBN10610 & \num{10610}&24 &   1&72 &   20 & \cog2 &     1234   \\
     DIBN7407  & \num{ 7406}&82 &   1&68 &   16 &     1 &     123    &      DIBN8041  & \num{ 8040}&73 &   2&10 &   18 &     1 &     1234   &      DIBN10615 & \num{10615}&05 &   1&57 &    6 &     1 & \cma1234   \\
     DIBN7411  & \num{ 7410}&69 &   1&69 &    4 &     1 & \cma1234   &      DIBN8099  & \num{ 8098}&84 &   2&10 &    7 &     1 & \cma1234   &      DIBN10627 & \num{10627}&29 &   0&91 &    2 &     1 & \cma12     \\
     DIBN7415  & \num{ 7414}&60 &   0&89 &    1 &     1 & \cma123456 &      DIBN8105  & \num{ 8104}&76 &   1&59 &   12 &     1 & \cma1234   &      DIBN10631 & \num{10630}&71 &   1&35 &    3 &     1 & \cma1234   \\
     DIBN7419  & \num{ 7419}&00 &   0&57 &    4 &     1 &     1234   &      DIBN8126  & \num{ 8125}&92 &   1&15 &    9 &     1 & \cma1234   &      DIBN10634 & \num{10633}&93 &   1&09 &    5 &     1 & \cma1256   \\
\cgr DIBB7430  & \num{ 7430}&4  &  16&3  &  213 &     1 &     123456 &      DIBN8155  & \num{ 8154}&69 &   1&69 &    4 &     1 &     234    &      DIBN10643 & \num{10643}&15 &   1&24 &    2 &     1 & \cma124    \\
     DIBN7451  & \num{ 7451}&23 &   0&99 &    9 &     1 &     123456 & \cre DIBI8215  & \num{ 8215}&3  &   8&3  &   42 &     1 & \cma14     &      DIBN10646 & \num{10646}&17 &   0&91 &    3 &     1 & \cma126    \\
     DIBN7458  & \num{ 7458}&21 &   1&10 &    3 &     1 &     123456 & \cre DIBI8258  & \num{ 8257}&9  &   9&8  &  159 &     1 & \cma14     &      DIBN10650 & \num{10650}&25 &   1&32 &    6 &     1 & \cma1234   \\
     DIBN7463  & \num{ 7462}&81 &   2&77 &   15 &     1 & \cma12345  & \cre DIBI8278  & \num{ 8277}&7  &   8&0  &  115 &     1 & \cma14     &      DIBN10698 & \num{10697}&61 &   4&41 &  181 & \cog4 &     1246   \\
     DIBN7470  & \num{ 7470}&30 &   0&71 &    8 &     1 &     124    &      DIBN8283  & \num{ 8283}&14 &   1&42 &   38 &     1 &     1234   &      DIBN10703 & \num{10703}&18 &   1&90 &   21 &     1 & \cma12346  \\
     DIBN7472  & \num{ 7472}&46 &   0&67 &    5 &     1 &     1234   & \cre DIBI8314  & \num{ 8314}&2  &   6&4  &   60 &     1 & \cma146    &      DIBN10731 & \num{10730}&91 &   1&35 &   10 &     1 & \cma1234   \\
     DIBN7476  & \num{ 7476}&43 &   1&22 &    6 &     1 &     1234   &      DIBN8409  & \num{ 8408}&68 &   4&51 &   25 &     1 & \cma145    &      DIBN10735 & \num{10734}&72 &   1&19 &   22 &     1 &     123456 \\
     DIBN7478  & \num{ 7478}&39 &   1&07 &    7 &     1 &     12345  &      DIBN8420  & \num{ 8419}&55 &   1&05 &   12 &     1 & \cma12346  &      DIBN10754 & \num{10753}&57 &   1&48 &    5 &     1 & \cma12346  \\
     DIBN7483  & \num{ 7482}&91 &   0&87 &   10 &     1 &     12345  &      DIBN8422  & \num{ 8422}&09 &   0&97 &    2 &     1 & \cma12346  &      DIBN10757 & \num{10756}&61 &   0&76 &    8 & \cog2 & \cma12346  \\
     DIBN7484  & \num{ 7483}&99 &   0&55 &    5 &     1 &     12345  &      DIBN8439  & \num{ 8439}&37 &   0&70 &   29 &     1 &     123456 &      DIBN10772 & \num{10772}&19 &   3&17 &   25 &     1 &     123456 \\
     DIBN7493  & \num{ 7493}&05 &   0&84 &    3 &     1 &     12345  &      DIBN8456  & \num{ 8456}&09 &   0&78 &    3 &     1 & \cma1236   &      DIBN10775 & \num{10775}&10 &   2&12 &    7 &     1 & \cma1236   \\
     DIBN7495  & \num{ 7494}&92 &   0&59 &   14 &     1 &     123456 &      DIBN8458  & \num{ 8457}&57 &   0&63 &    2 &     1 & \cma123    &      DIBN10781 & \num{10781}&48 &   1&46 &  152 & \cog4 &     12346  \\
     DIBN7520  & \num{ 7520}&46 &   0&92 &    6 &     1 &     123    &      DIBN8472  & \num{ 8471}&63 &   0&43 &    2 &     1 & \cma146    &      DIBN10792 & \num{10792}&47 &   1&75 &   54 & \cog2 &     12346  \\
     DIBN7532  & \num{ 7532}&40 &   0&65 &    3 &     1 &     123    &      DIBN8530  & \num{ 8530}&04 &   2&03 &   10 &     1 &     12456  &      DIBN10814 & \num{10813}&71 &   1&71 &   14 &     1 &     12346  \\
     DIBN7535  & \num{ 7535}&35 &   1&43 &    7 &     1 & \cma123    &      DIBN8571  & \num{ 8570}&89 &   0&45 &    2 &     1 & \cma1234   & \cre DIBI10815 & \num{10815}&1  &   6&8  &   56 &     1 & \cma12346  \\
     DIBN7544  & \num{ 7543}&90 &   2&96 &   12 &     1 &     123    &      DIBN8572  & \num{ 8572}&40 &   0&55 &    2 &     1 & \cma1234   & \cre DIBI10860 & \num{10859}&7  &   6&0  &   39 &     1 & \cma1456   \\
\cre DIBI7558  & \num{ 7558}&2  &   6&2  &   46 &     1 & \cma1234   &      DIBN8580  & \num{ 8580}&03 &   1&59 &   13 &     1 & \cma14     &      DIBN10869 & \num{10868}&79 &   1&53 &    6 &     1 & \cma12346  \\
     DIBN7558  & \num{ 7558}&28 &   1&09 &    8 &     1 &     12346  &      DIBN8621  & \num{ 8620}&99 &   3&99 &  314 & \cog2 &     123456 &      DIBN10877 & \num{10877}&30 &   2&49 &   12 &     1 &     12346  \\
     DIBN7559  & \num{ 7559}&27 &   0&71 &   13 &     1 &     2346   & \cre DIBI8646  & \num{ 8645}&5  &   7&1  &  103 &     1 &     12346  &      DIBN10884 & \num{10884}&07 &   1&30 &   16 &     1 &     12346  \\
     DIBN7562  & \num{ 7562}&15 &   1&72 &   79 & \cog2 &     12356  &      DIBN8764  & \num{ 8763}&54 &   0&70 &    2 &     1 &     36     &      DIBN10894 & \num{10893}&97 &   2&66 &   25 &     1 &     12346  \\
     DIBN7564  & \num{ 7564}&09 &   0&95 &    7 &     1 &     234    & \cre DIBI8781  & \num{ 8780}&8  &   7&7  &  108 &     1 & \cma12345  &      DIBN11691 & \num{11691}&18 &   1&98 &   44 &     1 &     123456 \\
     DIBN7568  & \num{ 7567}&83 &   0&78 &    3 &     1 &     1234   &      DIBN8936  & \num{ 8935}&56 &   1&92 &   20 &     1 & \cma124    &      DIBN11695 & \num{11695}&15 &   1&10 &   17 &     1 &     1234   \\
\cre DIBI7569  & \num{ 7568}&7  &   7&1  &   34 &     1 &     1234   &      DIBN9057  & \num{ 9057}&30 &   2&02 &   16 &     1 & \cma1234   &      DIBN11700 & \num{11699}&56 &   2&35 &   68 &     1 & \cma123456 \\
     DIBN7570  & \num{ 7570}&04 &   1&27 &    6 &     1 &     1234   &      DIBN9366  & \num{ 9365}&81 &   2&13 &  119 & \cog2 &     123    &      DIBN11709 & \num{11708}&79 &   2&90 &   29 &     1 &     12456  \\
     DIBN7571  & \num{ 7571}&45 &   0&54 &    4 &     1 &     1234   &      DIBN9429  & \num{ 9428}&89 &   1&85 &  194 & \cog2 &     2      &      DIBN11721 & \num{11721}&02 &   2&66 &   94 &     1 &     123456 \\
     DIBN7580  & \num{ 7579}&56 &   1&07 &   17 &     1 &     123456 &      DIBN9577  & \num{ 9577}&41 &   3&33 &  284 & \cog2 &     1234   &      DIBN11793 & \num{11793}&02 &   2&25 &   33 &     1 &     12346  \\
     DIBN7581  & \num{ 7581}&19 &   1&37 &   27 &     1 &     123456 &      DIBN9632  & \num{ 9631}&94 &   2&37 &  190 & \cog2 &     1      & \cre DIBI11796 & \num{11796}&3  &   5&6  &   65 &     1 &     123    \\
     DIBN7585  & \num{ 7585}&47 &   1&18 &    5 &     1 &     23     &      DIBN9674  & \num{ 9673}&51 &   1&25 &   13 & \cog2 &     123    &      DIBN11798 & \num{11797}&78 &   1&60 &  147 & \cog2 &     12346  \\
\cre DIBI7687  & \num{ 7687}&2  &   6&9  &  116 &     1 &     12346  &      DIBN9880  & \num{ 9880}&37 &   0&96 &   37 & \cog2 &     1234   &      DIBN11971 & \num{11970}&78 &   1&68 &  137 & \cog2 &     14     \\
     DIBN7696  & \num{ 7695}&75 &   0&64 &    7 &     1 &     123    &      DIBN9888  & \num{ 9887}&85 &   1&25 &   17 &     1 & \cma12345  &      DIBN12201 & \num{12200}&83 &   1&95 &   14 &     1 &     12346  \\
\cbl DIBU7697  & \num{ 7696}&5  & 151&6  & 5445 &     1 &     1234   &      DIBN9892  & \num{ 9892}&14 &   1&25 &   13 &     1 & \cma1234   &      DIBN12230 & \num{12230}&18 &   1&41 &   18 &     1 &     12346  \\
     DIBN7704  & \num{ 7704}&37 &   3&72 &   56 &     1 &     123456 &      DIBN9986  & \num{ 9986}&25 &   1&08 &    6 &     1 &     1234   &      DIBN12268 & \num{12268}&41 &   2&81 &   10 &     1 & \cma1234   \\
\midrule
\end{tabular}
\addtolength{\tabcolsep}{1mm}
\end{table*}

\addtocounter{table}{-1}
\begin{table*}
\caption{(Continued).}
\addtolength{\tabcolsep}{-1mm}
\vspace{-3mm}
\hspace{-10mm}\begin{tabular}{lr@{.}lr@{.}l@{}rc@{}c@{\hspace{5mm}}lr@{.}lr@{.}l@{}rc@{}c@{\hspace{5mm}}lr@{.}lr@{.}l@{}rc@{}c}
\midrule
Name      & \mcii{$\lambda_0$} & \mcii{$\!\!$FWHM}  & \mci{$\!\!\!\!$EW}     & $n_{\rm G}$ & $\!\!$Stars & Name      & \mcii{$\lambda_0$} & \mcii{$\!\!$FWHM}  & \mci{$\!\!\!\!$EW}     & $n_{\rm G}$ & $\!\!$Stars & Name      & \mcii{$\lambda_0$} & \mcii{$\!\!$FWHM}  & \mci{$\!\!\!\!$EW}     & $n_{\rm G}$ & $\!\!$Stars \\
          & \mcii{(\AA)}       & \mcii{$\!\!$(\AA)} & \mci{$\!\!\!\!$(m\AA)} &             & $\!\!$used  &           & \mcii{(\AA)}       & \mcii{$\!\!$(\AA)} & \mci{$\!\!\!\!$(m\AA)} &             & $\!\!$used  &           & \mcii{(\AA)}       & \mcii{$\!\!$(\AA)} & \mci{$\!\!\!\!$(m\AA)} &             & $\!\!$used  \\
\midrule
     DIBN12288 & \num{12287}&68 &    2&84 &   21 &     1 & \cma12346  &      DIBN12878 & \num{12878}&47 &    2&54 &   23 &     1 &     123456 &      DIBN15637 & \num{15636}&99 &    2&96 &   18 &     1 & \cma1236   \\
     DIBN12294 & \num{12294}&05 &    1&62 &   24 &     1 &     12346  &      DIBN12993 & \num{12992}&57 &    1&41 &   19 &     1 & \cma124    & \cre DIBI15647 & \num{15647}&3  &    5&2  &  100 & \cog2 &     234    \\
     DIBN12314 & \num{12313}&82 &    1&44 &   10 &     1 &     12346  &      DIBN12996 & \num{12995}&56 &    3&03 &   39 &     1 & \cma124    &      DIBN15932 & \num{15932}&32 &    3&23 &   13 &     1 & \cma12345  \\
     DIBN12407 & \num{12406}&70 &    1&87 &   15 &     1 & \cma1234   &      DIBN13021 & \num{13020}&70 &    1&58 &   32 &     1 &     124    &      DIBN16568 & \num{16568}&32 &    3&67 &   23 &     1 &     123456 \\
     DIBN12412 & \num{12412}&39 &    1&40 &    4 &     1 & \cma123    &      DIBN13026 & \num{13025}&92 &    2&58 &   44 &     1 &     124    &      DIBN16580 & \num{16580}&05 &    3&44 &   12 &     1 &     123    \\
     DIBN12519 & \num{12519}&27 &    3&00 &   28 &     1 &     123456 & \cre DIBI13028 & \num{13028}&3  &    6&4  &   88 &     1 & \cma124    &      DIBN16633 & \num{16633}&48 &    3&22 &    6 &     1 & \cma123    \\
     DIBN12538 & \num{12538}&11 &    3&25 &   61 & \cog2 & \cma1234   &      DIBN13052 & \num{13051}&58 &    2&29 &   12 &     1 &     1234   & \cre DIBI16918 & \num{16917}&5  &    5&5  &   79 &     1 & \cma123456 \\
     DIBN12624 & \num{12624}&25 &    1&87 &   48 &     1 &     1246   &      DIBN13176 & \num{13176}&29 &    3&89 &  559 & \cog2 &     12346  &      DIBN17063 & \num{17063}&42 &    3&42 &   31 &     1 & \cma123456 \\
     DIBN12650 & \num{12649}&94 &    3&85 &   28 &     1 &     126    &      DIBN15268 & \num{15268}&02 &    3&33 &  268 &     1 &     123456 &      DIBN17072 & \num{17072}&39 &    2&35 &   23 &     1 & \cma1234   \\
     DIBN12692 & \num{12692}&13 &    1&27 &   16 &     1 &     12     &      DIBN15612 & \num{15611}&66 &    4&57 &   65 & \cog3 &     23     &                &        \mcii{} & \mcii{} &      &       &            \\
     DIBN12799 & \num{12799}&12 &    1&44 &   29 &     1 &     1234   &      DIBN15620 & \num{15620}&46 &    1&67 &   14 &     1 & \cma123    &                &        \mcii{} & \mcii{} &      &       &            \\
\midrule
\textit{Other notes:} & \multicolumn{23}{l}{Names are in black for narrow DIBs, in {\cre red} for intermediate ones, in {\cgr green} for broad ones, and in {\cbl blue} for the ultrabroad one.} \\
                      & \multicolumn{23}{l}{DIBs with more than one Gaussian component are in {\cog orange} in the $n_{\rm G}$ column.} \\
                      & \multicolumn{23}{l}{New DIBs in this paper are in {\cma magenta} in the ``Stars used'' column.} \\
\vspace{-5mm}
\end{tabular}
\addtolength{\tabcolsep}{1mm}
\end{table*}

\begin{figure}
 \vspace{-2mm}
 \centerline{\includegraphics*[width=\linewidth]{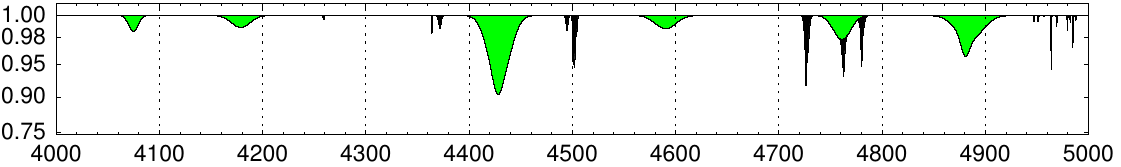}}
 \centerline{\includegraphics*[width=\linewidth]{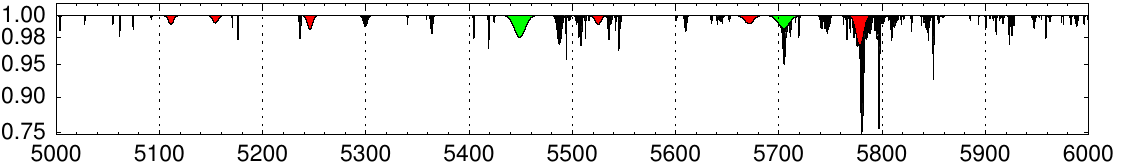}}
 \centerline{\includegraphics*[width=\linewidth]{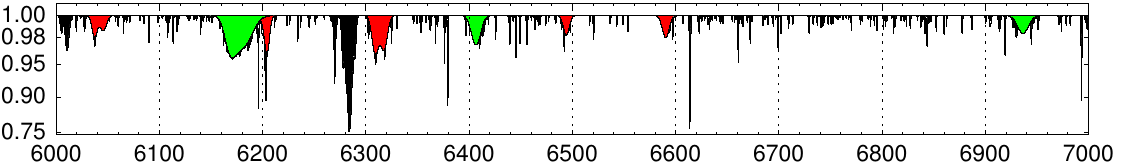}}
 \centerline{\includegraphics*[width=\linewidth]{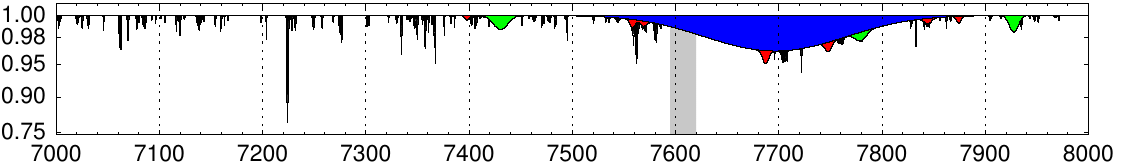}}
 \centerline{\includegraphics*[width=\linewidth]{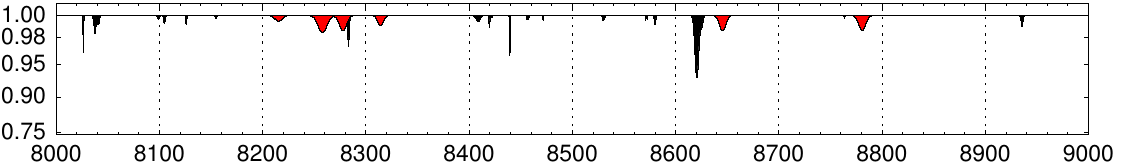}}
 \centerline{\includegraphics*[width=\linewidth]{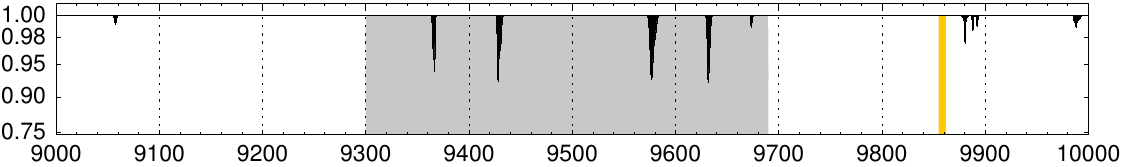}}
 \centerline{\includegraphics*[width=\linewidth]{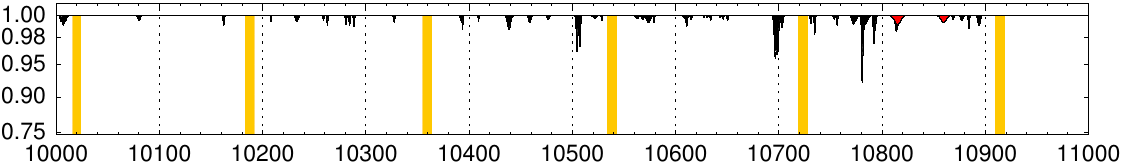}}
 \centerline{\includegraphics*[width=\linewidth]{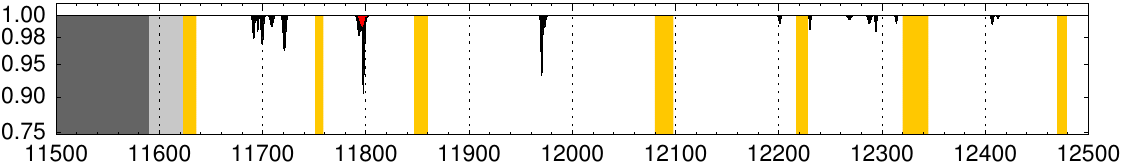}}
 \centerline{\includegraphics*[width=\linewidth]{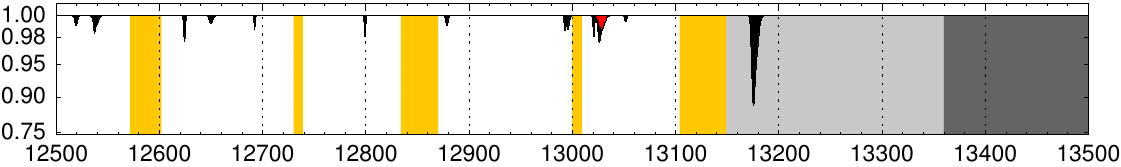}}
 \centerline{\includegraphics*[width=\linewidth]{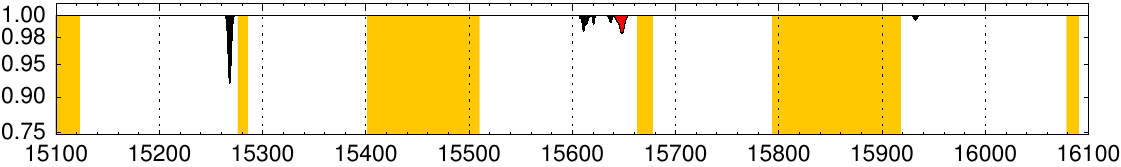}}
 \centerline{\includegraphics*[width=\linewidth]{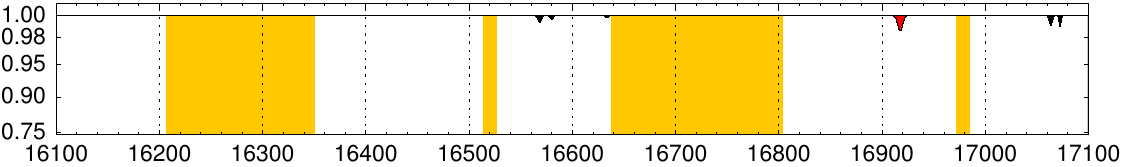}}
 \caption{Rectified DIB spectrum for GLS~\num{11448}. The horizontal scale is in \AA\ and the non-linear vertical scale is used to 
          emphasise weak DIBs. The DIB colour code is the same as in Table~\ref{dibs_all}. Light gray areas indicate where telluric
          absorption severely hampers the detection of DIBs and dark gray areas where it makes it impossible. Light orange areas are
          gaps (detectors or orders) in the CARMENES coverage.}
 \label{dib_spectrum}   
 \vspace{-3mm}   
\end{figure}

\begin{figure}
 \vspace{-2mm}
 \centerline{\includegraphics*[width=0.49\linewidth]{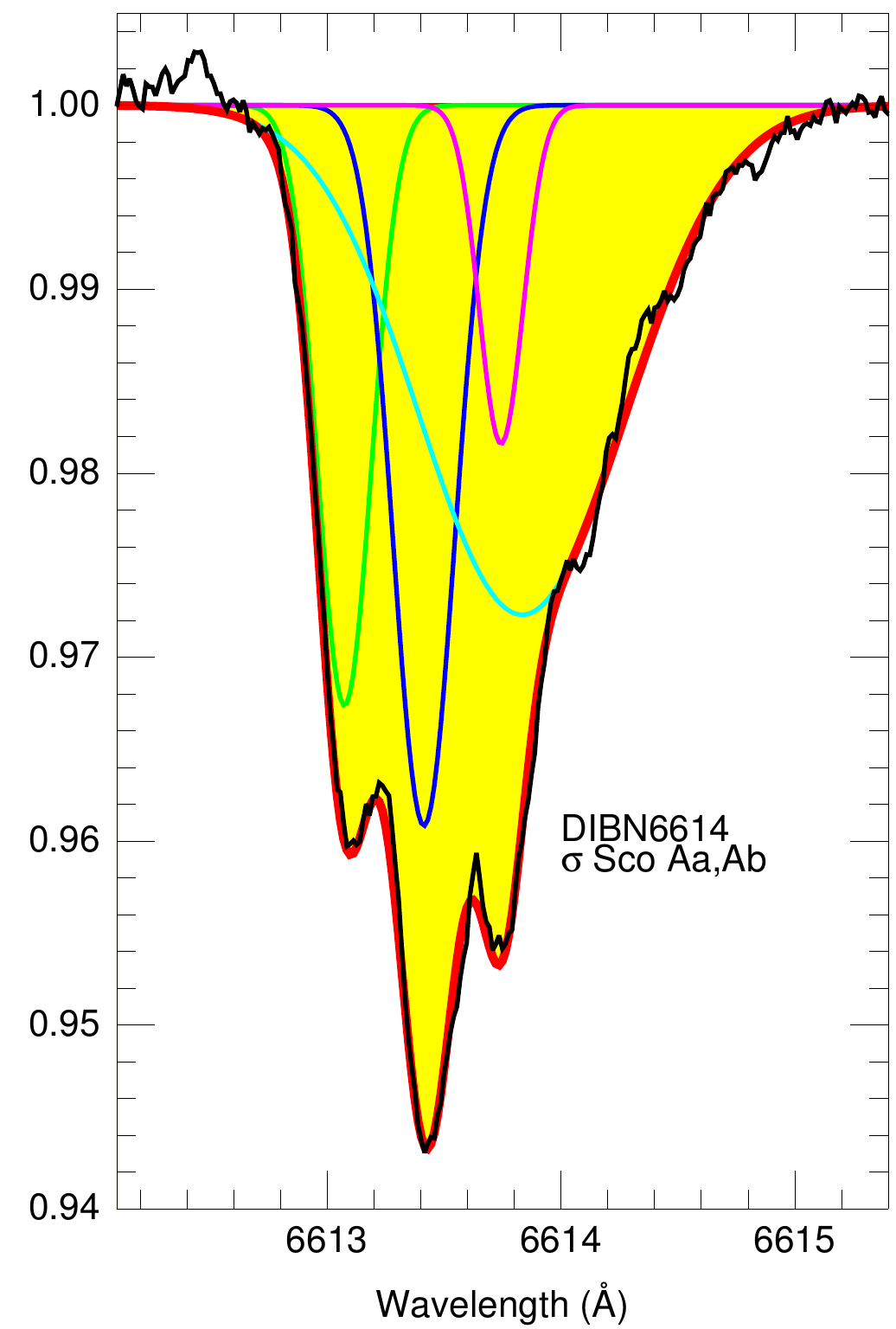}
             \includegraphics*[width=0.49\linewidth]{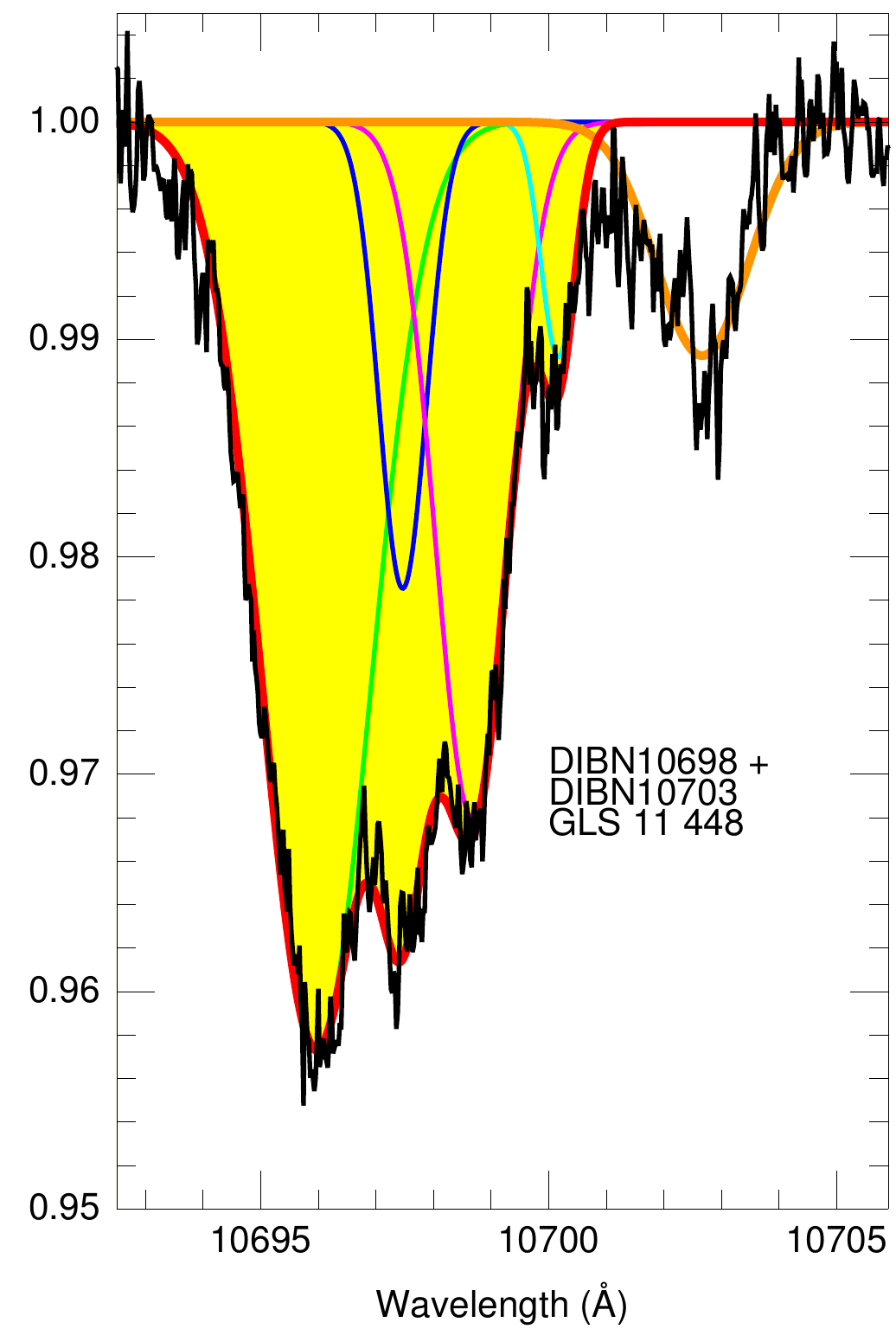}}
 \caption{Two narrow DIBs (black) with multiple Gaussian components (in different colours) and 
          the global fit in red filled in yellow. The right panel shows an additional DIB (DIBN10703).}
 \label{dib_multiple}   
\end{figure}

\begin{figure}
 \vspace{-2mm}
 \centerline{\includegraphics*[width=\linewidth]{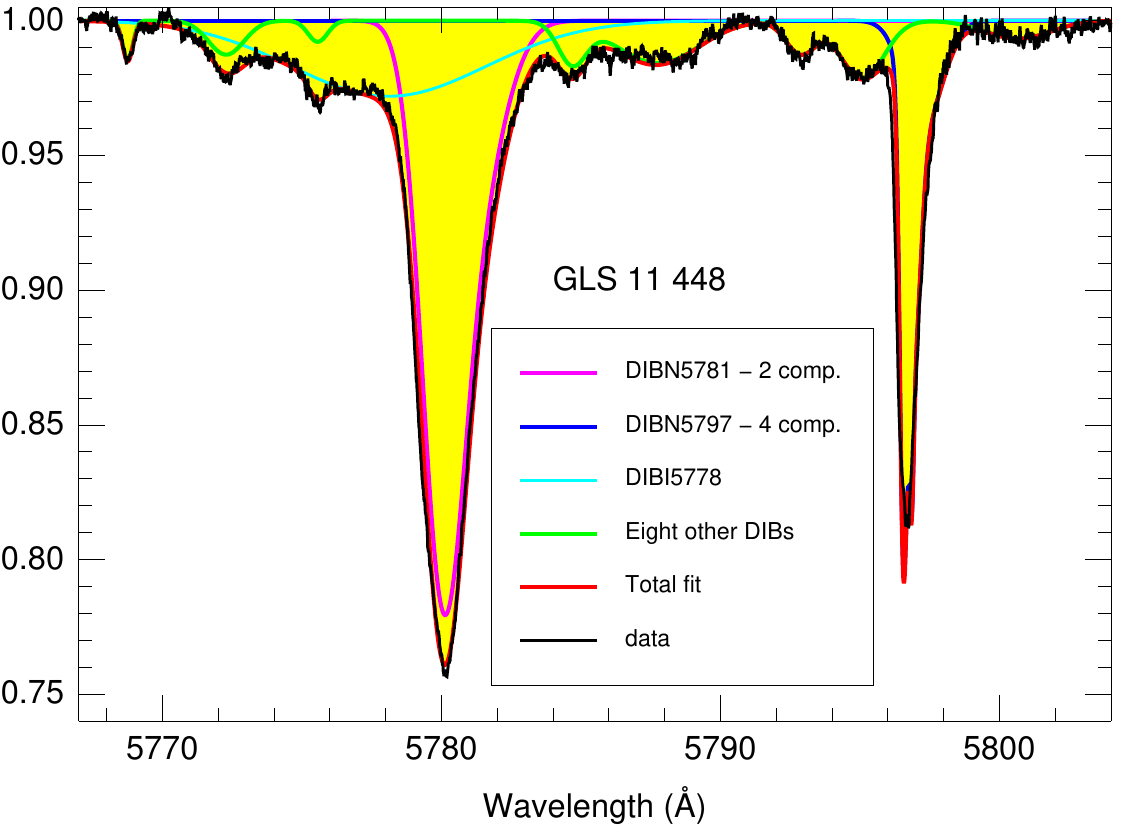}}
 \caption{The 11 DIBs in the 5790~\AA\ region for GLS~\num{11448} including the two-component DIBN5781, 
          the four-component DIBN5797, and DIBI5778. As DIBN5797 is kinematically broadened
          (Fig.~\ref{ISM_RV}), the data are slightly broader than the fit.}
 \label{reg_5790}   
\end{figure}

\begin{figure}
 \vspace{-2mm}
 \centerline{\includegraphics*[width=\linewidth]{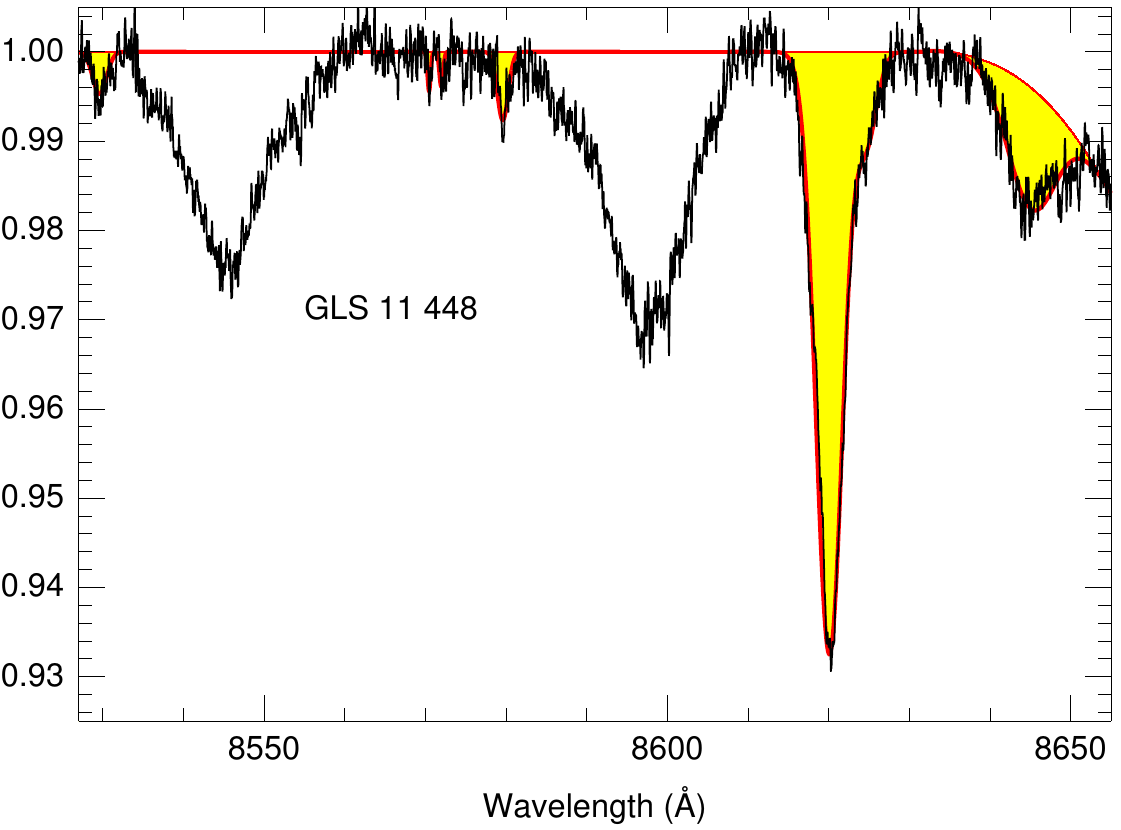}}
 \caption{Six DIBs in the \Gaia\ region for GLS~\num{11448} including the two-component DIBN8621  
          (another three weak DIBs are outside the plotted area close to the edge of the RVS window).}
 \label{reg_Gaia}   
\end{figure}

Given the different wavelength ranges of the \lili\ spectra as well as for practical purposes, we divided the full wavelength range
to be analysed (4000-\num{17100}~\AA) into smaller ranges of $\sim$100~\AA\ each (increasing or decreasing the value as convenient
and setting the limits at regions of featureless continuum) and processed them independently with UNWIND. We inspected each range to
determine the main rectification points away from stellar, ISM, and telluric features and the additional rectification points (if 
needed) inside stellar lines with ISM absorption features. We first cleaned the \lili\ spectra (previously cleaned from telluric
absorption by the pipeline for each telescope) from additional telluric features (when needed) and then we simultaneously fitted the
standard ISM lines and DIBs. 

The DIB fitting was done iteratively, fixing RV to that of the ISM rest frame but leaving $\lambda_0$, FWHM and the EW
free. For DIBs with asymmetries or internal level structure (e.g. DIBN6614) we fitted the profiles with multiple 
(2 to 4) Gaussians. We tried fitting each DIB in as many stars as possible, explicitly excluding the DIB-star 
combinations where the DIB was too weak or contaminated by stellar features. For cases with multiple Gaussians we were even more 
strict in the stars we selected, excluding those where some circumstance (e.g. kinematic broadening, see above) led to a poor fit.
Finally, the $\lambda_0$ and FWHM of each fitted Gaussian was averaged among the selected stars in each case to yield the final
parameters. The fact that a star was not selected to build the DIB parameters does not necessarily mean that the DIB is
not detected for that star in our data, just that it does not meet the requirements described above.

We identify a total of 631 DIBs in the 4000-\num{17100}~\AA\ range (Table~\ref{dibs_all}). Each DIB is given a DIBXY name, 
with X identifying its width category [N(arrow): FWHM $< 5$~\AA, I(ntermediate): between 5~and~10~\AA, B(road): between 
10~and~100~\AA, and U(ltrabroad): FWHM~$> 100$~\AA] and Y being a 4-5 digits number with its central (air) wavelength in \AA\ 
rounded to the nearest integer. A colour code is included in the DIB names to facilitate the identification of each category. For 
each DIB we give its preferred FWHM and $\lambda_0$ as well as the EW for GLS~\num{11448}. We also give the number of Gaussians used to 
fit each DIB (with a colour code to identify cases with multiple Gaussians, see below) and the stars used to build each profile (with a 
colour code to identify new DIBs, see below). 

Of the 631 DIBs in Table~\ref{dibs_all}, 37 are fitted with multiple Gaussians (examples in Fig.~\ref{dib_multiple})
and their information is given in 
Table~\ref{dibs_mult}. For each of those, we list the name, $\lambda_0$, and FWHM from Table~\ref{dibs_all} and also give the 
skewness and excess kurtosis (both zero for single Gaussians). In addition, we give the $\lambda_0$, FWHM, and flux fraction for 
each component. We note that, excluding multiple kinematic components, DIB profiles are relatively stable other than for a few 
exceptions \citep{Okaetal13}.  Therefore, in the standard use of UNWIND (getting rid of the ISM to study the stellar spectra) we fix
$\lambda_0$, FWHM, and the flux fraction of each Gaussian component and fit just the EWs and the overall DIB RV. The flexibility 
of UNWIND allows the user to modify the input DIB catalog in order to change (or fit) the $\lambda_0$ and FWHM and to 
leave each component free or fixed, thus allowing it to be used as a tool to study the ISM.

\subsection{A first analysis}

\begin{table*}
\caption{Additional information for DIBs fitted with multiple Gaussians: skewness, excess kurtosis, and $\lambda_0$, FWHM, and flux fraction for each component.}
\addtolength{\tabcolsep}{-1.2mm}
\vspace{-3mm}
\hspace{-2mm}\begin{tabular}{lr@{.}lr@{.}lrrr@{.}lr@{.}lc@{\hspace{6mm}}lr@{.}lr@{.}lrrr@{.}lr@{.}lc}
\midrule
Name      & \mcii{$\lambda_0$} & \mcii{$\!\!\!$FWHM}  & skew & kurt & \mcii{$\lambda_0$} & \mcii{$\!\!$FWHM}  & frac & Name      & \mcii{$\lambda_0$} & \mcii{$\!\!\!$FWHM}  & skew & kurt & \mcii{$\lambda_0$} & \mcii{$\!\!$FWHM}  & frac \\
          & \mcii{(\AA)}       & \mcii{$\!\!\!$(\AA)} &      &      & \mcii{(\AA)}       & \mcii{$\!\!$(\AA)} &      &           & \mcii{(\AA)}       & \mcii{$\!\!\!$(\AA)} &      &      & \mcii{(\AA)}       & \mcii{$\!\!$(\AA)} &      \\
\midrule
\cgr DIBB4429  & \num{ 4428}&8  & \num{ 17}&7  &    0.05 &    0.28 & \num{ 4427}&70 & \num{  9}&20 & 0.135 &      DIBN9366  & \num{ 9365}&81 & \num{  2}&13 & $-$0.42 & $-$0.40 & \num{ 9365}&43 & \num{  1}&87 & 0.655 \\
           --- &     \mcii{---} &   \mcii{---} &     --- &     --- & \num{ 4429}&00 & \num{ 22}&70 & 0.865 &            --- &     \mcii{---} &   \mcii{---} &     --- &     --- & \num{ 9366}&54 & \num{  0}&90 & 0.345 \\
     DIBN4502  & \num{ 4501}&93 & \num{  2}&99 &    0.42 & $-$0.42 & \num{ 4500}&35 & \num{  0}&72 & 0.141 &      DIBN9429  & \num{ 9428}&89 & \num{  1}&85 &    0.54 & $-$0.46 & \num{ 9428}&02 & \num{  1}&55 & 0.588 \\
           --- &     \mcii{---} &   \mcii{---} &     --- &     --- & \num{ 4501}&87 & \num{  2}&21 & 0.741 &            --- &     \mcii{---} &   \mcii{---} &     --- &     --- & \num{ 9430}&13 & \num{  2}&32 & 0.412 \\
           --- &     \mcii{---} &   \mcii{---} &     --- &     --- & \num{ 4504}&20 & \num{  1}&35 & 0.118 &      DIBN9577  & \num{ 9577}&41 & \num{  3}&33 &    0.54 &    0.02 & \num{ 9576}&70 & \num{  3}&01 & 0.782 \\
     DIBN4727  & \num{ 4726}&85 & \num{  2}&42 &    0.22 &    0.29 & \num{ 4726}&13 & \num{  0}&91 & 0.162 &            --- &     \mcii{---} &   \mcii{---} &     --- &     --- & \num{ 9579}&96 & \num{  3}&11 & 0.218 \\
           --- &     \mcii{---} &   \mcii{---} &     --- &     --- & \num{ 4726}&99 & \num{  3}&99 & 0.838 &      DIBN9632  & \num{ 9631}&94 & \num{  2}&37 &    0.29 & $-$0.15 & \num{ 9631}&51 & \num{  1}&82 & 0.700 \\
     DIBN4763  & \num{ 4762}&57 & \num{  2}&10 & $-$0.06 &    0.31 & \num{ 4762}&54 & \num{  3}&00 & 0.875 &            --- &     \mcii{---} &   \mcii{---} &     --- &     --- & \num{ 9632}&94 & \num{  1}&92 & 0.300 \\
           --- &     \mcii{---} &   \mcii{---} &     --- &     --- & \num{ 4762}&76 & \num{  0}&90 & 0.125 &      DIBN9674  & \num{ 9673}&51 & \num{  1}&25 &    0.43 & $-$0.29 & \num{ 9673}&16 & \num{  0}&75 & 0.538 \\
     DIBN4780  & \num{ 4780}&44 & \num{  1}&56 &    0.39 &    1.25 & \num{ 4780}&11 & \num{  1}&21 & 0.400 &            --- &     \mcii{---} &   \mcii{---} &     --- &     --- & \num{ 9673}&91 & \num{  1}&07 & 0.462 \\
           --- &     \mcii{---} &   \mcii{---} &     --- &     --- & \num{ 4780}&66 & \num{  3}&92 & 0.600 &      DIBN9880  & \num{ 9880}&37 & \num{  0}&96 & $-$0.25 &    2.44 & \num{ 9880}&27 & \num{  3}&49 & 0.486 \\
\cgr DIBB4884  & \num{ 4883}&7  & \num{ 16}&7  &    0.20 &    0.49 & \num{ 4880}&50 & \num{  9}&00 & 0.211 &            --- &     \mcii{---} &   \mcii{---} &     --- &     --- & \num{ 9880}&47 & \num{  0}&83 & 0.514 \\
           --- &     \mcii{---} &   \mcii{---} &     --- &     --- & \num{ 4884}&60 & \num{ 28}&60 & 0.789 &      DIBN10393 & \num{10393}&31 & \num{  1}&07 & $-$0.87 &    0.78 & \num{10392}&66 & \num{  1}&66 & 0.312 \\
     DIBN5720  & \num{ 5719}&60 & \num{  0}&66 &    0.37 & $-$0.59 & \num{ 5719}&39 & \num{  0}&59 & 0.706 &            --- &     \mcii{---} &   \mcii{---} &     --- &     --- & \num{10393}&61 & \num{  0}&95 & 0.688 \\
           --- &     \mcii{---} &   \mcii{---} &     --- &     --- & \num{ 5720}&09 & \num{  0}&53 & 0.294 &      DIBN10610 & \num{10610}&24 & \num{  1}&72 & $-$0.87 &    0.95 & \num{10609}&48 & \num{  3}&11 & 0.400 \\
     DIBN5766  & \num{ 5765}&91 & \num{  0}&74 & $-$0.14 &    0.62 & \num{ 5765}&88 & \num{  1}&34 & 0.762 &            --- &     \mcii{---} &   \mcii{---} &     --- &     --- & \num{10610}&74 & \num{  1}&48 & 0.600 \\
           --- &     \mcii{---} &   \mcii{---} &     --- &     --- & \num{ 5766}&00 & \num{  0}&45 & 0.238 &      DIBN10698 & \num{10697}&61 & \num{  4}&41 &    0.14 & $-$0.74 & \num{10696}&49 & \num{  2}&24 & 0.558 \\
     DIBN5781  & \num{ 5780}&69 & \num{  2}&06 &    0.46 &    0.32 & \num{ 5780}&34 & \num{  1}&78 & 0.709 &            --- &     \mcii{---} &   \mcii{---} &     --- &     --- & \num{10698}&01 & \num{  0}&97 & 0.122 \\
           --- &     \mcii{---} &   \mcii{---} &     --- &     --- & \num{ 5781}&53 & \num{  2}&34 & 0.291 &            --- &     \mcii{---} &   \mcii{---} &     --- &     --- & \num{10699}&21 & \num{  1}&48 & 0.276 \\
     DIBN5797  & \num{ 5797}&23 & \num{  0}&70 &    1.11 &    1.98 & \num{ 5796}&85 & \num{  0}&30 & 0.265 &            --- &     \mcii{---} &   \mcii{---} &     --- &     --- & \num{10700}&73 & \num{  0}&70 & 0.044 \\
           --- &     \mcii{---} &   \mcii{---} &     --- &     --- & \num{ 5797}&14 & \num{  0}&19 & 0.069 &      DIBN10757 & \num{10756}&61 & \num{  0}&76 &    0.84 &    0.73 & \num{10756}&39 & \num{  0}&65 & 0.625 \\
           --- &     \mcii{---} &   \mcii{---} &     --- &     --- & \num{ 5797}&21 & \num{  0}&60 & 0.317 &            --- &     \mcii{---} &   \mcii{---} &     --- &     --- & \num{10756}&98 & \num{  1}&23 & 0.375 \\
           --- &     \mcii{---} &   \mcii{---} &     --- &     --- & \num{ 5797}&56 & \num{  1}&61 & 0.349 &      DIBN10781 & \num{10781}&48 & \num{  1}&46 &    1.69 &    2.99 & \num{10780}&40 & \num{  1}&02 & 0.408 \\
\cgr DIBB6176  & \num{ 6176}&2  & \num{ 23}&9  &    0.35 & $-$0.27 & \num{ 6168}&60 & \num{ 11}&50 & 0.306 &            --- &     \mcii{---} &   \mcii{---} &     --- &     --- & \num{10781}&03 & \num{  0}&83 & 0.224 \\
           --- &     \mcii{---} &   \mcii{---} &     --- &     --- & \num{ 6179}&50 & \num{ 21}&30 & 0.694 &            --- &     \mcii{---} &   \mcii{---} &     --- &     --- & \num{10781}&76 & \num{  1}&41 & 0.105 \\
     DIBN6203  & \num{ 6202}&83 & \num{  1}&31 & $-$0.39 & $-$0.15 & \num{ 6202}&57 & \num{  1}&24 & 0.583 &            --- &     \mcii{---} &   \mcii{---} &     --- &     --- & \num{10783}&44 & \num{  5}&13 & 0.263 \\
           --- &     \mcii{---} &   \mcii{---} &     --- &     --- & \num{ 6203}&20 & \num{  0}&79 & 0.417 &      DIBN10792 & \num{10792}&47 & \num{  1}&75 &    0.53 & $-$0.13 & \num{10791}&96 & \num{  1}&33 & 0.630 \\
     DIBN6270  & \num{ 6269}&83 & \num{  1}&27 &    0.13 &    0.98 & \num{ 6269}&71 & \num{  0}&89 & 0.314 &            --- &     \mcii{---} &   \mcii{---} &     --- &     --- & \num{10793}&34 & \num{  1}&82 & 0.370 \\
           --- &     \mcii{---} &   \mcii{---} &     --- &     --- & \num{ 6269}&88 & \num{  2}&85 & 0.686 &      DIBN11798 & \num{11797}&78 & \num{  1}&60 &    0.49 & $-$0.00 & \num{11797}&31 & \num{  0}&99 & 0.415 \\
     DIBN6284  & \num{ 6284}&04 & \num{  3}&31 &    0.13 & $-$0.12 & \num{ 6281}&63 & \num{  1}&74 & 0.120 &            --- &     \mcii{---} &   \mcii{---} &     --- &     --- & \num{11798}&12 & \num{  1}&78 & 0.585 \\
           --- &     \mcii{---} &   \mcii{---} &     --- &     --- & \num{ 6283}&59 & \num{  1}&74 & 0.231 &      DIBN11971 & \num{11970}&78 & \num{  1}&68 &    1.13 &    1.00 & \num{11970}&27 & \num{  1}&65 & 0.810 \\
           --- &     \mcii{---} &   \mcii{---} &     --- &     --- & \num{ 6284}&65 & \num{  3}&63 & 0.649 &            --- &     \mcii{---} &   \mcii{---} &     --- &     --- & \num{11972}&95 & \num{  2}&37 & 0.190 \\
     DIBN6614  & \num{ 6613}&71 & \num{  0}&96 &    0.52 & $-$0.04 & \num{ 6613}&18 & \num{  0}&27 & 0.164 &      DIBN12538 & \num{12538}&11 & \num{  3}&25 &    0.52 &    0.24 & \num{12536}&96 & \num{  1}&91 & 0.328 \\
           --- &     \mcii{---} &   \mcii{---} &     --- &     --- & \num{ 6613}&52 & \num{  0}&31 & 0.226 &            --- &     \mcii{---} &   \mcii{---} &     --- &     --- & \num{12538}&67 & \num{  4}&94 & 0.672 \\
           --- &     \mcii{---} &   \mcii{---} &     --- &     --- & \num{ 6613}&85 & \num{  0}&22 & 0.075 &      DIBN13176 & \num{13176}&29 & \num{  3}&89 &    0.66 &    0.36 & \num{13175}&24 & \num{  3}&17 & 0.606 \\
           --- &     \mcii{---} &   \mcii{---} &     --- &     --- & \num{ 6613}&94 & \num{  1}&04 & 0.535 &            --- &     \mcii{---} &   \mcii{---} &     --- &     --- & \num{13177}&90 & \num{  5}&20 & 0.394 \\
     DIBN6993  & \num{ 6993}&14 & \num{  0}&80 &    0.27 & $-$0.16 & \num{ 6992}&91 & \num{  0}&49 & 0.319 &      DIBN15612 & \num{15611}&66 & \num{  4}&57 &    0.16 & $-$0.26 & \num{15607}&53 & \num{  1}&48 & 0.061 \\
           --- &     \mcii{---} &   \mcii{---} &     --- &     --- & \num{ 6993}&24 & \num{  0}&74 & 0.681 &            --- &     \mcii{---} &   \mcii{---} &     --- &     --- & \num{15610}&54 & \num{  2}&45 & 0.477 \\
     DIBN7224  & \num{ 7224}&06 & \num{  1}&24 &    0.24 &    0.06 & \num{ 7223}&81 & \num{  0}&94 & 0.364 &            --- &     \mcii{---} &   \mcii{---} &     --- &     --- & \num{15613}&37 & \num{  3}&89 & 0.462 \\
           --- &     \mcii{---} &   \mcii{---} &     --- &     --- & \num{ 7224}&20 & \num{  1}&32 & 0.636 & \cre DIBI15647 & \num{15647}&3  & \num{  5}&2  & $-$0.31 & $-$0.20 & \num{15643}&15 & \num{  3}&29 & 0.140 \\
     DIBN7562  & \num{ 7562}&15 & \num{  1}&72 & $-$0.01 & $-$0.14 & \num{ 7560}&90 & \num{  0}&43 & 0.025 &            --- &     \mcii{---} &   \mcii{---} &     --- &     --- & \num{15648}&03 & \num{  4}&96 & 0.860 \\
           --- &     \mcii{---} &   \mcii{---} &     --- &     --- & \num{ 7562}&18 & \num{  1}&71 & 0.975 &                &        \mcii{} &      \mcii{} &         &         &        \mcii{} &      \mcii{} &       \\
     DIBN8621  & \num{ 8620}&99 & \num{  3}&99 &    0.57 &    0.39 & \num{ 8620}&53 & \num{  3}&94 & 0.901 &                &        \mcii{} &      \mcii{} &         &         &        \mcii{} &      \mcii{} &       \\
           --- &     \mcii{---} &   \mcii{---} &     --- &     --- & \num{ 8625}&15 & \num{  2}&98 & 0.099 &                &        \mcii{} &      \mcii{} &         &         &        \mcii{} &      \mcii{} &       \\
\midrule
\multicolumn{24}{l}{Names are in black for narrow DIBs, in {\cre red} for intermediate ones, and in {\cgr green} for broad ones.} \\
\end{tabular}
\addtolength{\tabcolsep}{1.2mm}
\vspace{-5mm}
\label{dibs_mult}
\end{table*}

$\,\!$\indent As mentioned in the main text, the DIB catalog was built for the practical purpose of eliminating the effects of the
ISM in stellar spectra. Yet, it is valuable for ISM studies. An in-depth analysis is left for future papers,
here we provide a quick overview.

The new DIB catalog is more extensive in number and wavelength coverage than previous ones 
\citep{JennDese94,Galaetal00,Hobbetal08,Hobbetal09,Maizetal15c,Elyaetal17b,Fanetal19,Hamaetal22,Ebenetal22,Ebenetal24}. 
The whole DIB spectrum between 4000~\AA\ and 
\num{17100}~\AA\ is shown in Fig.~\ref{dib_spectrum}. Our coverage is pretty complete in that region except for the atmospheric
gaps between the $Y$ and $J$ and between the $J$ and $H$ bands and some holes in the CARMENES coverage in the near infrared. The 
detection threshold changes due to telluric absorption (ranging from total blockage to little or no effect, allowing for the detection 
of only strong DIBs in between) and S/N changes (at short wavelengths the highly extinguished stars where weak DIBs are visible
are faint). Nevertheless, 
Fig.~\ref{dib_spectrum} shows that DIBs are more common in the 5400-8000~\AA\ range than at longer wavelengths. The situation
is less clear for shorter wavelengths, as the S/N is lower and OB stars have more absorption lines that can hide weak DIBs.
Narrow DIBs outnumber those with larger FWHMs but in terms of EW the situation is different, with broad and ultrabroad DIBs being
more significant. In particular, broad DIBs clearly dominate the blue region but this may be partially a sensitivity
issue due to the faint character of the first four standards there.

For 116 out of the 631 DIBs here we have not found equivalents
in previous catalogs. The number of new discoveries is especially large in the 9800-\num{10900}~\AA\ ($Y$ band) region, though we
missed two DIBs seen by \citet{Hamaetal22} because they fall on gaps of the CARMENES wavelength coverage. Among the new
DIBs there are some particularly strong ones: DIBB6407 (hindered by stellar absorption in B stars but easily visible in
obscured O stars), DIBB6937, and DIBI8258 (both in regions of moderate telluric absorption).

We end up this Appendix discussing some specific DIBs.

\textit{DIBB4429.} This is the most prominent broad DIB in the blue region and has been studied for a long time \citep{Duke51}. If
one rectifies the continuum far away from the DIB itself it may be possible that the profile is a Lorentzian rather than a Gaussian
\citep{Snowetal02} but such an endeavour is not practical for our primary purpose of analysing the stellar spectra. Therefore, we
set the rectification points around 4400~\AA\ and 4460~\AA\ (the precise values depending on the star) and we find, after
subtracting the stellar features, that the profile is slightly asymmetric and requires two Gaussians.

\textit{DIBB4884 (a.k.a. the H$\beta$ DIB).} As this DIB is located right next to the (almost always present in early-type stars) 
H$\beta$ line, its properties are difficult to determine. We did so by forcing the stellar profile to be symmetric with respect to 
its central wavelength. The DIB profile has to be fitted with two Gaussians that we combine into a single profile
for convenience, as in most stars it is difficult to fit both independently with precision, but it is possible that they exhibit
variations from star to star. Note, however, than in the two relatively different sightlines of GLS~\num{11448} and GLS~\num{11449}
they have EWs consistent (within the uncertainties) with the same profile \citep{Maizetal15c}.

\textit{The 5790~\AA\ region.} This region includes the two DIBs that were first discovered, DIBN5781 and DIBN5797 \citep{Hege22},
and that are used as the basis for distinguishing $\sigma$ and $\zeta$ sightlines. It also includes an intermediate-width DIB with
typical EWs comparable to the two main ones, DIBI5778, as well as another eight weaker narrow DIBs (Fig.~\ref{reg_5790}) that should
be fitted to yield accurate EWs of DIBN5781 and DIBN5797, especially at low/intermediate spectral resolutions. We fit DIBN5781 with
two Gaussians and DIBN5797 with four, with the latter shows kinematic broadening in our first four standard stars.

\textit{DIBU7697.} This ultrabroad band was discovered by \citet{Maizetal21a} in HST and ground-based spectra and later confirmed by
\citet{Weiletal23} using \Gaia\ XP spectra. These bands are sometimes called intermediate-scale structures and several others
appear to exist in the optical spectrum but those are somewhat broader and harder to detect \citep{Zhanetal24b,SaydGree25}. The
single-Gaussian fit yields a central wavelength of 7696.5~\AA\ and a FWHM of 151.6~\AA, the first compatible with the (vacuum) value
of $7699.2 \pm 1.3$~\AA\ of \citet{Maizetal21a} but the second somewhat smaller than the $176.6\pm 3.9$~\AA\ value there. The likely
explanations for that small discrepancy are the lower extinction and resolution of the original STIS data and the addition of 
several new DIBs in the present analysis that detract from the ultrabroad profile.

\textit{The Gaia RVS window.} The 8450-8720~\AA\ region covered by the \Gaia\ RVS spectrometer \citep{Prusetal16} has become 
popular for DIB studies thanks to the data availability in \Gaia~DR3, soon to increase significantly in DR4. Most such
studies concentrate on the strongest DIB, DIBN8621 (e.g. \citealt{Schuetal23a}), with some including the second one in EW for most
sightlines, DIBI8646 (e.g. \citealt{Zhaoetal24}). A third previously known DIB, DIBN8530 \citep{JennDese94},
has been ignored in recent studies to our knowledge. In addition to those, we detect three new weak DIBs close to the bluemost
window limit (DIBN8456, DIBN8459, and DIBN8472, possibly excluded by the processing pipeline) and another three in the central part
of the window (DIBN8571, DIBN8572, and DIBN8580, Fig.~\ref{reg_Gaia}). We suspect the reason they have not been discussed before 
despite the availability of \Gaia\ data is that most current analyses in the RVS window are done with late-type stars, where 
stellar lines more easily hide ISM bands. Of the not studied ones, DIBN8530 and DIBN8580 have the best potential for detection. The 
well-studied DIBN8621 is clearly asymmetric in our data, a fact not considered in some of the recent studies, and has to be fitted
with a double Gaussian profile that yields a significant skewness of 0.57. The intermediate-width DIBI8646 is more difficult to
measure due to the presence of a nearby Paschen line.

\textit{The fullerene DIBs.} DIBN9366, DIBN9429, DIBN9577, and DIBN9632 are the only four DIBs with an identified carrier,
C$_{60}^+$ \citep{Campetal15}. All four are in a region with severe telluric absorption, especially the first two. Nevertheless, we
are able to detect them thanks to our multi-epoch technique that moves telluric lines with respect to the ISM rest
frame. All four require two Gaussians due to their asymmetric nature.

\section{Additional table}

\begin{table*}
\caption{Spectroscopic epochs for GLS~\num{11448}, see Table~\ref{spectrographs} for the code meanings}
\addtolength{\tabcolsep}{-1.5mm}
\centerline{
\begin{tabular}{cccr@{.}l@{$\pm$}r@{.}lr@{.}l@{$\pm$}r@{.}l@{\hspace{6mm}}cccr@{.}l@{$\pm$}r@{.}lr@{.}l@{$\pm$}r@{.}l}
\midrule
Code & Evening date             & BRJD & \mciv{RV$_{\rm Aa}$}                 & \mciv{RV$_{\rm Ab}$}                 & Code & Evening date             & BRJD & \mciv{RV$_{\rm Aa}$}                 & \mciv{RV$_{\rm Ab}$}                 \\
     &                          &      & \mciv{\ion{He}{ii}~$\lambda$5411.53} & \mciv{\ion{He}{ii}~$\lambda$5411.53} &      &                          &      & \mciv{\ion{He}{ii}~$\lambda$5411.53} & \mciv{\ion{He}{ii}~$\lambda$5411.53} \\
     & {\scriptsize (YYYYMMDD)} & (d)  & \mciv{(km/s)}                        & \mciv{(km/s)}                        &      & {\scriptsize (YYYYMMDD)} & (d)  & \mciv{(km/s)}                        & \mciv{(km/s)}                        \\
\cmidrule(r){1-11}\cmidrule(r){12-22}
HET-B  & 20110621 & \num{55734.77} &  $-$45&3& 3&3 &     22&9& 1&8 & Merc   & 20160926 & \num{57658.41} & $-$166&3& 5&9 &    147&0& 5&8 \\
HET-R  & 20110626 & \num{55739.78} &  $-$18&5& 5&5 &      2&6& 4&5 & Merc   & 20160926 & \num{57658.43} & $-$163&3&10&6 &    151&8& 2&6 \\
HET-R  & 20110913 & \num{55818.79} & $-$116&3& 2&3 &    104&0& 2&4 & Merc   & 20160927 & \num{57659.38} & $-$166&1& 6&8 &    149&4& 2&6 \\
FIES   & 20110915 & \num{55820.49} &  $-$95&8& 5&4 &     87&6& 2&6 & FIES   & 20171001 & \num{58028.43} &     64&2& 2&5 &  $-$88&9& 3&7 \\
FIES   & 20110917 & \num{55822.36} &  $-$83&5& 5&6 &     84&6& 3&5 & Merc   & 20171019 & \num{58046.40} & $-$172&5&10&0 &    145&4& 6&4 \\
HET-B  & 20110928 & \num{55833.75} &  $-$27&0& 7&4 &     16&3& 9&6 & Merc   & 20171019 & \num{58046.42} & $-$168&3& 7&6 &    152&1& 2&0 \\
HET-R  & 20111002 & \num{55837.72} &  $-$11&4& 6&7 &   $-$3&8& 6&7 & Merc   & 20171020 & \num{58047.50} & $-$168&4& 7&1 &    152&8& 1&7 \\
HET-R  & 20111003 & \num{55838.72} &  $-$15&6& 2&6 &   $-$0&2& 2&0 & Merc   & 20171020 & \num{58047.52} & $-$164&2& 5&1 &    151&3& 1&7 \\
HET-R  & 20111006 & \num{55841.71} &  $-$14&3& 4&1 &   $-$4&2& 6&4 & Merc   & 20171020 & \num{58047.55} & $-$163&5& 5&8 &    155&3& 8&4 \\
HET-R  & 20111007 & \num{55842.71} &   $-$7&3& 2&2 &   $-$3&3& 4&0 & Merc   & 20171025 & \num{58052.37} & $-$128&0& 4&8 &    113&0& 4&0 \\
HET-R  & 20111010 & \num{55845.72} &   $-$5&6& 7&2 &  $-$14&0& 6&3 & Merc   & 20171025 & \num{58052.39} & $-$129&4&25&0 &    118&6&25&2 \\
HET-R  & 20111031 & \num{55866.65} &     38&4& 2&2 &  $-$57&6& 2&3 & Merc   & 20171026 & \num{58053.34} & $-$114&2&13&7 &    108&8& 6&5 \\
HET-R  & 20111115 & \num{55881.60} &     57&5& 1&1 &  $-$75&7& 3&5 & Merc   & 20171026 & \num{58053.36} & $-$112&6& 9&9 &    104&0& 7&4 \\
HET-R  & 20111127 & \num{55893.59} &     63&5& 1&7 &  $-$80&1& 2&6 & Merc   & 20181018 & \num{58410.48} &     48&6&16&5 &  $-$87&6&14&6 \\
HET-R  & 20120411 & \num{56029.95} &  $-$21&1& 6&7 &      8&5& 7&2 & Merc   & 20181025 & \num{58417.39} &     69&7& 5&6 &  $-$80&8& 7&3 \\
HET-R  & 20120515 & \num{56063.88} &     43&4& 1&8 &  $-$64&6& 0&9 & CAR    & 20190824 & \num{58720.40} &  $-$18&9& 7&4 &      7&2& 8&4 \\
HET-R  & 20120528 & \num{56076.82} &     61&4& 1&3 &  $-$79&3& 3&4 & FIES   & 20190831 & \num{58727.59} & $-$168&4& 9&0 &    139&5& 4&0 \\
HET-R  & 20120610 & \num{56089.82} &     54&7& 2&5 &  $-$76&8& 2&1 & FIES   & 20190901 & \num{58728.55} & $-$160&0& 4&3 &    149&8& 3&0 \\
HET-B  & 20120827 & \num{56167.81} &     54&3& 0&9 &  $-$75&1& 1&0 & FIES   & 20190902 & \num{58729.49} & $-$151&1& 3&3 &    143&3& 2&2 \\
CAF\'E & 20120907 & \num{56178.54} &     69&6& 4&7 &  $-$87&6& 5&1 & Merc   & 20200613 & \num{59014.65} &  $-$97&0& 6&5 &     88&8& 5&9 \\
HET-R  & 20121002 & \num{56203.71} & $-$149&8& 2&4 &    129&6& 5&5 & Merc   & 20200613 & \num{59014.67} & $-$100&1& 5&4 &     74&4& 4&2 \\
HET-R  & 20121004 & \num{56205.73} & $-$132&9& 2&3 &    115&3& 2&4 & Merc   & 20200629 & \num{59030.70} &  $-$75&6& 3&7 &     60&1& 3&9 \\
HET-R  & 20121005 & \num{56206.73} & $-$125&0& 2&8 &    106&7& 2&6 & CAR    & 20210622 & \num{59388.61} &     60&5& 4&1 &  $-$80&3& 5&9 \\
CAF\'E & 20121022 & \num{56223.34} & \mciv{---}    & \mciv{---}    & CAR    & 20210622 & \num{59388.63} &     60&9& 4&4 &  $-$75&3& 4&5 \\
HET-R  & 20121204 & \num{56266.55} &     51&2& 4&6 &  $-$78&1& 1&5 & CAR    & 20210720 & \num{59416.52} &  $-$85&0&11&1 &     83&3& 4&7 \\
CAF\'E & 20121205 & \num{56267.38} &     71&6& 5&8 &  $-$71&0& 8&2 & CAR    & 20210918 & \num{59476.47} &     58&2& 3&4 &  $-$75&3& 3&6 \\
HET-R  & 20121207 & \num{56269.53} &     58&2& 1&9 &  $-$81&2& 0&9 & FIES   & 20211020 & \num{59508.37} & $-$140&4& 2&5 &    124&6& 3&1 \\
CAF\'E & 20121230 & \num{56292.28} &  $-$41&9& 8&1 &     33&5& 7&3 & CAR    & 20211023 & \num{59511.39} & $-$102&5&10&3 &    105&9& 5&2 \\
CAF\'E & 20130930 & \num{56566.34} &     76&1& 9&7 &  $-$80&4&15&3 & CAR    & 20220519 & \num{59719.54} &  $-$19&6&10&2 &      3&1&10&8 \\
Merc   & 20131028 & \num{56594.33} & $-$130&0&11&2 &    116&4& 7&0 & FIES   & 20220809 & \num{59801.63} & $-$122&5& 4&2 &    114&4& 2&7 \\
CAF\'E & 20140413 & \num{56761.62} &     58&4&11&2 &  $-$88&8&13&9 & CAR    & 20221106 & \num{59890.40} & $-$138&6& 6&1 &    119&4& 3&0 \\
CAF\'E & 20140414 & \num{56762.61} &     58&4&31&6 & $-$105&7&19&7 & FIES   & 20221106 & \num{59890.42} & $-$134&5& 3&7 &    119&4& 3&4 \\
CAF\'E & 20140805 & \num{56875.51} &  $-$58&4& 4&2 &     32&8& 6&9 & FIES   & 20221107 & \num{59891.30} & $-$152&6& 4&8 &    135&9& 4&8 \\
CAF\'E & 20140806 & \num{56876.37} &  $-$80&9& 7&0 &     72&9& 6&7 & FIES   & 20221108 & \num{59892.35} & $-$165&9& 5&0 &    149&2& 2&4 \\
CAF\'E & 20140807 & \num{56877.45} & $-$108&2&12&5 &    101&5& 8&8 & FIES   & 20221109 & \num{59893.45} & $-$163&1& 2&7 &    154&1& 2&0 \\
CAF\'E & 20140814 & \num{56884.38} & $-$136&2&11&0 &    134&9& 5&5 & FIES   & 20221111 & \num{59895.34} & $-$152&4& 2&7 &    150&5& 2&6 \\
CAF\'E & 20140816 & \num{56886.53} & $-$121&6& 7&3 &    118&6& 5&7 & CAR    & 20231211 & \num{60290.32} & $-$108&3&14&2 &     85&3& 9&2 \\
CAF\'E & 20140818 & \num{56888.54} & $-$115&2&12&6 &     89&3&10&2 & HAR-N  & 20240720 & \num{60512.65} &      0&9& 8&5 &  $-$21&1& 7&7 \\
Merc   & 20150529 & \num{57172.72} & $-$167&3& 7&0 &    145&1& 9&3 & Merc   & 20251013 & \num{60962.39} & $-$165&8& 4&3 &    154&7& 3&5 \\
Merc   & 20150531 & \num{57174.72} & $-$160&4& 9&5 &    135&0& 9&9 & Merc   & 20251018 & \num{60967.44} & $-$124&4& 3&8 &    110&9& 4&3 \\
CAF\'E & 20150818 & \num{57253.56} &     64&6& 6&5 &  $-$90&6&16&1 & Merc   & 20251019 & \num{60968.38} & $-$122&1& 4&6 &    103&6& 3&2 \\
CAF\'E & 20150828 & \num{57263.56} & \mciv{---}    & \mciv{---}    & Merc   & 20251020 & \num{60969.49} & $-$105&3& 5&1 &     89&9& 7&1 \\
CAF\'E & 20150831 & \num{57266.51} & $-$156&6& 7&6 &    124&9& 7&7 & Merc   & 20251021 & \num{60970.41} & $-$106&3& 4&5 &     74&0& 9&6 \\
CAF\'E & 20160327 & \num{57475.70} &  $-$78&9& 8&8 &     61&2&13&4 & Merc   & 20251022 & \num{60971.49} &  $-$96&0& 4&6 &     71&1&14&2 \\
\midrule
\end{tabular}
}
\addtolength{\tabcolsep}{1.5mm}
\label{epochs}
\end{table*}

\end{appendix}

\end{document}